\documentclass[longauth]{aa}  

\usepackage{graphicx}
\usepackage{txfonts,xcolor}
\usepackage{multirow}

\begin{document} 

   \title{The GRAVITY Young Stellar Object survey}
   \subtitle{VIII. Gas and dust faint inner rings in the hybrid disk of HD141569}
   
   \author{GRAVITY Collaboration\thanks{GRAVITY is developed in a collaboration by the Max Planck Institute for Extraterrestrial Physics, LESIA of Paris Observatory and IPAG of Université Grenoble Alpes / CNRS, the Max Planck Institute for Astronomy, the University of Cologne, the Centro de Astrofísica e Gravitação and the European Southern Observatory.}:
          V. Ganci\inst{1,2}
          \and L. Labadie\inst{1}
          \and L. Klarmann\inst{3}
          \and A. de Valon\inst{4}
          \and K. Perraut\inst{4}
          \and M. Benisty\inst{4,5}
          \and W. Brandner\inst{3}
          \and A. Caratti o Garatti\inst{3,6,7,16}
          \and C. Dougados\inst{4}
          \and F. Eupen\inst{1}
          \and R. Garcia Lopez\inst{3,6,7}
          \and R. Grellmann\inst{1}
          \and J. Sanchez-Bermudez\inst{3,8}
          \and A. Wojtczak\inst{1}
          \and P. Garcia\inst{9,10}
          \and A. Amorim\inst{9,11} 
          \and M. Baub\"ock\inst{12}
          \and J.-P. Berger\inst{4}  
          \and P. Caselli\inst{12} 
          \and Y. Cl\'enet\inst{13} 
          \and V. Coud\'e du Foresto\inst{13} 
          \and P.T. de Zeeuw\inst{12,14} 
          \and A. Drescher\inst{12} 
          \and G. Duvert\inst{4} 
          \and A. Eckart\inst{1,2} 
          \and F. Eisenhauer\inst{12}  
          \and M. Filho\inst{9,10}
          \and F. Gao\inst{12} 
          \and E. Gendron\inst{13} 
          \and R. Genzel\inst{12} 
          \and S. Gillessen\inst{12}
          \and G. Heissel\inst{13} 
          \and T. Henning\inst{3}
          \and S. Hippler\inst{3}
          \and M. Horrobin\inst{1} 
          \and Z. Hubert\inst{4} 
          \and A. Jim\'enez-Rosales\inst{12}
          \and L. Jocou\inst{4}
          \and P. Kervella\inst{13} 
          \and S. Lacour\inst{13} 
          \and V. Lapeyr\`ere\inst{13} 
          \and J.-B. Le Bouquin\inst{4}
          \and P. L\'ena\inst{13}
          \and T. Ott\inst{12}
          \and T. Paumard\inst{13} 
          \and G. Perrin\inst{13} 
          \and O. Pfuhl\inst{15}
          \and G. Hei{\ss}el\inst{13} 
          \and G. Rousset\inst{13}
          \and S. Scheithauer\inst{3}
          \and J. Shangguan\inst{12} 
          \and T. Shimizu\inst{12} 
          \and J. Stadler\inst{12} 
          \and O. Straub\inst{12} 
          \and C. Straubmeier\inst{1} 
          \and E. Sturm\inst{12}
          \and E. van Dishoeck\inst{12,14}
          \and F. Vincent\inst{13}
          \and S.D. von Fellenberg\inst{12} 
          \and F. Widmann\inst{12} 
          \and J. Woillez\inst{15}
}

    \institute{I. Physikalisches Institut, Universit\"at zu K\"oln, Z\"ulpicher Str.              77, 50937, K\"oln, Germany\\
    \email{ganci@ph1.uni-koeln.de}
    \and
    Max-Planck-Institute for Radio Astronomy, Auf dem H\"ugel 69, 53121 Bonn, Germany
    \and
    Max Planck Institute for Astronomy, K\"onigstuhl 17, 69117 Heidelberg, Germany
    \and
    Univ. Grenoble Alpes, CNRS, IPAG, 38000 Grenoble, France
    \and 
    Unidad Mixta Internacional Franco-Chilena de Astronomía (CNRS, UMI 3386), Departamento de Astronomía, Universidad de Chile, Camino El Observatorio 1515, Las Condes, Santiago, Chile
    \and 
    Dublin Institute for Advanced Studies, 31 Fitzwilliam Place, D02 XF86 Dublin, Ireland
    \and
    School of Physics, University College Dublin, Belfield, Dublin 4,Ireland
    \and
    Instituto de Astronom\'ia, Universidad Nacional Aut\'onoma de M\'exico, Apdo. Postal 70264, Ciudad de M\'exico, 04510, M\'exico
    \and
    CENTRA, Centro de Astrof\'isica e Gravita\c c\~{a}o, Instituto Superior T\'ecnico, Avenida Rovisco Pais 1, 1049 Lisboa, Portugal
    \and
    Universidade do Porto, Faculdade de Engenharia, Rua Dr. Roberto Frias, 4200-465 Porto, Portugal
    \and
    Universidade de Lisboa - Faculdade de Ci\^{e}ncias, Campo Grande, 1749-016 Lisboa, Portugal
    \and
    Max Planck Institute for Extraterrestrial Physics, Giessenbachstrasse, 85741 Garching bei M\"unchen, Germany
    \and
    LESIA, Observatoire de Paris, PSL Research University, CNRS, Sorbonne Universit\'es, UPMC Univ. Paris 06, Univ. Paris Diderot,Sorbonne Paris Cit\'e, France
    \and
    Sterrewacht Leiden, Leiden University, Postbus 9513, 2300 RA Leiden, The Netherlands
    \and
    European Southern Observatory, Karl-Schwarzschild-Str. 2, 85748, Garching bei M\"unchen, Germany
    \and
    INAF -- Osservatorio Astronomico di Capodimonte, via Moiariello 16, 80131 Napoli, Italy
    }

\date{Received xx, xxxx; accepted xx, xxxx}

% \abstract{}{}{}{}{} 
% 5 {} token are mandatory
 
  \abstract
   {The formation and evolution of planetary systems impact the evolution of the primordial accretion disk in its dust and gas content. HD\,141569 is a peculiar object in this context as it is the only known pre-main sequence star characterized by a  hybrid disk. Observations with 8\,m class telescopes probed the outer-disk structure showing a complex system of multiple rings and outer spirals. Furthermore, interferometric observations attempted to characterize its inner 5\,au region, but derived limited constraints. 
   %HD\,141569  is  the  only  known  pre-main  sequence  star  characterized  by  a  hybrid  disk. Previous observations probed the outer disk features showing a complex system of multiple rings and outer spirals. There are no spatially resolved observations of the inner region out to 5 au.
   }
   {
   The goal of this work was  to explore with new high-resolution interferometric observations the geometry, properties, and dynamics of the dust and gas in the internal regions of HD\,141569.
%   The goal of this work is to exploit new interferometric observations at the milliarcsecond-scale with GRAVITY/VLTI in the near-IR to % 
   %to reveal the geometry and dynamics of HD\,141569 internal structure.
   }
   {
   We observed HD\,141569 on milliarcsecond scales with GRAVITY/VLTI in the near-infrared (IR) at low ($R\sim$20) and high ($R\sim$4000) spectral resolution. 
   % We modelled HD\,141569 data obtained with the GRAVITY/VLTI instrument at the Very Large Telescope Interferometer operating in the K-band, with 
   %a spatial resolution down to 1.7 mas 
   %milli-arcsecond-scale spatial resolution in the continuum and at a spectral resolution of 4000. 
   We interpreted the interferometric visibilities and spectral energy distribution with geometrical models and through radiative transfer techniques using the code MCMax to constrain the dust emission. We analyzed the high spectral resolution quantities (visibilities and differential phases) to investigate the properties of the Brackett-$\gamma$ (Br$\gamma$) line emitting region.}
   {Thanks to the combination of three different epochs, GRAVITY resolves the inner dusty disk in the K band with squared visibilities down to $V^2$\,$\sim$\,0.8. 
   A differential phase signal is also detected in the region of the Br$\gamma$ line along most of the six baselines.
   %for one single epoch. 
   %The low spectral resolution data show that the system was partially resolved for the first time. 
   Data modeling shows that an IR excess of about 6\% is spatially resolved 
   %in the K\,band,
   %We reproduce the observations at best by considering 
    and that the origin of this emission is confined in a ring of material located at a radius of $\sim$1\,au from the star with a width $\lesssim \,$0.3\,au. The MCMax modeling suggests that this emission could originate from a small amount ($1.4\times 10^{-8}\,$M$_{\oplus}$) of quantum-heated particles, while large silicate grain models cannot reproduce at the same time the observational constraints on the properties of near-IR and mid-IR fluxes. 
   %on keeping low the MIR emission while having a NIR excess consistent with the GRAVITY observations. 
   The high spectral resolution differential phases in the Br$\gamma$ line clearly show an S-shape that can be best reproduced with a gaseous disk in Keplerian rotation, 
   %co-aligned with the continuum emission, 
   confined within 0.09\,au (or 12.9\,R$_\star$). This is also hinted at by the double-peaked Br$\gamma$ emission line shape, known from previous observations and confirmed by GRAVITY. The modeling of the continuum and gas emission shows that the inclination and position angle of these two components are consistent with a system showing relatively coplanar rings on all scales.}
   %The 0\,km/s signal of the pure-line differential phases suggests an ongoing phenomenon that produces a geometrical asymmetry on the gas region.}
   {With a new and unique observational dataset on HD\,141569, we show that the complex disk of this source is composed of a multitude of rings on all scales. This aspect makes HD\,141569 a potentially unique source to investigate planet formation and disk evolution in intermediate-mass pre-main sequence stars.}


   \maketitle
%
%-------------------------------------------------------------------
\section{Introduction}
\label{sec:Int}

The formation and evolution of protoplanetary disks are directly linked to planet formation. The outer disk features of young stellar objects (YSOs) have been thoroughly studied in the past through scattered light imaging (e.g., with SPHERE/VLT, \citealt{Beuzit2019}) and with ALMA in the (sub-)millimeter range \citep{ALMA2015}. With both techniques, disk observations have shown rings, gaps, and asymmetric structures up to a few hundred au \citep[e.g.,][]{Lodato2019, Benisty2017}. The inner regions (at $\sim$au scale) of such disks are also of prime interest since key processes like gas accretion flows, winds, outflows, and dust sublimation take place. All these processes affect the dynamics and evolution of the first few au regions where terrestrial planets may form and/or migrate over few million years. Constraints on these processes can be derived indirectly through spectroscopic studies, but at typical distances of a few hundred parsecs only observations with milliarcsecond (mas) resolution, which are required to probe sub-au scales, can discriminate between competing models. 
%Therefore, it is particularly interesting to use infrared long baseline interferometry to perform such observations. 
Numerous interferometric studies have been conducted in the past in the near- and the mid-infrared (IR), for instance with IOTA \citep{MillanGabet2001}, the Palomar Testbed Interferometer \citep[PTI;][]{Eisner2004}, the Keck Interferometer \citep{Monnier2005,Eisner2014}, and the VLTI \citep{Menu2015,Lazareff2017,Perraut2019}.
%through observations with instruments such as AMBER, MIDI, or PIONIER combining up to four telescopes \citep{Lebouquin2011}. 
To date, the new generation of  four-telescope instruments including GRAVITY, operating in the K band \citep{Gravity2017}, and MATISSE, operating in L to N band \citep{Lopez2014}, are pushing further the achievable spectral coverage, the sensitivity, and precision of interferometric measurements. 
%which opens new perspectives in the field of protoplanetary disks.
Even though statistical studies of large YSO samples are of high relevance \citep{Lazareff2017,Perraut2019}, some of these objects require a more in-depth study enabled by the improved data quality of recent interferometric observations. Some objects, like HD\,141569, are unusual in their evolutionary sequence and require dedicated studies. For such systems the exact nature and properties of the very inner regions are still a matter of debate and can be solved in part by modeling the distribution of the warm dust traced in the K band. Furthermore, it is still unclear how to characterize the star--disk mechanisms traced by emission lines like the hydrogen Brackett-$\gamma$ (Br$\gamma \, \sim \,$8000-10000 K) and CO (T$\, \sim \,$2000-3000 K) bandheads, and where this emission occurs. The high-quality spectroscopic capabilities of the GRAVITY instrument allows us to study in detail the gas phase and its spatial morphology thanks to the interferometric visibilities and differential phase signals \citep{Gravity2020IRS2,Gravity2020TWH, Gravity2021-51Oph}.\\
\noindent HD\,141569 is a Herbig star classified as a B9-A0 spectral type \citep{Augereau2004}, with an effective temperature of $9750 \pm 250$\,K, an estimated age of $7.2 \pm 0.02$\,Myr, a luminosity between $16.60 \pm 1.07\,$L$_\odot$ \citep{Vioque2018} and $27.0 \pm 3.6\,$L$_\odot$ \citep{DiFolco2020}, a mass of $2.14 \pm 0.01$\,M$_{\odot}$, and a GAIA distance of $110 \pm 1$\,pc \citep{Arun2019}\footnote{The recent distance estimate in the new EDR3 of Gaia suggests 111.6$\pm$0.4\,pc. Considering the very close value to DR2, we decided to keep the former distance throughout the paper.}. It is a non-flaring disk system with little mid-IR excess classified as a group-II source \citep{Meus2001}. It is the only known pre-main sequence star characterized by a  hybrid disk \citep{Wyatt2015, Pericaud2017, DiFolco2020}, an evolutionary disk state between the protoplanetary and debris-disk regimes. 
%Previous observations from optical to (sub-)millimetric wavelength probed in detail the outer disk features. 
Near-IR imaging spatially 
% By imaging near-infrared (near-IR) scattered light using HST-NICMOS, \cite{Augereau1999} and \cite{Weinberger1999} 
resolved an optically thin disk consisting of two rings located at about $\sim 280$ and $\sim 455$\,au from the star \cite{Augereau1999,Weinberger1999,Biller2015}. 
%The same rings were seen by \cite{Biller2015} with Gemini/NICI. 
A more complex system is shown in the visible, consisting of multiple rings and outer spirals that could be explained through perturbations by two nearby ($\sim 7.5$\,arcsec) M dwarfs, or by planetary perturbations 
%may also be required 
\citep{Augereau2004, Wyatt2005, Reche2009}. 
%Exploring the inner 100\,au region of the HD\,141569 disk, 
\cite{Fisher2000} found a warm disk component up to 110\,au at 10.8 and $18.2\, \mu$m, %, using OSCIR on the Keck observatory, 
later confirmed by \cite{Marsh2002}. 
The short-wavelength counterpart of this component was detected by \cite{Mawet2017} through L$^{\prime}$ imaging and ranging between 20 and 85\,au. 
Emission at $8.6 \, \mu$m 
%overlapping the above disk 
was detected by \cite{Thi2014}, and was interpreted as emission from polycyclic aromatic hydrocarbons (PAHs).  NOEMA and ALMA observations in the millimeter range showed continuum emission equally shared between a compact ($\lesssim 50$\,au) and a smooth extended dust component ($\sim 350$\,au), with large millimeter grains dominating the inner regions and smaller grains in the outer ones \citep{DiFolco2020}.  %Another disk component was found by \cite{Konishi2016} at 60-95\,au by analysing an optical scattered light HST/STIS image. 
%This component was shown by \cite{Mawet2017} to not be the scattered light counterpart of the one found by \cite{Fisher2000}, \cite{Marsh2002} and \cite{Mawet2017}, but rather a distinct fourth disk component and dust population. 
Finally, inner disk features were detected by SPHERE in the Y, J, H, and K bands \citep{Perrot2016} and by Keck/NIRC2 in the L' band \citep{Currie2016} at physical separations of 45, 61, and 88\,au. These results point out   the high morphological complexity of the outer disk in the HD\,141569 system.\\
\noindent Little is known about the central astronomical units of the system. The spectral energy distribution (SED) of HD\,141569 alone does not help us in this sense since the IR excess is very small (see Fig.~9 of \citealt{Thi2014}). The majority of the K-band measurements  listed in Table \ref{tab:K-flux} reflect a featureless SED in the near-IR. 
%except for one 2MASS measurement that appears as outlier. 
Moreover, pure SED fits obtained by different authors \citep{Li2003, Merin2004, Thi2014} may suggest at first that the near-IR emission is exclusively photospheric in nature and 
% t the star emits nearly all the NIR radiation and 
that the disk contributes only at longer wavelengths.
The object was observed at milliarcsecond resolution in the K band with the PTI and the Keck interferometer, 
%(KI, $\lambda_0 = 2.2\, \mu$m, $\Delta \lambda = 0.4\, \mu$m), 
but it was spatially unresolved 
%unable to spatially resolve the near-IR emission 
\citep{Eisner2004, Eisner2009}. 
%From such measurements, 
\cite{Monnier2005} derived a 10\,mas upper limit in radius for the spatial extension of the K-band emission. 
Therefore, trustworthy information on the first $5\,$au of the system are scarce, and the question arises of whether the inner region of the disk could be already in a debris disk stage, where the SED fits in the near-IR are not accurate enough to detect such a faint excess. \\
%However, the disk-averaged gas-to-dust ratio is estimated to be $>100$ \citep{DiFolco2020}, larger than the initial interstellar value. \\
The circumstellar gas 
% around the star was 
has been observed in both atomic and molecular form, which suggests the system has not yet reached the gas-depleted stage characteristic of a debris disk system. 
\cite{Mendigutia2017} 
%, by analysing HD\,141569 through H$\alpha$ spectro-interferometry with a spectral resolution of 6000, 
set an upper limit of $\sim 0.11$\,au for the gas region responsible for the spatially unresolved  double-peaked H$\alpha$ emission. 
%, since the double peaked line shown by the object is spectrally resolved but the emitting region is spatially unresolved. 
A comparable upper limit of 
% , the unresolved K band emission observed by the KI sets an upper limit of 
$\sim 0.13$\,au for the Br$\gamma$ line emitting region is suggested by 
%gas region 
\cite{Eisner2009}. Both lines are observed to be not variable over timescales of days and years \citep{Eisner2015, Mendigutia2011a}. In addition to hydrogen, CO ro-vibrational emission ($v \geq 1$, $\Delta\, v = 1$)  was observed by many authors extending from 10 to 275\,au \citep{Dent2005, Goto2006, Brittain2007, Flaherty2016, White2016, Miley2018, DiFolco2020}. \\
% Again, what happens inside the innnermost astronomical units remains unclear. 
We present here the first GRAVITY interferometric observations of this disk, with  the goal %of our data analysis is 
of revealing the geometry and dynamics of the internal structure of HD\,141569, and of gaining insights about the dust and gas properties. Section\,\ref{sec:Observations} describes the observations; Section\,\ref{sec:Data} and Section\,\ref{sec:Methodology} present the observational data and the adopted methodology; 
% we take a first look at the data; Sections \ref{sec:Methodology} presents the methodology; 
Section\,\ref{sec:Results} describes the results of the possible scenarios along with the corresponding modeling; A discussion is developed in Section\,\ref{sec:Discussion}.
% we discuss the results; and Section \ref{sec:Summary} summarizes the main conclusions.

\begin{figure*}[!htbp]
\centering
\includegraphics[width=\textwidth]{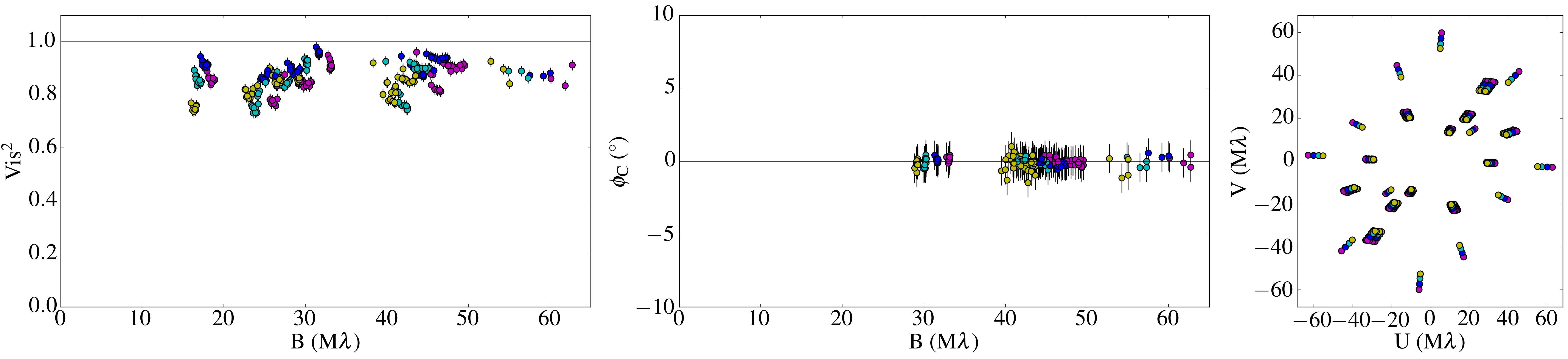}
\caption{HD\,141569 FT data; squared visibilities (left panel), closure phases (central panel), and U-V plane  coverage (right panel), from all the observation epochs. Colors refer to the different GRAVITY spectral channels.}
\label{fig:FT-Data}
\end{figure*}

\section{Observations}
\label{sec:Observations}

% Observations of 
HD\,141569 was observed with VLTI/GRAVITY \citep{Gravity2017}
%, the first second generation instrument for the VLTI \citep{Merand2014} 
% at ESO’s Paranal observatory 
using the four 1.8\,m Auxiliary Telescopes (ATs), 
on March 18 and July 12, 2019, 
% with the four 1.8 m Auxiliary Telescopes (ATs) 
in the intermediate D0-G2-J3-K0 configuration, and on May 23,  2019, in the large A0-G1-J2-J3 configuration. The observations span a spatial frequency range between about 15 M$\lambda$ and 65 M$\lambda$ (see Fig.~\ref{fig:FT-Data}, left panel) with a maximum angular resolution of $\lambda/2B$, about 1.7 mas for the longest baseline (B) of 130\,m, which corresponds to about 0.19\,au at a distance of 110\,pc.
The data consist in high spectral resolution (R$\, \sim \, $4000) observables recorded by the science channel (SC) detector over the whole K band with individual integration times of 30\,s 
% \ll{why 10 to 30s? I think our individual frames are only 30s IT.}
and in low spectral resolution (R$\, \sim \, $20) observables recorded by the fringe tracker (FT) detector (five spectral channels over the K band at 1.908, 2.058, 2.153, 2.256, 2.336\,$\mu$m) at frame rates of $\approx$\,300 and $\approx$\,900 Hz \citep{Lacour2019}.
% \ll{where the FT data acquired with 0.85ms or 3ms IT? Put down the right value.}. \\
Each observation block 
%file 
corresponds to 5 minutes on the object. 
In total, three files were acquired in March 2019, one in May 2019, and eight in July 2019.
HD\,141569 observations were preceded by the observation of a point-source calibration star, 
close to our object on the sky, and of 
similar spectral type and brightness 
in order to calibrate the atmospheric and instrumental transfer function. We observed the calibrator HD\,137006 in March and July 2019, and the calibrator HD\,141977 in May 2019. Further details on the executed runs and observing conditions are given in Table ~\ref{tab:ObsLog}. \\

\section{Data}
\label{sec:Data}

All the data were reduced and calibrated using the GRAVITY data reduction software \citep{Lapeyrere2014}.
For the low-resolution FT data we discarded the first spectral channel, which can typically be affected by the metrology laser operating at 1.908\,$\mu$m. Figure~\ref{fig:FT-Data} shows the U-V plane coverage and the FT calibrated squared visibilities and closure phases (right, left, and center panel, respectively). Following \cite{Perraut2019}, we applied a floor value on the error bars of 2\% for the squared visibilities and 1$^{\circ}$ on the closure phases as the error bars computed by the pipeline might be underestimated or correlated. We observe that GRAVITY partially resolved the near-IR emission in HD\,141569 with squared visibilities between 
% down to 
0.8 and 1.0. Therefore, the data can be used to estimate the characteristic size of the dust environment (see Section \ref{sec:FT-Results}). Moreover, 
with the inclusion of the May 2019 large configuration data, 
we observe that the visibility reaches a plateau at almost all spatial frequencies,
allowing us to constrain near-IR flux contributions of the star and environment.

\begin{figure}[!htbp]
\centering
\includegraphics[width=\columnwidth]{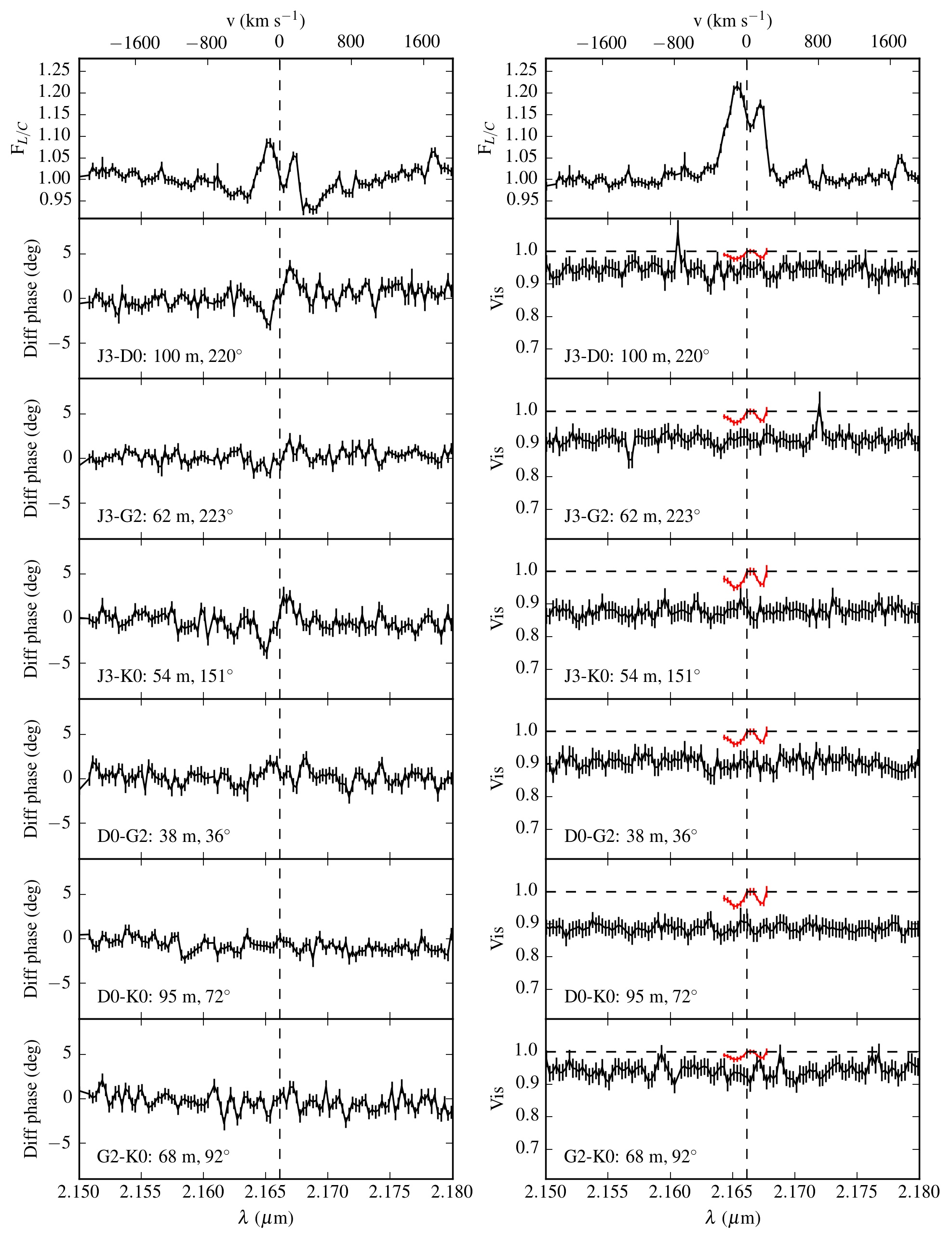}
\caption{
% HD\,141569 July 
Science channel data of July 2019. Top left: Wavelength-calibrated continuum-normalized spectrum, corrected for telluric lines. Top right: Same as top left, but corrected for the Br$\gamma$ photospheric absorption. Differential phases (left column) and visibilities (right column) along the six GRAVITY baselines. The red lines in the visibility plots show the pure-line visibilities.}
\label{fig:SC-Data}
\end{figure}

Since the closure phases are consistent with 0$^{\circ}$ at all baselines and for all the epochs, we can confidently consider the emission to be centro-symmetric on the spatial scale of our observations, and we therefore discard the hypothesis of a close companion as the origin of the resolved emission, at least within the 250\,mas ($\sim$28\,au) field of view of GRAVITY with the ATs. 
%of the GRAVITY ATs is 250\,mas, which translates to a linear field of view of 27.5\,au, 
% the type of companion we can exclude is one having an orbit solely inside this region.
\\
For the high-resolution SC data, we concentrated on the July 2019 dataset only since this is the epoch where we gathered the highest number of files. In order to optimally exploit the SC dataset, the eight files from July 2019 have been merged in order to increase the signal-to-noise ratio per spectral channel. Considering that the maximum span in position angle between the two extreme positions of the UV coverage is only $\sim$4\,$^{\circ}$, we do not expect any visibility smearing of the data due to differences in hour angles. The error bars were computed as the standard deviation between the eight files of the corresponding differential quantity (visibility or differential phase) in each spectral channel. For instance, we derived the absolute error on the visibility to about 1.8\%. Figure~\ref{fig:SC-Data} shows the visibility amplitude (left panels) and the differential phase (right panels) for the six baselines in the region of interest of the Br$\gamma$ line,  between 2.15 and 2.18\,$\mu$m. The top panels show the object spectrum normalized to the continuum, corrected for telluric lines (left plot), and corrected for both telluric lines and photospheric absorption (right plot). 
The visibilities appear to be spectrally flat with no clear signature at 2.16612\,$\mu$m. They are measured to vary between 0.9 and 0.96 as a function of the baseline, which is indicative of a compact region well inside the dusty disk. 
Interestingly, the differential phase signal is more marked at the position of the Br$\gamma$ line. % , but the continuum contribution has to be taken out.
We observe a clear S-shaped signal through baselines J3-D0 (100.1\,m, 220$^{\circ}$) and J3-K0 (53.6\,m, 152$^{\circ}$), and a weaker signal through baselines J3-G2 (62.1\,m, 223$^{\circ}$) and D0-G2 (38.2\,m, 36$^{\circ}$). No clear signal is detected within the error bars through the baselines D0-K0 (95.1\,m, 72$^{\circ}$) and G2-K0 (68.1\,m, 92$^{\circ}$).\\
%--------------------------------------------------------------------------------------------
In order to study the  Br$\gamma$ line gas region using the SC high spectral resolution data, 
% the SC data and to get information on the Br$\gamma$ line gas region, 
we need precise measurements for the line-to-continuum flux ratio. It is essential to perform a proper wavelength calibration of the spectrum and to take out the contribution by the telluric lines. We describe the whole procedure in Appendix \ref{apx:spctr-corr}. Errors are from the original data, reduced and calibrated through the GRAVITY data reduction software.\\
The high-resolution GRAVITY spectrum of HD\,141569 (see top panels of Fig.~\ref{fig:SC-Data}) shows a double-peaked Br$\gamma$ emission line.
%centered at 2.1665\,$\mu$m as already reported in \cite{Garcia2006} and \cite{Brittain2007}. 
Since the error associated with the GRAVITY wavelength calibration is $\sim 3\,$\AA{}, the two peak positions can be considered to be symmetric  with respect to the Br$\gamma$ wavelength rest position.
Both the double-peaked emission line and the S-shaped  feature in the differential phase suggest a scenario where the gas emitting in the Br$\gamma$ line could be in Keplerian rotation. We  explore this hypothesis further in the following sections.

\section{Methodology}
\label{sec:Methodology}

The properties of the spatially resolved continuum emission were first  investigated with the help of  low spectral resolution FT data (see Sects.  \ref{sec:FT-Methodology} and \ref{sec:FT-Results}). The squared visibility curve was modeled through chromatic geometrical models accounting for a point-like central star, and simple geometrical shapes (Gaussian geometrically thin rings, Gaussian-convolved infinitesimally thin rings) representing the circumstellar environment. Useful information was obtained such as the star-to-dust flux ratio, the dust spectral index, and the spatial distribution of the dust. To gain further information on the dust emission properties, we investigated through radiative transfer (RT) modeling the impact of such  a component on the SED (see Section \ref{sec:MCMax}). We used the RT code MCMax \citep{Min2009}, which solves 2D RT (e.g., \citealt{Bjorkman2001}) to calculate the dust density and temperature structure of a given disk setup. In Sects. \ref{sec:SC-Methodology} and \ref{sec:SC-Results} we discuss the high-spectral resolution SC data in the Br$\gamma$ region used to constrain the spatial scale of the hot gas emitting component through modeling of the visibility curves, and through the analysis of the Br$\gamma$ spectrum under the assumption of a gas disk in Keplerian rotation. Further information on the gas region size and its dynamical properties were derived through the analysis of the differential phases and the resulting photocenters shifts. Finally, an analytical axisymmetric Keplerian disk model is compared to our observations.

\subsection{Dust continuum: low spectral resolution data}
\label{sec:FT-Methodology}
% \ll{add the following}\\
%The properties of the spatially resolved continuum emission are investigated with the help of the low spectral resolution FT data. 
Following the work of \cite{Lazareff2017} and \cite{Perraut2019}, we used geometric models that consist of a point-like central star, assumed to be unresolved at all observed baselines, and a circumstellar environment in order to fit the observed visibilities. The complex visibility of the system at spatial frequencies $(u,\varv)$ and at wavelength $\lambda$ is therefore described by a linear combination of the two components as  \begin{equation}\label{eq:Visibility}
    V(u,\varv,\lambda) = \frac{F_{\rm s}(\lambda)\,V_{\rm s}(u,\varv,\lambda) + F_{\rm c}(\lambda)\, V_{\rm c}(u,\varv,\lambda)}{F_{\rm s}(\lambda) + F_{\rm c}(\lambda)},
\end{equation}
where $V_{\rm c}$ is the visibility of the circumstellar environment, and  $F_{\rm s}$ and $F_{\rm c}$ the specific fractional flux contributions of the star and of 
the circumstellar environment, respectively. The visibility of the star $V_{\rm s}$ is equal to 1 since we assume it to be unresolved. Since our GRAVITY FT data contains six visibility measurements and four closure phases for each of the four 
% 5 
spectral channels 
% (\ll{do we have still 5 channels after flagging of the first spectral channel due to the laser?}) 
and for each file, we can derive the spectral dependence of the circumstellar environment by modeling it as a power law, defined by its spectral index $k_{\rm c}$, where $k = d \, \mathrm{log} \, F_{\lambda} / d \, \mathrm{log} \, \lambda$, and by describing the complex visibility of the system as 
\begin{equation}\label{eq:Visibility-k}
    V(u,\varv,\lambda) = \frac{F_{\rm s}\,(\lambda/\lambda_0)^{k_{\rm s}} + F_{\rm c}\,(\lambda/\lambda_0)^{k_{\rm c}}\,V_{\rm c}(u,\varv)}{F_{\rm s}\,(\lambda/\lambda_0)^{k_{\rm s}} + F_{\rm c}\,(\lambda/\lambda_0)^{k_{\rm c}}},
\end{equation}
where $\lambda_0 = 2.15 \, \mu$m is the wavelength of the central spectral channel of the FT, and $k_{\rm s}$ the spectral index of the star derived assuming that it radiates as a black body at the star effective temperature $T_{\mathrm{eff}}$=9750\,K,  
% T from Vioque et al. 2018 
which translates into s spectral index $k_{\rm s} = -3.62$ at $\lambda_0$ for the central star.\\
We chose to fit our visibility data with three different geometric models that differ only by the $V_{\rm c}$ term in Eq. \ref{eq:Visibility-k}. Since the closure phases are basically zero for every baseline, as shown in the central panel of Fig. \ref{fig:FT-Data}, we do not consider any azimuthal modulation in our models. Therefore, the resulting brightness distributions are centro-symmetric.\\
The first model consists of a Gaussian disk whose visibility is described, from \cite{Berger2007}, as 
\begin{equation}\label{eq:Visibility-Gauss}
    V_{\rm c}(u,\varv) = V_{\mathrm{gauss}}(u,\varv) = {\rm exp}\left( - \frac{(\pi \Theta r)^2}{4\, \mathrm{ln}2} \right),
\end{equation}
where $\Theta$ is the Gaussian full width at half maximum (FWHM) and $r = \sqrt{u^2 + \varv^2} =  B/\lambda$, 
with $u$ and $v$ the spatial frequency coordinates, and B the projected baseline. \\
The second model consists of a geometrically thin ring whose visibility is described by subtracting an inner smaller uniform disk from a larger one:
\begin{equation}\label{eq:Visibility-Ring-general}
    V_{\rm c}(u,\varv) = V_{\mathrm{ring}}(u,\varv) = \frac{F_{\rm outer}\, V_{\rm outer} - F_{\rm inner}\, V_{\rm inner}}{F_{\rm outer} - F_{\rm inner}}
.\end{equation}
Here the subscript \textit{outer} refers to the larger 
% bigger 
disk and the subscript \textit{inner} refers to the 
inner 
hole. 
After normalization 
% Following \cite{Berger2007} 
and knowing that $F_{\rm disk} = \pi \, D^2/4$, where $D$ is the diameter of the disk, we can express Eq.~\ref{eq:Visibility-Ring-general} as
\begin{equation}\label{eq:Visibility-Ring}
    V_{\mathrm{ring}}(u,\varv) = \frac{D_{\mathrm{outer}}\, 2\, \frac{J_1(\pi\, D_{\mathrm{outer}}\, r)}{\pi\, r} - D_{\mathrm{inner}}\, 2\, \frac{J_1(\pi\, D_{\mathrm{inner}}\, r)}{\pi\, r}}{D_{\mathrm{outer}}^2 - D_{\mathrm{inner}}^2},
\end{equation}
where $J_1$ is the first-order  Bessel function.
Finally, our last model consists of an infinitesimally thin ring convolved by a Gaussian whose visibility is described as the product between $V_{\mathrm{gauss}}(u,\varv)$ and
\begin{equation}\label{eq:Visibility-inf_Ring}
    V_{\mathrm{inf-ring}}(u,\varv) = J_0(2\pi a r)
,\end{equation}
where $J_0$ is the zero-order Bessel function and  $a$ is the radius of the infinitesimally thin ring \citep{Berger2007}. Since in the Fourier space the convolution of two functions is simply their multiplication, we have
\begin{equation}\label{eq:Visibility-GaussRing}
    V_{\rm c}(u,\varv) = V_{\mathrm{gauss-ring}}(u,\varv) = V_{\mathrm{gauss}}(u,\varv) \cdot V_{\mathrm{inf-ring}}(u,\varv).
\end{equation}
The inner hole radius and ring width are defined as $r_i = D_{\rm inner}/2$ and $w = (D_{\rm outer}-D_{\rm inner})/2$, respectively, for the geometrically thin ring model while they are defined as $r_i = a - \Theta/2$ and $w = \Theta$, respectively, for the Gaussian-convolved infinitesimally thin ring.
Inclination and position angle of the circumstellar environment are taken into account through the parameter $r = \sqrt{u^2 + \varv^2}$ following \cite{Berger2007}.
%As a final step, 
The model fitting is based on a Markov chain Monte Carlo  (MCMC, \citealt{Foreman-Mackey2013}) numerical approach 
% (\ll{reference needed here}) 
and was implemented on the combined
dataset 
% squared visibilities 
of all three epochs in order to maximize the number of experimental points against 
% have a statistically robust data set against 
the number of free parameters. This assumes that the near-IR emission and the disk structures are not variable over a five-month period, which
%do not vary. %This assumption 
is strengthened by the fact that the star is not variable, either spectroscopically in the optical \citep{Mendigutia2011} or photometrically in the mid-IR \citep{Kospal2012}. 
Once a global solution was identified, we further checked how well the parameters are 
% well 
indeed constrained by the data.  
% or not, 
For this purpose, we performed 
% made a 
a series of squared visibility fits to the 
% geometrically thin ring 
model by fixing the tested parameter to different values and left the other parameters free in the subsequent minimization. In this way we obtain a $\chi^2$ curve as a function of the tested parameter value.\\
%, as shown in Fig.~\ref{fig:GTRing-Chi2r_maps}.\\
% , gaining their $\chi^2$ maps (\ll{Maps? Short exchange needed here}). 
Finally, the error on the $\chi^2$ was estimated by treating the quantity T$_{\rm i}$ as a stochastic variable with
\begin{equation}
    \rm{T}_i = \rm{N}\, (\frac{\rm{y_i}-\rm{y_{model}}}{\rm{\sigma_i}})^2
,\end{equation}
where $\rm{N}$ is the number of points in the dataset, $\rm{y_i}$ the individual measurement, $\rm{y_{model}}$ the value of the model, and $\rm{\sigma_i}$ the error associated to the measurement. The $\chi^2$ value is given by the mean of T$_{\rm i}$,
\begin{equation}
    \chi^2 = \frac{1}{\rm{N}}\sum_{\rm{i}=1}^{\rm{N}}\rm{T_i} = \sum_{\rm{i}=1}^{\rm{N}}(\frac{\rm{y_i}-\rm{y_{model}}}{\rm{\sigma_i}})^2
,\end{equation}
and the $\chi^2$ error is given by the error on the mean of T$_{\rm i},$
\begin{equation}\label{chi2_err}
    \sigma_{\chi^2} = \sqrt{\frac{\sum_{\rm{i}=1}^{\rm{N}} \rm{T}_i^2}{\rm{N}(\rm{N}-1)}-\frac{(\chi^2)^2}{\rm{N}-1}}
.\end{equation}
% 
% The fitting procedure, based on a Monte-Carlo Markov Chain (MCMC) numerical approach, was done on the combined squared visibilities of all three epochs in order to have a statistically robust data set against the number of free parameters, assuming that the NIR emission and the disk structures do not vary. %This assumption is strengthened by the fact that the star is not variable, either spectroscopically in the optical \citep{Mendigutia2011}, or photometrically in the mid-IR \citep{Kospal2012}. 
This applies to the reduced $\chi^2$ as well. 
% Since the error bars on our data points can be underestimated by the pipeline and/or correlated, we apply floor error bars of 2$\%$ on the squared visibilities, following \cite{Perraut2019}.

\subsection{Gas: High spectral resolution data}
\label{sec:SC-Methodology}

To estimate the gas region size from the SC visibility we 
%need to 
extrapolated the pure-line contribution from the total visibility (line+continuum)  displayed in Fig. \ref{fig:SC-Data}. To do this we modeled the total visibility with a three-component model that accounts for the contributions from the star, the circumstellar dust, and the line emitting gas. The total visibility is given therefore by 
\begin{eqnarray}\label{eq:Visibility-3C}
    % \lefteqn{
    V_{\rm  tot} (u,\varv,\lambda) = \nonumber \\
    \frac{\alpha(\lambda)\,F_{\rm s}(\lambda)\,V_{\rm s}(u,\varv,\lambda) + F_{\rm  c}(\lambda)\, V_{\rm  c}(u,\varv,\lambda) + F_{\rm L}(\lambda)\, V_{\rm L}(u,\varv,\lambda)}{\alpha(\lambda)\,F_{\rm s}(\lambda) + F_{\rm  c}(\lambda) + F_{\rm L}(\lambda)
    % \nonumber && \\ 
    }
,\end{eqnarray}
where $\alpha(\lambda)$ is the science star continuum-normalized photospheric absorption (see Appendix \ref{apx:spctr-corr} for more details), the subscript $c$  refers to the dust 
% ring, 
component, and the subscript L  refers to the Br$\gamma$ line gas. From Eq. \ref{eq:Visibility-3C} it can be proven (see Appendix \ref{apx:PL-Vis}) that the pure-line visibility is given by
\begin{eqnarray}\label{eq:Visibility-PL}
    V_{\rm L}(u,\varv,\lambda) = \nonumber \\
    \frac{V_{\rm  tot}(u,\varv,\lambda)\,\{[1+\beta(\lambda)]\,F_{\rm L/C}-1-\beta(\lambda)\}-\alpha(\lambda)+1}{[1+\beta(\lambda)]\,F_{\rm L/C}-\alpha(\lambda)-\beta(\lambda)}
,\end{eqnarray}
where $F_{\rm L/C}$ is the line-to-continuum flux ratio
\begin{equation}\label{eq:F_L/C}
    F_{\rm L/C} = \frac{\alpha(\lambda)\,F_{\rm s}(\lambda) + F_{\rm  c}(\lambda) + F_{\rm L}(\lambda)}{F_{\rm s}(\lambda) + F_{\rm  c}(\lambda)}
\end{equation}
and $\beta(\lambda)$ is the disk-to-star flux ratio outside the line
\begin{equation}\label{eq:beta}
    \beta(\lambda) = \frac{F_{\rm  c}}{F_{\rm s}}(\lambda) = \frac{F_{\rm  c}}{F_{\rm s}}\, \left( \frac{\lambda}{\lambda_0} \right)^{k_{\rm c} - k_{\rm s}}.
\end{equation}
For clarity, we note that, outside the line emitting region, $\alpha(\lambda)$ and $F_{\rm L}(\lambda)$ tend  toward 1 and 0, respectively. 
Equation~\ref{eq:F_L/C} corresponds to the line-to-continuum ratio including the photospheric absorption. 
Finally, we estimated the gas region size by modeling it with an infinitesimal ring model given by Eq.~\ref{eq:Visibility-inf_Ring}. % (\ll{Here you should say a bit more about we why limit ourselves to a ring model}). 
\\
In the same way, we needed to take out the continuum contribution from the total differential phases (Fig. \ref{fig:SC-Data}). Following \cite{Weigelt2011}, the pure-line differential phase is given by
\begin{equation}\label{eq:Diff-phase}
\mathrm{sin}\,\phi_{\rm L} = \mathrm{sin}\,\phi\, \frac{F_{\rm  tot}\,V_{\rm  tot}}{F_{\rm L}\,V_{\rm L}} = \mathrm{sin}\,\phi\, \frac{V_{\rm  tot}}{V_{\rm L}}\, \frac{F_{\rm L/C}}{F_{\rm L/C}-(\frac{\alpha+\beta}{1+\beta})},
\end{equation}
where $\phi_{\rm L}$ is the pure-line differential phase, $\phi$ the total differential phase, $F_{\rm tot}$ the total flux (star, dust, and gas), and we write the ratio $F_{\rm tot}/F_{\rm L}$ through Eq. \ref{eq:Apx-FL-FT}. Following \cite{Bouquin2009}, we can derive wavelength-dependent photocenter displacements along each baseline from the 
%continuum-corrected 
pure-line differential phases by
\begin{equation}\label{eq:Photo-shift}
\mathbf{p} = \frac{- \phi_{\rm L}}{2 \pi} \cdot \frac{\lambda}{\mathbf{B}},
\end{equation}
where $\mathbf{p}$ is the projection on the baseline $\mathbf{B}$ of the 2D photocenter vector with origin on the central star.
The error bars on the pure-line differential quantities are computed through error propagation in Eq.~\ref{eq:Visibility-PL} and Eq.~\ref{eq:Diff-phase}.
% (\ll{I removed the subscript i in Eq. 16 as it kind of comes out from nowhere.})

\section{Results}
\label{sec:Results}
\subsection{Disk component inside 2\,au }\label{sec:FT-Results}
% -----------------------------------------
% -----------------------------------------
% -----------------------------------------
% -----------------------------------------
% -----------------------------------------
\begin{table*}[t]
\centering
\begin{small}
\centering
\caption{\centering FT squared visibility best-fit solution for the geometrically thin ring model and the Gaussian-convolved ring model.}
\begin{tabular}{lcclccc} 
\hline
\hline
Geometrically thin ring & & & Gaussian-convolved ring & & & \\
\hline
Parameter & Unit & Fit solution & Parameter & Unit & Fit solution & Scan range \\
\hline
$F_{\rm s}$& [\%]        & $93.81 \pm 0.07$         & $F_{\rm s}$& [\%]        & $93.81 \pm 0.07$         & [0 ; 100\%] \\
$F_{\rm c}$& [\%]        & $6.19  \pm 0.07$         & $F_{\rm c}$& [\%]        & $6.19  \pm 0.07$         & 1 - $F_s$   \\
$k_{\rm c}$&             & $-0.35 \pm 0.21$         & $k_{\rm c}$&             & $-0.35 \pm 0.21$         & [-4 ; 4]    \\
$r_i$      & [mas\,(au)] & $ 7.40 \pm 0.21$\,(0.81) & $r_i$      & [mas\,(au)] & $ 7.35 \pm 0.21$\,(0.81) & [0 ; 60]    \\
$w$        & [mas\,(au)] & $ 0.35 \pm 0.35$\,(0.04) & $w$        & [mas\,(au)] & $ 0.24 \pm 0.24$\,(0.03) & [0 ; 100]   \\
$i$        & [deg]       & $58.47 \pm 1.55$         & $i$        & [deg]       & $58.45 \pm 1.55$         & [30 ; 80]   \\
$PA$       & [deg]       & $-1.77 \pm 1.11$         & $PA$       & [deg]       & $-1.78 \pm 1.11$         & [-40 ; 40]  \\
$\chi_r^2$ &             & $4.67  \pm 0.33$         & $\chi_r^2$ &             & $4.67  \pm 0.33$         &             \\
\hline
\end{tabular}
\tablefoot{$F_{\rm s}$ is the stellar flux contribution, $F_{\rm c}$ the dusty circumstellar environment flux contribution, $k_{\rm c}$ the dust spectral index at 2.15 $\mu$m, $r_i$ the ring inner hole radius, $w$ the ring width ($r_i$ and $w$ defined in different ways for the two models, see Section \ref{sec:FT-Methodology}), $i$ the ring inclination from face-on, $PA$ the northeast position angle, and $\chi_r^2$ the reduced chi-square. The uncertainties on the fitted parameter correspond to the 1$\sigma$ error. Scan ranges refer to both models.
}
\label{tab:FT-fit-solutions}
\end{small}
\end{table*}
% -----------------------------------------
% -----------------------------------------
% -----------------------------------------
% -----------------------------------------
Of the three models discussed in Sect.~\ref{sec:FT-Methodology}, the fit of the squared visibilities to the Gaussian disk model (Eq.~\ref{eq:Visibility-Gauss}) did not converge to a solution (i.e., the marginal posterior distributions for the Gaussian width, inclination, and position angle are flat). Therefore, we 
% do not show the result and 
discarded this model in the rest of the work.
 Solutions were found for the geometrically thin ring model (Eq.~\ref{eq:Visibility-Ring}) and the Gaussian-convolved ring model (Eq.~\ref{eq:Visibility-GaussRing}). The two models converge basically toward the same solution, with a reduced $\chi_r^2 $= 4.7 for both models. \\
% \ll{Here we must add a sentence like: the minimization followed a larger parameter scan phase first, followed by a refinement around the identified global minimum.}\\
We performed a wide scan range of the fitted parameters 
% of Table~\ref{tab:FT-fit-solutions} 
to find convergence toward a global solution. The results of the minimization are presented in 
% shown in 
% Fig.~\ref{fig:Continuum results} and 
Fig. \ref{fig:Continuum results} and Table~\ref{tab:FT-fit-solutions} for the ring model along with the parameter scan range and the 1$\sigma$ uncertainties. The resulting MCMC posterior distribution is presented in Fig.~\ref{fig:Continuum results} (top plots) and allows us to identify an optimal global solution for the six parameters.
The fitting process leads to a photospheric near-IR flux contribution of $\sim93.8 \%$, and therefore to a dust ring flux contribution of $\sim6.2 \%$ for both models.
%consistent with the low-excess SED of the source shown in Fig. \ref{fig:SED_MCMax}. 
Interestingly, the degeneracy typically found between the disk's flux and the characteristic size in V$^{\rm 2}$ is broken here because the constant plateau as a function of spatial frequencies unambiguously determines the level of the disk's flux contribution. Both models predict a spectral index $k_{\rm c}$ for the dust ring with a value of $-0.35 \pm 0.2$.
To better illustrate the visibility plateau and the expected modulation due to the modeled thin ring, we show in the bottom plots of Fig. \ref{fig:Continuum results} three visibility curves corresponding to the best model for three selected baseline orientations and for a fixed wavelength value of 2.15\,$\mu$m.
%\footnote{$k_{\rm c}$ is the spectral index of the specific flux $d(\mathrm{log} \, F_{\lambda})/d(\mathrm{log} \, \lambda)$.} 
%meaning that the ring SED has a spectral index d\,$(\mathrm{log} \, \lambda F_{\lambda})/$d\,$(\mathrm{log} \, \lambda)$ of 0.75 at 2.15 $\mu$m.
Regarding the geometrical shape of the circumstellar environment, both models lead to a ring inclined from face-on by $58.5^{\circ}$, with a northeast position angle $PA\sim0^{\circ}$. 
% This orientation is comparable to the inclination , e.g. \cite{Mawet2017}
The inner hole radius is estimated to be $r_i \approx7.4$\,mas ($0.8\,$au) from both models, while the ring width is estimated to be $\sim 0.24-0.35$ mas ($0.03-0.04\,$au, for the Gaussian-convolved ring model and the geometrically thin ring model, respectively).\\
%within the error bars. 
%and the outer radius $r_o\approx7.75$ mas, i.e., a ring width of $\sim 0.35$ mas ($0.04\,$au).
Importantly, Fig.~\ref{fig:GTRing-Chi2r_maps} shows the $\chi_r^2$ curves of each parameter for the geometrically thin ring model, which helps us to evaluate how well each parameter is constrained by our data. The star near-IR flux contribution is very well constrained as we expected from the plateau seen in the squared visibility curve. 
% at the longest baselines. 
The ring % $2.15\, \mu$m 
spectral index is more loosely constrained, since the best value is consistent with values ranging between $-2$ and $2$. 
The ring inclination is constrained between $\sim$\,45$^{\circ}$ and $\sim$75$^{\circ}$, while the position angle is less well constrained with two possible minima at $\sim$\,0$^{\circ}$ and 120$^{\circ}$, the former considered to be the absolute minimum. 
The inner hole radius is constrained to be inside the first $2\,$au of the system, with a global minimum found around $0.8\,$au (7.4\,mas) and a second (almost equally possible) solution, at $\sim$1.7\,au (15.4\,mas). Taking into account the upper limit of 10\,mas for the radius of the K-band emission found by \cite{Monnier2005}, we decided to adopt $r_i$=0.8\,au. 
According to Fig.~\ref{fig:GTRing-Chi2r_maps}, the ring width $w$ tends toward small values, not larger than $\sim\,$0.3\,au. This is  discussed further in Sect.~\ref{sec:Discussion_continuum}.

% zero probably due to the very small ring flux contribution that does not let us obtained accurate information on the geometrical shape of the circumstellar environment.

\subsection{Dust properties through radiative transfer modeling}
\label{sec:MCMax}

% \ll{I need to better verify in this section if some information on the temperature of the dust is given. The radiative transfer delivers this information}\\
To strengthen the obtained results and to assess the scenario of an inner ring 
%prove the possible presence of dust 
as close as $0.8\,$au, gaining further information on the emission properties, we investigated through RT modeling the impact of such a component on the SED. 
We note  here that our  aim was not to perform a detailed 
% rigorous 
mineralogy study of the system, %therefore our goal is not a perfect fit of the SED 
but rather to understand 
% check 
how the detected inner dust % sas close as $0.8\,$au 
is consistent with both the near-IR flux and overall SED of the system. \\
We modeled the multiple and complex outer rings with only three rings based on
the results from \cite{Thi2014}. 
In their model, the lower limit particle size was set at 0.5\,$\mu$m and the upper one at 0.5-1\,cm for the two outermost rings and the innermost one, respectively. The grain size follows a distribution $\propto\,a^{-3.5}$, the surface density profile is a modified version from 
%a modified version of the surface density profile of 
\cite{Li2003}, and the flaring index is $\gamma = 1$.
The three rings peak at $\sim 15$, 185, and 300\,au, with the first two rings separated by a 75\,au gap.\\
Our initial disk setup consists of a disk structure similar to \cite{Thi2014}, but with updated stellar parameters (see Sect.~\ref{sec:Int}) and  a grain population based on DIANA standard dust grains \citep{Woitke2016} containing 75\%  amorphous silicates (e.g., Mg$_{0.7}$Fe$_{0.3}$SiO$_3$), 25\% porosity, and no amorphous carbon. 
Our modified grain size distribution and surface density is described in next paragraph.
The computed SEDs account for both thermal emission and scattered light contributions.\vspace{0.2cm}\\
{\bf A silicate dust ring: }
% From this point, 
First, we attempted to reproduce the near-IR excess detected by GRAVITY by including dust grains close to the star, %  flux emission of the dust detected by GRAVITY located 
%at $\sim$\,0.8\,au from the star according to our best-fit model,
but also taking into account the fact that HD\,141569 does not show a silicate emission feature at 10\,$\mu$m \citep{Seok2017}, which is in part connected to the grain size distribution. 
% One should not exclude the case of a ring composed only by large silicate grains which would have a quenched emission in the $10\, \mu$m wavelength range and still have a flux contribution in the NIR. 
We tested several models of the inner ring with the following properties: a varying 
% few models which differ in their 
lower-limit dust grain size of 0.6, 1.2, 2.5, 5.0, 10, 20, 40, 80, 158, and 316
%10, 20, 40, 80, 158, 316, 631, and 1259
\,$\mu$m; an equal upper-limit grain size of 1\,cm, with a size distribution $\propto$\,$a^{-3.5}$; a surface density $\propto$\,$r^{-1}$; and an inner ring radius fixed at 0.8\,au with a width of 
% , 
% and an outer radius tested at
% respectively, 1 and 2\,au
0.04\,au according to our best-fit model.
\\
% -------------------------------
All models with grains smaller than $\sim$\,20\,$\mu$m in the inner $\sim$\,1\,au region can be tuned to reproduce the $\sim$\,6\% near-IR excess, but at the same time they still exhibit a clear 10\,$\mu$m silicate feature, which is not in adequacy with the observations. \\
% -------------------------------
% ---- Proposed by Valerio -------
Models accounting only for grains larger than 40\,$\mu$m result in an almost complete quenching of the 10\,$\mu$m silicate feature. 
When testing the mass at $\sim$\,1\,au required not to exceed the mid-IR flux, we find that 10$^{-10}$\,M$_\odot$ (or 3.3$\times$10$^{-5}$\,M$_\oplus$) would be compliant with this condition, but only a $\sim$\,1\% near-IR excess is generated. On the other hand, the dust mass required to reach a $\sim$6\% near-IR excess is well beyond 8.5$\times 10^{-10}$M$_\odot$ (or 2.8$\times$10$^{-4}$\,M$_\oplus$), but this then produces a too large mid-IR emission inconsistent with the known SED. \\
Finally, decreasing the percentage of silicates and increasing the carbon percentage in the dust grains up to 25\,\% did not improve the SED fit.
% (shown in Fig. D.1)
% -------------------------------
% For the tested model involving an inner ring made of silicate dust, a possible solution involves the presence of dust grains with a minimum size of $\sim$40\,$\mu$m at 0.8\,au and a total dust mass of 3.5$\times$10$^{-10}$\,$M_\odot$ (or 1.2$\times$10$^{-4}$\,$M_\oplus$). This would produce a 5.7\% excess at 2.2\,$\mu$m and a featureless mid-infrared excess compatible with the total SED. 
Obtaining a more precise fit to the global SED would require further analysis and tuning of the outer ring contribution in the mid-IR, which is beyond the immediate goal of the paper. However, our modeling seems to point out that it is difficult to reconcile the level of near- and mid-IR excess reported with a model of solely silicate dust in the disk ring at $\sim$\,1\,au. 
Three representative cases of our modeling are presented in Fig.~\ref{fig:MCMax-Sil-model}.
% -------
% -------
\begin{table*}[t]
\centering
\caption{\centering List of parameters relevant to the RT modeling of the HD\,141569 disk. This model only describes the case of a QHP-dominated inner ring.}
\begin{tabular}{llcccc} 
\hline
\hline
Star parameters & & & & & \\
\hline
Distance        & [pc]          & 110   & & &\\
Temperature     & [K]           & 9750  & & &\\
Luminosity      & [L$_{\odot}$] & 19 & (see Table notes) & &\\
%Radius         & [R$_{\odot}$] & 1.6   & & &\\
A$_{\rm V}$     & [mag]         & 0.095 & & &\\
\hline
\hline
\multicolumn{6}{l}{Rings parameters}\\
\hline
Parameter & Unit & 1$^{st}$ & 2$^{nd}$ & 3$^{rd}$ & 4$^{th}$ \\
\hline
Inner rim                        & [au]           & 0.8  & 5    & 185 & 300 \\
Outer rim                        & [au]           & 0.84 & 110  & 500 & 500 \\
Particle type                        &                & QHP & Silicate & Silicate & Silicate\\
Dust mass                        & [M$_{\oplus}$] & $1.4\times 10^{-8}$ & 0.020 & 1.665 & 0.233 \\
Smallest particle size           & [$\mu$m]       & 0.006 & 1.26 & 0.63 & 20 \\
Biggest particle size            & [$\mu$m]       & 0.006 & 10000 & 10000 & 10000 \\
Inner surface density power 
law  &                & -1 & -1 & -1 & -1 \\
Size distribution power law      &                & -3.5 & -3.5 & -3.5 & -3.5 \\
Scale-height at 100\,au          & [au]           & 1.4 & 1.4 & 7.5 & 7.5 \\
Ring inclination                 & [degree]       & 59 & 59 & 59 & 59 \\
\hline
\end{tabular}
\tablefoot{The stellar luminosity has been revisited in this work as follows. We determined a lower and upper limit of that value by matching in the K band the photospheric flux {plus} the near-IR excess with the 2\,MASS photometry within its 5\% uncertainty (see Table~\ref{tab:K-flux}). This provided a stellar luminosity between 18 and 19\,L$_\odot$, in agreement with the revised value by \cite{Vioque2018}.}
\label{tab:MCMax}
\end{table*}
% -------
% -------
\vspace{0.2cm}\\
{\bf A ring of quantum heated particles: }
Another way to produce near-IR emission consistent with the absence of the prominent 10\,$\mu$m silicate feature and with the presence of the mid-IR PAH bands is to consider quantum heated particles (QHPs, \citealt{Purcell1976, Draine2001}). 
In the context of interferometric observations, this  scenario was  invoked for HD\,100453 where QHPs were detected in the disk gap \citep{Klarmann2017}, and for HD\,179218 with the presence of hot QHPs inside the disk cavity \citep{Kluska2018}.\\
We tested different amounts of QHPs (from $10^{-13}$ to $10^{-15}\,$M$_{\odot}$), and two different QHP particles sizes ($10^5$ and $5 \times 10^5$ carbon atoms), based on the results obtained by \cite{Klarmann2017}. 

\begin{figure}[!htpb]
\includegraphics[width=\columnwidth]{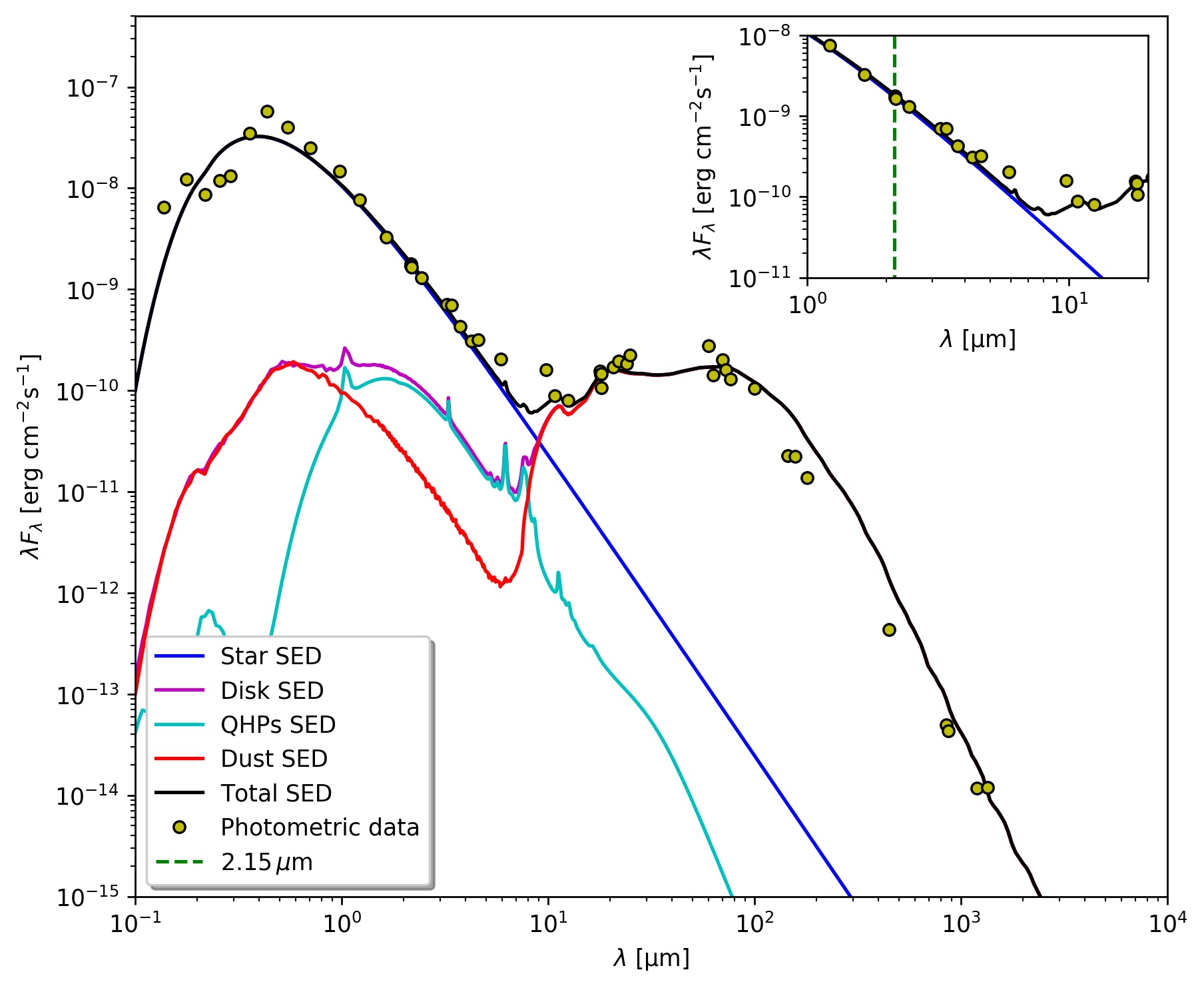}
\caption{MCMax SED (black line) of the model described in Table \ref{tab:MCMax}. The yellow circles are HD\,141569 photometric data listed in Table \ref{tab:K-flux}. The blue line represents the star SED modeled as a black body. The cyan line represents the QHPs SED and the red line the silicates SED. The magenta line is the total disk SED accounting for both QHP and silicate emission. The top right plot focuses on the first $10\,\mu$m wavelength range of the SED.}
\label{fig:SED_MCMax}
\end{figure}

The highest masses ($\sim$\,10$^{-13}\,$M$_{\odot}$) produce an 
% optically thick 
inner rim emission which results in a near-IR excess that is too large and  inconsistent with the GRAVITY measurement. The smallest value ($\sim$\,$10^{-15}\,$M$_{\odot}$) corresponds instead to an optically thin disk with negligible excess at 2\,$\mu$m. 
% Moreover, the mass that would correspond to a 6\,\% NIR excess depends on the width of the ring. 
%, which we did not constrained. 
For a ring geometry in agreement with our best fit of the GRAVITY data, 
% extending from 0.8 to 2\,au, 
a sweet spot is found for a mass of 4.3$\times$10$^{-14}\,$M$_{\odot}$ (or  1.4$\times$10$^{-8}\,$M$_{\oplus}$) for which the resulting disk produces a near-IR excess of $\sim$\,7\,\%. 
Smaller particle sizes (e.g., 10$^{2}$ carbon atoms) would require a larger mass reservoir of QHPs to reach a near-IR excess of $\sim$\,6\%. As a consequence, a higher mass would result in stronger mid-IR PAH features overestimating the HD\,141569 SED  observed in the Spitzer IRS spectrum \citep{Sloan2005}. The order of magnitude of $\sim 10^5$ carbon atoms per particle appears consistent with the SED profile  estimated by GRAVITY and IRS/Spitzer.\\
The resulting disk SED shows a spectral index (d\,$\mathrm{log}F_{\lambda}$/d\,$\mathrm{log}\lambda$) 
of -1.4 at 2.15\,$\mu$m 
which is consistent with the minimization curve of the 
% Looking at the $\chi_r^2$ curve of 
spectral index in Fig.~\ref{fig:GTRing-Chi2r_maps}.
% ) we note that the values are consistent within the error bar with the value found through the GRAVITY data analysis. 
The details of the QHP model parameters is shown in Table \ref{tab:MCMax}. In Fig. \ref{fig:MCMax-model_DT} we show the density and temperature structure of the full-disk model. For the silicate dust in the outer second ($\sim 15\,$au), third ($\sim 185\,$au), and fourth ring ($\sim 300\,$au), the equilibrium temperature varies between 125\,K at $\sim$5\,au 
% for the innermost region 
to 20\,K in the outermost disk regions. The QHPs used to model the  innermost optically thin ring are not in thermal equilibrium. Their temperature distribution depends on their size (the smaller, the hotter) and on the strength of the local ultraviolet (UV) radiation field as they absorb UV photons to quickly re-emit in the near-IR, changing their temperature drastically and very fast. The dark red color in Fig. \ref{fig:MCMax-model_DT} shows the location of the QHPs, and according to our RT simulations their temperatures are in the  range between 1865\,K and 95\,K.\\
\\
We find that the strongest constraint on the ring width is set by our interferometric measurement. When considering widths from 0.04 to 0.3\,au in our RT modeling, the observed tendency remains the same: the population of silicate dust grains does not satisfactorily reproduce the near- and mid-IR excess,  contrary to a population of a stochastically heated small grains.\\
\\
Beside the ring's width, the important result of this analysis suggests that a tenuous, QHP-dominated, optically thin inner ring may provide a suitable description of the close circumstellar environment of HD\,141569 in agreement with existing detailed modeling of the outer disks. On the contrary, a silicate-dominated dusty inner ring, heavier by about four orders of magnitude and composed of large grains, fails to provide a satisfying description of the system, in particular in terms of flux contribution in the mid-IR spectral range. Figure~\ref{fig:SED_MCMax} presents the final result of our MCMax modeling with the parameters of Table~\ref{tab:MCMax}. We note that the strong PAH feature at 7.8\,$\mu$m in HD\,141569 is not seen in this model because the corresponding opacity has not been added to our models of the outer rings, unlike the models of  \cite{Thi2014}.

\subsection{Spatial scale of the Br$\gamma$-line emitting region
%Gas: high spectral resolution data results
}
\label{sec:SC-Results}

\begin{figure}[thbp!]
\includegraphics[width=\columnwidth]{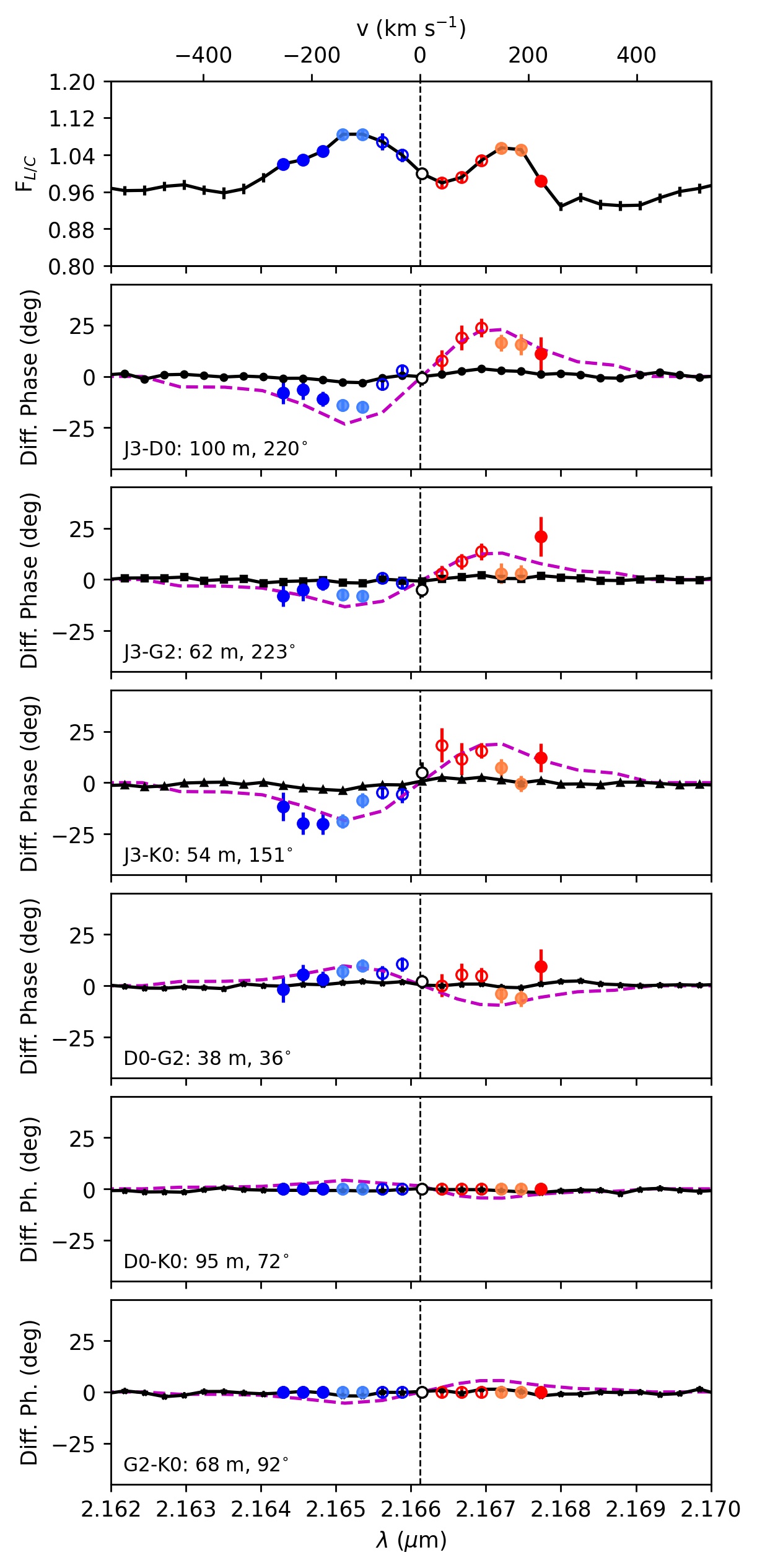}
\caption{HD\,141569 spectrum (top), total (black circles and line), and pure-line differential phases (colored signs) along the different baselines. D0-K0 and G2-K0 baseline pure-line differential phases are set to zero. The colors refer to the different spectral channels. Dashed magenta lines represent the pure-line differential phases of the analytical Keplerian disk model described in Section \ref{sec:SC-Results}.}
\label{fig:PureL_Diff_ph}
\end{figure}

\begin{figure}[thpb!]
\includegraphics[width=\columnwidth]{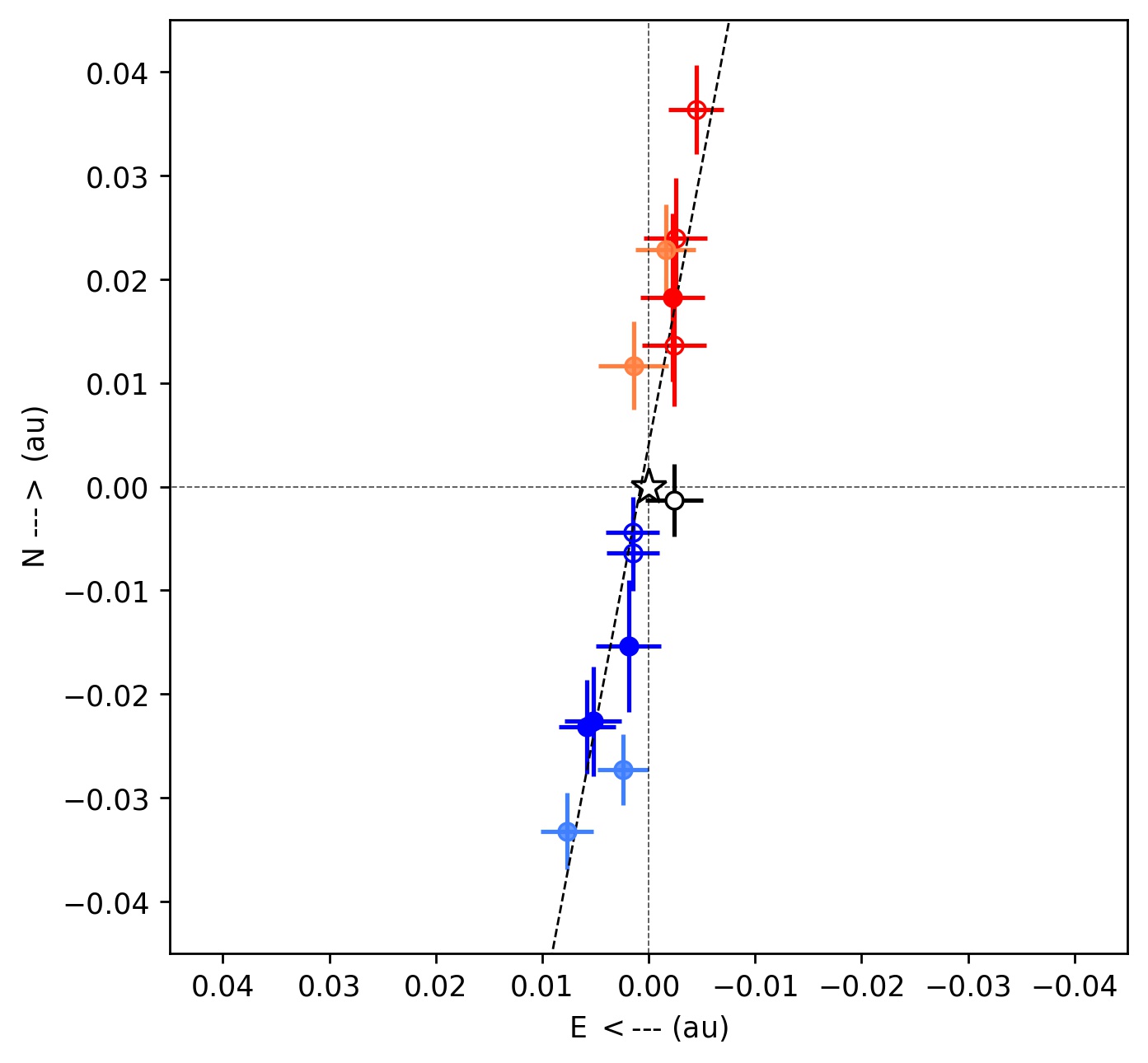}
\caption{Deprojected photocenter shifts. The colors refer to the different spectral channels and velocities, as shown in Fig. \ref{fig:PureL_Diff_ph}. The dashed black line, derived through a linear fit of the photocenter shifts, represents the gas region position angle.}
\label{fig:Photo-shifts}
\end{figure}

We exploited the high-spectral resolution data of GRAVITY in the Br$\gamma$ region to constrain the spatial scale of the hot gas emitting component, following the formalism in Sect.~\ref{sec:SC-Methodology}. Conservatively, the resulting pure-line visibilities plotted in red in Fig.~\ref{fig:SC-Data} are very close to 1 for all the baselines. Considering the error bars we can say that the gas region is at the limit of spatially unresolved emission. We therefore propose to constrain the size of the gas emitting region 
%at the Br$\gamma$ line wavelength peaks 
by considering the gas emitting at the Br$\gamma$ line wavelength peaks 
(i.e., 2.1654\,$\mu$m and 2.1672\,$\mu$m) using an infinitesimally thin ring model and estimating the upper-limit size that would exceed the error bar of $\sim$2\% on the pure-line visibilities. In this way, we estimated a maximum radius of $\sim$\,0.35\,mas (0.0385\,au) for the gas emitting region. \\
While the analysis of the visibility amplitudes only provides us an estimate of the size scale of the gas emitting region, further information on the spatial and kinematic properties of the gas component is found in the differential phase signal. Typically, differential phases provide information on photocenter displacements along the baselines on angular scales that can surpass the nominal resolution of the interferometer. 
Figure \ref{fig:PureL_Diff_ph} shows the GRAVITY pure-line differential phases. After removal of the continuum contribution (cf. Sect.~\ref{sec:SC-Methodology}), the S-shape becomes clearly visible for the baselines J3-D0 (100\,m, 220$^{\circ}$) and J3-K0 (54\,m, 151$^{\circ}$) around the Br$\gamma$ line, while a weaker trend is seen for the baselines J3-G2 (62\,m, 223$^{\circ}$) and D0-G2 (38\,m, 36$^{\circ}$). The strongest signature shows an amplitude in the differential phase exceeding $\sim$20$^{\circ}$ for J3-K0. Since the differential phase signals for the baselines D0-K0 (95\,m, 72$^{\circ}$) and G2-K0 (68\,m, 92$^{\circ}$) are consistent with zero at all wavelengths, we fixed the pure-line differential phases to 0$^{\circ}$. The typical uncertainties after correcting for the continuum subtraction are $\sim 4^{\circ}$. The resulting deprojected photocenter shifts per spectral channel, with reference frame fixed to the star location, are shown in Fig. \ref{fig:Photo-shifts}. We clearly observe that all points are aligned along the same direction with an angle of -10$^{\circ} \pm 7^{\circ}$.  The redshifted points are located along northwest, while the blueshifted points are located toward the  southeast.
%, which would indicate \textbf{a} counterclock-wise rotating gas structure. 
Based on the redshifted maximum extent of the photocenter shifts, we estimated the radius of the gas region from the differential phases to be $0.333\pm0.039$\,mas or $0.037\pm0.004$\,au. This appears consistent with the less precise upper-limit size set through the analysis of the visibility amplitudes. 
%The characteristic sizes derived from 
We recall however that the size estimate derived through the photocenter shift does not correspond to the physical outer radius of the gas region, 
but to the size 
%since they only give the location 
where the gas emission is more intense for a given wavelength.\\

\begin{figure}[thbp!]
\includegraphics[width=\columnwidth]{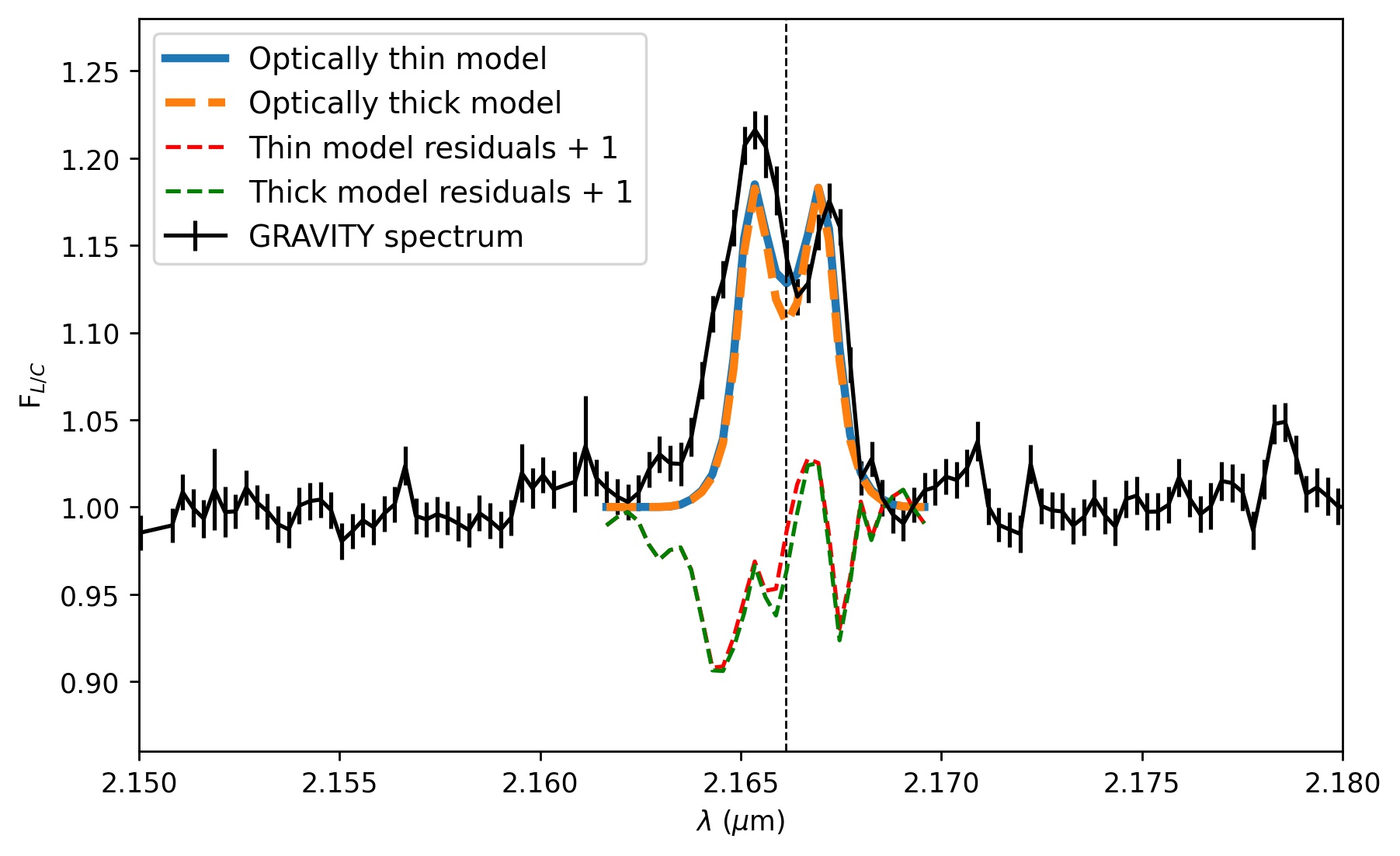}
\caption{HD\,141569 GRAVITY spectrum (black line), and the spectra of the Keplerian disk models described in Section \ref{sec:SC-Results}. The blue line represents the model in the optically thin scenario with residuals given by the red dashed line, while the orange line represents the model in the optically thick scenario with residuals given by the green dashed line.}
\label{fig:Alois_spectra}
\end{figure}

\begin{figure*}[thbp!]
\includegraphics[width=\textwidth]{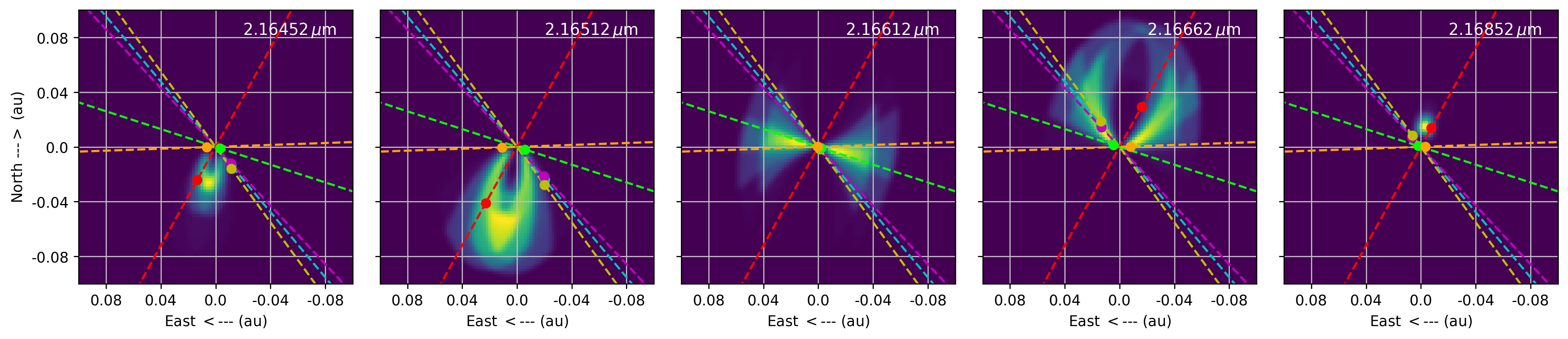}
\caption{Monochromatic images (from $2.16452$ to $2.16852\,\mu$m, i.e., from $221$ to $332\,$km/s) of the Keplerian ring model described in Section \ref{sec:SC-Results}. The colored dashed lines refer to the different GRAVITY baselines. Circles represent the 2D photocenter shift for each baseline.}
\label{fig:KM_Ph-shift}
\end{figure*}

The distribution of the 2D photocenter solution can be interpreted to the first order as being caused by a gas disk in Keplerian rotation orbiting HD\,141569. Under this assumption the analysis of the Br$\gamma$ emission line's shape provides further clues to the gas kinematics. 
We can indeed derive an estimate of the gaseous disk's radius from the separation between the two peaks of the line and the rest position. From \cite{Beckwith1993} we use 
\begin{equation}\label{eq:Beckwith_KD}
    R_g = \frac{\rm{G\,M}_\star}{v_{obs}^2}\, \rm{sin^2}i, % \, \rm{cos^2}\phi,
\end{equation}
where $R_g$ is the radius, 
%locus of coordinates within the disk of a given polar coordinate $\phi$, 
G is the gravitational constant, M$_{\star}$ the mass of the star, $v_{obs}$ the projected velocity at the line peaks, and $i$ the disk inclination. The $128\pm42$\,km/s average peaks shift of the Br$\gamma$ line with respect to the line rest position leads to an outer limit for the gas region of $0.766\pm0.554$\,mas in radius, or $0.084\pm0.061$\,au. The resulting error accounts for the uncertainty on the stellar mass (0.01\,M$_{\odot}$), on the distance (1\,pc), on the disk inclination (15$^{\circ}$), and on the peak position (3\,\AA{}), the last being the dominant one.\\
The three approaches presented above and based on the analysis of the visibility amplitudes, differential phases, and the spectrum all seem consistent with a gas component in Keplerian rotation confined within $\sim 0.8$\,mas ($\sim 0.09$\,au, $\sim 12.9$\,R$_{\star}$) in radius. 
We compared our differential phase signals, our strongest measurable quantity, to a simple geometrical model of an axisymmetric disk in Keplerian rotation built with two thin layers that account for the top and bottom sides of the disk, parameterized by an inner radius $r_{\rm in}$ (varying from 0.008 to 0.03\,au), an outer radius $r_{\rm out}$ (varying from 0.033 to 0.8\,au), and a  power-law exponent $\alpha$ (varying from 0 to 4.0) for the disk's intensity radial profile following $I(r)\propto\,r^{-\alpha}$. 
The inclination of the disk was fixed to 58.5$^{\circ}$, based on the results of the FT data analysis and under the assumption of coplanar dust-gas rings, and its position angle was fixed to -10$^{\circ}$, based on the photocenter shift analysis.
The hypothesis of an optically thin disk is made, which implies that the disk intensity does not strongly depend on the disk scale height, which is then fixed to ${\rm H/R}=0.1$ throughout the ring, but only on the surface area of the disk: $I \propto dS$ where $dS$ is an area element (for a complete description of the model, see de Valon et al. in preparation).
The model delivers a spectral profile that is eventually normalized and fitted to our experimental double-peaked spectrum in Fig.~\ref{fig:Alois_spectra}. A grid of 2700 models has been explored, and a best-fit model is found for $r_{\rm in}$=0.011\,au, $r_{\rm out}$=0.09\,au, and $\alpha$=0.5.
From this best-fit model, we produced 2D velocity maps (Fig.~\ref{fig:KM_Ph-shift}) and retrieved the theoretical pure-line differential phase signal (see Fig.~\ref{fig:PureL_Diff_ph}).
Our Keplerian disk model reproduces qualitatively well the observed pure-line differential phases both in the orientation of the sine wave relative to the wavelength and in amplitude reinforcing the Keplerian gas disk scenario. We also observe that the weakest signals indeed correspond  to the baselines D0-K0 and G2-K0 for which the peak amplitudes are $\sim$3--5$^{\circ}$.
On a final note, for the best-fit solution model we tested the optically thick hypothesis by accounting for projection effects on the emissivity law, 
$ I \propto \mathbf{dS} \cdot \mathbf{e_y} = dS \cdot (\mathbf{n} \cdot \mathbf{e_y}) $,
where $\mathbf{e_y}$ is the line-of-sight unit vector and $\mathbf{n}$ the normal unit vector of the area element considered (see de Valon et al. in preparation). The resulting model leads to a  spectrum that is similar to that derived from the optically thin model (a small difference is seen at the lowest velocities), and to pure-line differential phases and photocenters shifts consistent with those of the optically thin model.

\section{Discussion}
\label{sec:Discussion}

\subsection{A newly detected inner dust ring}
\label{sec:Discussion_continuum}

\begin{figure*}[!htbp]
    \centering
    \includegraphics[width=\textwidth]{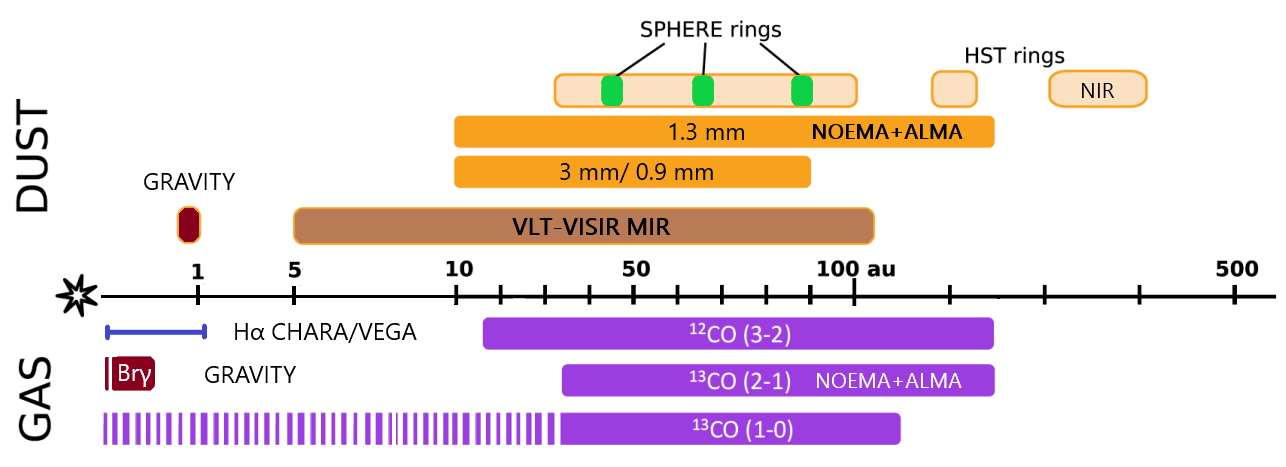}
    \caption{Visualization of dust and gas distribution in HD\,141569 (adapted from \citealt{DiFolco2020}). Shown in light
orange are the optical--IR dust rings  detected in scattered light by HTS \citep[e.g.,][]{Augereau1999, Clampin2003, Konishi2016}. Overplotted in green are the three near-IR dust ringlets detected by VLT/SPHERE \citep{Perrot2016}. The extended millimeter continuum emission detected by ALMA \citep{Miley2018} and NOEMA \citep{DiFolco2020} is shown in orange. The large brown ring represents the mid-IR continuum emission detected by VLT-VISIR and modeled by \cite{Thi2014}. In dark red is  shown the near-IR dust emission detected by GRAVITY and studied in this work. In the same color  is  the Br$\gamma$ line emitting gas region detected by GRAVITY, also analyzed in this work. In blue is depicted the H$\alpha$ line region based on the upper-limit size estimated by \cite{Mendigutia2017}. Finally, in purple is shown the CO gas region whose emissions were detected by ALMA \citep[e.g.,][]{White2016, Miley2018} and NOEMA \citep{DiFolco2020}. 
    %Dashed regions mean that their sizes are not constrained.
    }
    \label{fig:HD141569-visualization}
\end{figure*}

Thanks to GRAVITY, an additional inner
%, narrow, 
ring component at $\sim$\,1\,au from the star, narrower than $\sim\,$0.3\,au, has been discovered, which adds to the multi-ring picture identified for 
%and added to the complex 
HD\,141569. A schematic visualization of the dust and gas distribution obtained from multi-epoch and multi-instrument observations is depicted in Fig. \ref{fig:HD141569-visualization}. 
The low level of near-IR excess ($\sim$\,6\%) is comparable to the typical 5\% accuracy of K-band photometric data (e.g.,\,2\,MASS), which led to considering the near-IR flux of HD\,141569 as essentially photospheric in earlier studies. 
With this result, the presence of dust as close as $1\,$\,au from the star, but well beyond the sublimation radius $R_{\rm sub}$\,$\sim$\,0.25\,au is a more robust piece of  evidence.
The disk inclination and position angle estimated through the GRAVITY FT data analysis are consistent within the error bars with the values found for the outer rings that make up the circumstellar environment of the system (see Table \ref{tab:i-PA}), suggesting an almost coplanar system of circumstellar rings. 
Considering how many constraints can effectively be set on the position angle (see Fig.~\ref{fig:GTRing-Chi2r_maps}), it does not appear that any clear misalignment between the inner and outer disks could be claimed.\\
The inclination and position angle could be the reason why PTI and the Keck interferometer were not able to spatially resolve the near-IR emission of HD\,141569 in earlier measurements, since the alignment of their single baseline was around 42$^{\circ}$ from north to east, hence not far from the semi-minor axis of the disk. However, the upper limit of 10\,mas in radius for the location of the dust proposed by \cite{Monnier2005} is in agreement with our results. They also report a fractional excess of $\sim$5\%, but compatible within the error bars with pure photospheric flux in the K\,band. \\
%, that made it harder to detect the faint ring flux excess. 
Our results are substantially different from those obtained by \cite{Lazareff2017} with PIONIER in the H band. The reported flux excess is $\sim$\,50\% larger than in our case and their half-flux radius for the dust emission is only 0.03\,au,  well inside the sublimation radius of the system expected to be at $\sim$0.21-0.26\,au (for $L_{\star}=16.6-27.0\,$L$_{\odot}$, T$_{\rm subl}$\,$\sim$\,1470\,K and a cooling efficiency $\epsilon$\,$\sim$0.5). \cite{Lazareff2017} modeled the H-band excess emission using an ellipsoid distribution and not a ring (see their Table~B.2 and B.3). The choice of the model and the low-quality PIONIER data could potentially explain this unexpected result. \\
The values derived for the radius and width of this new innermost dusty component allows us to compare the system with other YSOs. The analysis made by \cite{Perraut2019} for a sample of 27 Herbig Ae/Be stars revealed dusty circumstellar environments with half-flux radii that range between 0.1 and 6\,au depending on the stellar luminosity with a median of 0.6\,au, and a width-to-radius ratio $w$ ranging from 0.1 to 1 with a median of 0.83, which is interpreted as  smooth and wide rings, even though with large error bars. \\
Our radius estimate for HD\,141569 ($\sim$0.8\,au) is within the range found by these authors, but its peculiarity is also reflected in the fact that its position in the size-luminosity diagram does not coincide with the bulk of the Herbig stars, 
reinforcing the idea that HD\,141569 is a unique system in terms of evolution. 
From Table~\ref{fig:GTRing-Chi2r_maps}, the width-to-size ratio is estimated to 0.05$\pm$0.05 for the best-fit model, but it is also noticeable from Fig.~\ref{fig:GTRing-Chi2r_maps} that the ring's width is difficult to clearly constrain in the region below 0.3\,au. This implies that the width-to-size ratio could be seven times larger. Therefore, our estimate of the width-to-size ratio goes on the lower end of the range found in \cite{Perraut2019}, but remains comparable to systems such as HD\,114981 (0.10$\pm$0.03) or HD\,190073 (0.14$\pm$0.03).\\
In addition to  the  disk's well-known flux-size degeneracy, which in our case is broken thanks to the constant plateau in the squared visibilities curve as a function of spatial frequencies, \cite{Lazareff2017} found a negative correlation between the ring width and its half-flux radius, meaning that it gets more difficult to detect with the VLTI baseline small-radius ring-like structures than small-radius ellipsoid structures for angular sizes of the K-band emission of $\sim$1\,mas or smaller. 
% for marginally resolved objects. 
In our case the size is well constrained being beyond the suggested limit, which is one argument favoring the robustness of our modeling as opposed to the model of a Gaussian brightness distribution that did not converge to a solution.\\
% From the analysis of our data, we are confident on the ring distribution of the dust since the star + Gaussian circumstellar environment  model (Eq. \ref{eq:Visibility-Gauss}) did not converge to a solution and the inner rim estimate for our ring model ($\sim 7.4\,$mas) is far enough from the limit value (1\,mas) where it gets difficult to distinguish between ellipsoid and ring like structure \citep{Lazareff2017}. 
% However, the analysis of \cite{Lazareff2017} and \cite{Perraut2019} favor ring with a wide extension ($\omega_{\rm norm} \geq 0.5$) in contrast with our estimate for HD\,141569 ring. 
To further test our model findings, first we tried to add the contribution of a fully resolved emission (also known as a {halo}) to our geometrically thin ring model following Eq.~4 of \cite{Lazareff2017}. Our best-fit solution ($\chi^2_{red} = 4.69 \pm 0.32$) led to results similar to our geometrically thin ring model with a halo contribution that actually converges toward zero ($0.103^{+0.144}_{-0.075}\,\%$). Interestingly, this is in full agreement with \cite{Lazareff2017} who found null halo flux contribution as well. Second, we also tried  a Lorentzian-convolved 
%ed our Gaussian-convolved 
infinitesimally thin ring 
%by convolving the ring with a Lorentzian-like function 
(see Table 5 of \citealt{Lazareff2017}), 
%instead of the Gaussian one. 
which leads to practically   the same solution ($\chi_r^2 = 4.68 \pm 0.32$). 
% and we cannot say which of the two model fit better the data. 
We conclude from this detailed analysis that the narrow ring-like shape with a width smaller than 0.3\,au is a good description of our observations.\\
% width is likely true.\\

\subsection{Nature and origin of the detected ring}\label{origin}
%------
%From the data and the analysis shown in this work we are not able to make conclusion on the origin of such ring shape.
We advance in Section~\ref{sec:MCMax} the scenario of a population of stochastically heated particles (e.g., PAH-like very small grains) as the cause for the near-IR excess. A number of arguments can be discussed in this context.\vspace{0.1cm}\\
Our best-fit chromatic model shows a spectral index $k_c$=-0.35$\pm$0.21 for the circumstellar emission. Following \cite{Lazareff2017} and \cite{Perraut2019}, we estimate from the parameter $k_c$ a temperature of the radiating dust under the  gray body hypothesis (i.e., wavelength-independent emissivity) and find $T_{\rm c}$=1460$\pm$70\,K. In the case of a silicate dust ring in thermal equilibrium at $\sim$0.8\,au, we would expect a cooler temperature\footnote{Eq.~14 in \cite{Lazareff2017}} of $T_{\rm c}$\,$\sim$\,650--850\,K using the stellar parameter of Table~\ref{tab:MCMax} and a cooling efficiency $\epsilon$ between 0.3 and 1. Therefore, we argue that the near-IR emission is not dominated by emission of dust in thermal equilibrium, which can be explained by the presence of these small particles that are  quantum heated by the stellar UV radiation. Even though the spectral index is found to be not very well constrained (see Fig.~\ref{fig:GTRing-Chi2r_maps}), the spectral index corresponding to dust in thermal equilibrium at 800\,K would be around +3.4 at $\lambda_0$, relatively far from the our best-fit model.\\
\cite{Maaskant2014} interpret the high intensity ratio between the PAH band at 6.2\,$\mu$m to the band at 11.3\,$\mu$m as a tracer of predominantly ionized PAH species located in a disk's gap and exposed to the intense ionizing UV radiation field of the central star.
%\textbf{indeed HD\,141569 is believed to be in the so called UV pumping regime \citep{Krotkov1980, Brittain2003, Banzatti2015}}. 
For instance, with a $I_{6.2}$/$I_{11.3}$ feature peak ratio of $\sim$3$-$4 \citep{Seok2017} the PAH sources IRS\,48 and HD\,179218 present emission from such predominantly ionized PAHs located in part inside the gap or disk cavity \citep{Maaskant2014,Klarmann2017,Kluska2018,Taha2018}. Interestingly, the $I_{6.2}$/$I_{11.3}$ peak ratio of HD\,141569 derived from \cite{Seok2017} is high as well,  estimated to $\sim$\,5$-$6. This may indicate the presence of PAH species close to the star, with a predominantly ionized state due to the direct irradiation by UV stellar flux,
%We believe this aspect brings further support to a possible scenario of a QHP-dominated inner ring in HD\,141569. 
bringing further support to our QHP-dominated inner ring model.\\
Comparing the IRS spectrum to our model, we find that the our model accounts for 28\%, 25\%, and 4\% of the observed PAH peak emission for the features at 6\,$\mu$m, 8\,$\mu$m, and 11\,$\mu$m, respectively. The remaining emission would come from PAHs located in the outer rings.
Comparing the outer ring PAH mass reservoir estimated by \cite{Thi2014} to that of our innermost ring model, we find that our estimate ($4.3 \times 10^{-14}$\,M$_\odot$) is smaller than their $\sim 15\,$au and $\sim 300\,$au rings by three orders of magnitude ($2.0\times10^{-11}$\,M$_\odot$ and $2.1\times10^{-11}$\,M$_\odot$, respectively), and by four orders of magnitude for their $\sim 185\,$au ring ($1.2\times10^{-10}$\,M$_\odot$) and the entire outer disk environment ($1.6\times10^{-10}$\,M$_\odot$). We recall however that in their model \cite{Thi2014} do not account for any dust located at $\sim$\,1\,au and suggest a $5\,$au dust-free inner gap, which could result in overestimated values on their side.\vspace{0.2cm} \\
Regarding the origin of the ring structure, the case of HD\,141569 is particularly interesting under the aspect of our proposed QHP-dominated inner component: one could question the presence of QHPs in an inner narrow ring since this kind of particle  is expected to be coupled to the gas component and to date have been mostly invoked in more extended emission  (e.g., \citealt{Klarmann2017, Kluska2018}). \\
Several authors have detected CO emission beyond $\sim$10\,au (see Introduction), whereas little information is available on the presence of CO within the first 10\,au. It is likely however that this inner region is not gas depleted. 
% even though CO emission was not detected inside the first $\sim 10\,$au by many authors (see Introduction) it does not mean that the region is depleted of gas. 
The [O$\, \rm _I$]\,$\lambda$6300 emission detected by \cite{Acke2005}, and suggested by these authors as a dissociation product of OH in the circumstellar disk, would originate between $\sim$0.05 and 0.8\,au under the assumption of a gas disk in Keplerian motion \citep{Brittain2007}. Hydrogen recombination and sodium lines were also detected \citep{Vanderplas2015}. Both \cite{Mendigutia2017} and our results (see Section~\ref{sec:Discussion_BrG}) find excited atomic hydrogen in the dust-free cavity, which implies the existence of a replenishment mechanism from the outer regions. 
Moreover, \cite{Brittain2007} 
% by modelling the $v=2-1$ R9 CO emission line through a UV fluorescence model 
set an upper limit on the column density of CO inside 6\,au 
%along the line of sight from the star to the inner edge of their CO disk model (6\,au) 
of N(CO)$\,<10^{15}$\,cm$^{-2}$, which translates into a gas mass $< 5.9\times10^{-13}\,$M$_{\odot}$. Considering our QHPs mass ($4.3 \times 10^{-14}$\,M$_\odot$), a gas-to-dust ratio of 100, and a [CO/H$_2$] ratio of $10^{-4}$, this would translate in a CO mass of $4.3\times10^{-16}\,$M$_\odot$, which is well below the upper limit and therefore not detected yet. 
These arguments suggest that a gaseous component may exist and coincide with the proposed QHP component to which it would be coupled.
%The fact that the UV fluorescence model can explain the CO emission while suggesting a CO mass consistent with our findings gives support to the hypothesis that the near-IR flux excess of HD\,141569 could be dominated by QHPs emission.
\\
Furthermore, even though QHPs have been invoked in more extended emission in past works, we cannot exclude the possibility of more compact or narrow components. For example, \cite{Khalafinejad2016} modeled its inner circumstellar region through an optically thin spherical halo extending from 0.1 to 1.7\,au in order to explain the near-IR flux of HD\,100453 and to fit simultaneously its Q-band flux. The choice of the spherical halo was based on the fact that their data poorly constrained the structure of the inner disk (and so it is the halo extension estimate) and the optically thin hypothesis was set in order to not affect the Q-band flux modeling. This component was also suggested by \cite{Klarmann2017}. Their QHP model for HD\,100453 underestimates the observed flux in the 1-5\,$\mu$m wavelength range by up to 30\%, and slightly overestimates the long-baseline visibility data, indicating that the missing flux is emitted on short spatial scales. Closer results to those we obtained for HD\,141569 were found by \cite{Maaskant2013}. These authors suggest a compact optically thin spherical halo for HD\,169142 (0.1-0.2\,au), HD\,135344\,B, and Oph IRS\,48 (0.1-0.3\,au) to reproduce the observed near-IR flux.
Several scenarios have been proposed to explain the different structures that protoplanetary disks exhibit, such as gaps, spirals, or rings. 
%, but the matter is still an open question. 
Fragmentation of wide rings into narrow ones by secular gravitational instability (e.g., \citealt{Tominaga2020}), self-induced pileup of particles by aerodynamical feedback (e.g., \citealt{Gonzalez2017}), and dust traps at local maxima in the gas density due to a reversal of the pressure gradient by dynamical clearing from a companion (e.g., \citealt{Pinilla2012}) could explain a structured nature of the disk. 
%the nature of HD\,141569.
Our observations leave this matter as an open question since GRAVITY informs us solely on the spatial properties of the detected K-band continuum emission.

\subsection{HD\,141569 Br$\gamma$-line emitting gas region}
\label{sec:Discussion_BrG}

\begin{figure}[t]
\includegraphics[width=\columnwidth]{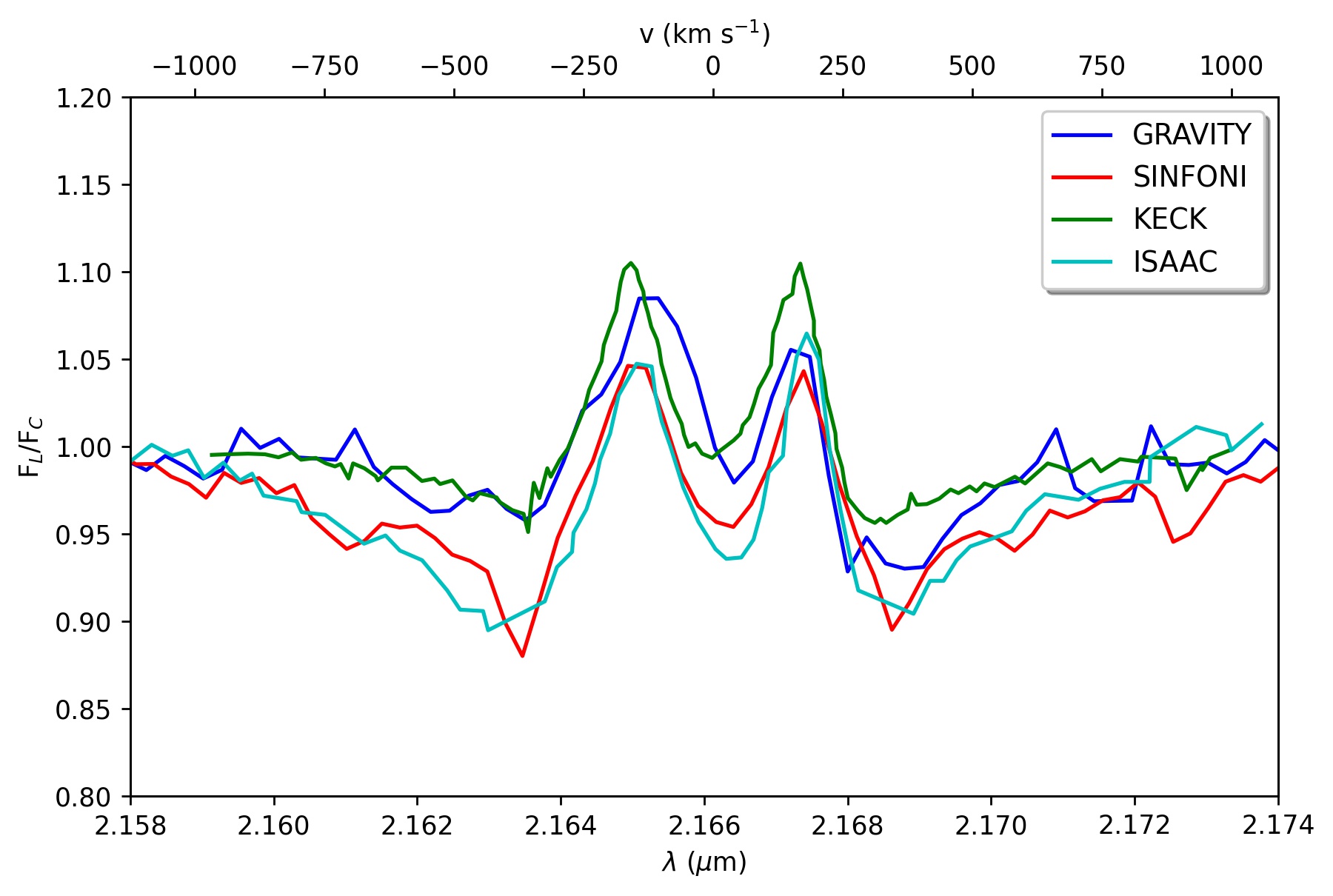}
\caption{HD\,141569 continuum-normalized spectrum taken at different epochs and with different instruments. In blue is depicted the data taken on July 12, 2019, by GRAVITY; in red data taken in June 2019 by SINFONI; in green data taken in 2002 by the KECK NIRSPEC \citep{Brittain2007}; and in cyan data taken in 2004 by the VLT/ISAAC \citep{Garcia2006}.}
\label{fig:BRG-Spectra}
\end{figure}

Our analysis of the kinematic and spatial distribution (via the differential phase) of the hot hydrogen gas is in line with a scenario of a Keplerian disk inside the dust-free cavity. 
The distribution of the photocenter shifts shown in  Fig.~\ref{fig:Photo-shifts} agrees well with the behavior expected from a Keplerian disk \citep{Mendigutia2015}. The position angle of the photocenter shifts distribution (-10$^{\circ} \pm 7^{\circ}$ north to east) is also found to be in overall agreement with the position angle of the inner ring responsible for the near-IR excess, and of the outer rings. Moreover, the photocenters of the Br$\gamma$ line emitting gas region are located as the photocenters of the outer CO regions, blueshifted ones along the southeast and redshifted ones along the northwest \citep{White2016}.
%meaning that the two gas regions share the same counterclock-wise rotation direction. 
The profiles of pure-line differential phase signals depart a bit from the perfect S-shaped signal expected for a pure Keplerian disk. We believe that it is  also limited ultimately by our spectral calibration. In this sense, more accurate measurement of the differential phases in HD\,141569 using GRAVITY with the 8\,m Unit Telescopes could certainly improve the accuracy of this analysis. In order to evaluate the quality of our spectral data, we chose to compare the GRAVITY profile measured with the ATs to other high-quality spectra obtained with the ISAAC spectrograph at the VLT \citep{Garcia2006}, the NIRSPEC echelle spectrograph at the Keck Observatory \citep{Brittain2007}, and with SINFONI/VLT from archival data. The comparison is shown in Fig.~\ref{fig:BRG-Spectra} and we observe that the GRAVITY spectrum exhibits a mild asymmetry between the blue and red peaks. This would suggest that our spectrum could be still affected by some calibration effects, either telluric or instrumental. We then further explored how far our resulting differential phases might be impacted by the slight spectrum asymmetry and tested the derivation of the pure-line differential phases using the SINFONI spectrum,  which has a very similar spectral resolution, instead of the GRAVITY spectrum. We found that the 2D distribution of the photocenter shifts remains unchanged within the error bars reported in Fig.~\ref{fig:Photo-shifts}. \\
The star is known to be a fast rotator ($222.0 \pm 7.0\,$km/s, \citealt{Folsom2012}), which results in a small co-rotation radius 
% , the distance from the star where the centrifugal force on a particle co-rotating with the star itself balances the gravitational attraction, which is 
around $2.38\pm0.53\,$R$_{\odot}$ ($0.011\pm0.002$\,au), assuming $R_{\star}=1.5\pm0.5\,$R$_{\odot}$ \citep{Fairlamb2015}. We cannot exclude that part of the Br$\gamma$ line emission comes from magnetospheric accretion flows, but the small co-rotation radius compared to the size of the Br$\gamma$ line emitting region estimated from the SC data analysis ($\sim 0.09\,$au) would not favor this scenario, as opposed to what has been recently found for TW\,Hya \citep{Gravity2020TWH}. 
% , could translate in a magnetospheric accretion flow whose emission volume could not be extended enough to explain the larger size observed with the GRAVITY observations. 
Interestingly, the scenario of magnetospheric accretion was also tested by \cite{Mendigutia2017} to explain the H$\alpha$ double-peaked emission line, but they were not able to reproduce the observed profile with any set of input parameters. 
Comparing the extent of the Br$\gamma$ emission to the continuum emission (R$_{\rm Br\gamma}$/R$_{\rm cont} \approx 0.1$), we find that the case of HD\,141569 is in contrast with the findings of \cite{Kraus2008}. These authors found for a small sample (5 objects) of Herbig Ae/Be stars that those showing a P Cygni H$\alpha$ line profile and a high mass-accretion rate ($>10^{-7}$\,M$_{\odot}\,$yr$^{-1}$) seem to show compact Br$\gamma$-emitting regions (R$_{\rm Br\gamma}$/R$_{\rm cont} < 0.2$), from which the emission stems from magnetospheric accretion or recombination line from ionized hydrogen, while stars showing a double-peaked or single-peaked H$\alpha$ line profile show a more extended Br$\gamma$-emitting region ($0.6 \leq$R$_{\rm Br\gamma}$/R$_{\rm cont}\leq 1.4$), which would trace a stellar or disk wind. Our system shows mixed features, a Br$\gamma$ and H$\alpha$ double-peaked emission line that originates from compact disks in Keplerian rotation where magnetospheric accretion is not the  most likely  main emission mechanism.
Therefore, recombination line emission from ionized hydrogen in an inner gaseous accretion disk as hinted by GRAVITY is a more supported scenario.
%However, 
If that is the case, considering the age of the system and the reported accretion rates between $10^{-7}-10^{-11}\,$M$_{\odot}$\,yr$^{-1}$ \citep{Merin2004, Garcia2006, Mendigutia2011, Thi2014, Fairlamb2015}, the inner gaseous disk requires some sort of replenishment mechanism to explain its presence and for it to survive. 
Replenishment flows,   planet-boosted or not \citep{Mendigutia2017},  connecting the inner and outer disk as already observed in other Herbig stars like HD\,142527 \citep{Casassus2013} could be investigated in the future.

\subsection{Hybrid or debris disk?}

As mentioned in the introduction, HD\,141569 is the only known pre-main sequence star characterized by a  hybrid disk. The main characteristic of a hybrid disk is the weak fractional excess of IR emission ($8.4 \times 10^{-3}$ for HD\,141569, \citealt{Sylvester1996}), that stems from the optically thin second-generation grains dust component, coupled with the presence of a significant   gaseous component, believed to be primordial. Unfortunately the other known best hybrid disk system candidates, which are 49 Cet, HD\,21997 \citep{Moor2011}, HD\,131835 \citep{Moor2015}, HD\,121617, HD\,131488 \citep{Moor2017}, and HD\,32297 \citep{Moor2019}, have not been studied in their hot dust content. As HD\,141569 does,  they show a featureless SED in the near-IR suggesting systems depleted of material inside the first 5\,au. However, as we saw in this work, interferometric observations could reveal as-yet-undetected dust, so this scenario should not be excluded for the other mentioned objects. The potential of long-baseline interferometry to detect small levels of hot circumstellar dust emission in supposedly dust-free systems has been indeed exploited for older main sequence stars \citep{Ertel2014}.\\
In this context it is interesting to compare our results for HD\,141569 with Vega, the most iconic debris disk system with a similar spectral type (A0V), but that is significantly older than our source (400\,$-$\,700\,Myr). Near-IR excess from Vega was detected and constrained to $\sim$\,1.29$\pm$0.19\% by interferometric observations with CHARA/FLUOR in the K band \citep{Absil2006}. 
These authors suggest by SED modeling that the excess comes from hot small grains starting at $\sim$0.2-0.3\,au with a total dust mass of 8$\times$10$^{-8}$\,M$_{\oplus}$. Knowing the age of Vega, it is clear that its circumstellar dust is of second generation, a characteristic of debris disks. 
% with the most credited origins being stellar radiation destruction of comets orbiting in the inner system or a late heavy bombardment kind-like event. 
In the scenario of a silicate-dominated inner ring for HD\,141569, the total mass required to induce a larger near-IR excess is, as expected, significantly larger ($>$\,10$^{-4}$\,M$_\oplus$). Considering the younger age of HD\,141569 ($\sim$7.2\,Myr), a more massive inner disk is compatible with a system at an earlier stage of disk evolution. 
Since the timescale for disk dissipation is known to be about $5-10\,$Myr \citep{Wyatt2008}, it is plausible that part of the dust in the inner region of  HD\,141569  is of first generation and a remnant of the primordial circumstellar environment. This would confirm that HD\,141569 is closer to a system in the final stage of the protoplanetary disk phase than a debris disk system.\\
\\
The fact that gas,  both H and CO, is detected in HD\,141569 is an factor in favor of a system in the (late) protoplanetary disk stage rather than in the debris disk stage. The system, similarly to other hybrid disks, does not follow the correlation between the CO flux density and the millimeter continuum emission followed by T Tauri, Herbig Ae, and debris disks, but lies systematically above the correlation line \citep{Pericaud2017}. The authors suggest that the dust and gas evolution are decoupled, with the dust evolving faster than the gas, leading to an unusual high gas-to-dust ratio (between 135 and 2370 for HD\,141569, \citealt{DiFolco2020}). 
Other than the primordial origin scenario, a secondary origin of CO was proposed \citep{Kral2019} in which the gas is self-shielded and shielded by accumulated neutral carbon produced through photodissociation of molecular gas released by planetesimals. This model is able to explain the estimated CO masses in all the hybrid disk candidates \citep{Kral2019, Moor2019} except for HD\,141569, for which it was not tested. Since CO molecule photodissociation occurs at UV wavelengths, we note that  C$^0$ could shield these molecules,  and also QHPs \citep{Woitke2016}, which are known to absorb UV photons and cool down very quickly by emitting photons in the near-IR. The new detected innermost ring, which we propose in this work to be dominated by a small amount of QHPs, could contribute to the shielding process.

\section{Summary}
\label{sec:Summary}

We presented the first GRAVITY interferometric observations of HD\,141569. Here we summarize the main conclusions of our work:
\begin{itemize}
    \item The system was resolved by GRAVITY with squared visibilities down to V$^2 \sim0.8$. If before these observations the near-IR flux contribution of a dust disk was considered absent because no feature was seen in the SED of the object, now, thanks to interferometry, the presence of dust in the first au of the system is a more robust piece of  evidence and the flux excess is clearly detected and constrained to be $\sim 6.2\, \%$ of the total flux.
    \item Data modeling suggests that the dust is located in a thin ring ($\lesssim 0.3\,$au in width) at a radius of $\sim$1\,au  from
    % between 0.8 and 2\,au from 
    the star. The ring shares, within the errors of Fig.~\ref{fig:GTRing-Chi2r_maps}, the same inclination ($\sim$\,$58^{\circ}$) and position angle ($\sim$\,$0^{\circ}$) as the outer rings observed in the past.
    \item MCMax SED modeling suggests that this innermost ring could be made  of a small amount ($1.4 \times 10^{-8}\,$M$_{\oplus}$) of QHPs. Large silicate grain models, with and without carbon, can reproduce the 6\% flux excess at 2.15\,$\mu$m, but at the same time they show a significant emission in the mid-IR not consistent with the SED of the system.
    \item The SC data analysis confirms the significant amount of Br$\gamma$ line emitting gas already observed in the past. The gas region is spectrally resolved, but spatially unresolved.
    \item The pure-line differential phases constrain the gas to be in a Keplerian disk-like structure, as hinted by the double-peaked line shape, confined within $\sim 0.09\,$au ($\sim 12.9$\,R$_{\star}$) and oriented in the same was as the outer rings (${\rm PA_{NE}} \sim -10^{\circ}$). 
    %Asymmetry in the photo-center shifts could be explained by replenishment flows, planet-boosted or not, but more accurate data are needed.
\end{itemize}
These results confirm the complexity of the HD\,141569 circumstellar environment also at milliarcsecond scale, making the system a unique astronomical laboratory to investigate the missing steps of disk evolution and planet formation theories.  

%The innermost regions of the disk appear to be very close to the magnetospheric truncation radius, suggesting that an inner clearing process is taking place as gas is embedded by the star. The scenario is consistent with the HD\,141569 doubled-peaked Br$\gamma$ line whose red peak is less intense than the blue one, suggesting constant infall of gas.

\section*{Acknowledgements}

We thank the anonymous referee for reviewing this work in detail and for providing useful comments that have improved the quality of the paper. 
This work is based on observations made with ESO Telescopes at the La Silla Paranal Observatory under program IDs 0102.C-0408(D), 0103.C-0347(C), and 0103.C-0347(A). V.G was supported for this research through a stipend from the International Max Planck Research School (IMPRS) for Astronomy and Astrophysics at the Universities of Bonn and Cologne, and from the Bonn-Cologne Graduate School of Physics and Astronomy (BCGS). 
A. C. G. received fundings from the European Research Council (ERC) under the European Union's Horizon 2020 research and innovation programme (grant agreement No. 743029). 
L. K. acknowledges the helpful discussions that took place during the ISSI meeting "Zooming In On Rocky Planet Formation" (team 482).
We thank Michiel Min for allowing us to use his MCMax code and for his help on understanding the code itself.
We thank Anne-Marie Lagrange, Micka\"el Bonnefoy and Florian Peissker for providing the SINFONI spectrum of HD\,141569. 
We acknowledge the Gemini Observatory for the use of the IR spectrum model of the atmospheric transmission above Cerro Pachon.
This research has made use of the model atmosphere grid NeMo, provided by the Department of Astronomy of the University of Vienna, Austria (\url{http://www.univie.ac.at/nemo/}). NeMo was funded by the Austrian Science Fonds. 

\bibliographystyle{aa}
\bibliography{sample} % if your bibtex file is called example.bib

\begin{thebibliography}{102}
\expandafter\ifx\csname natexlab\endcsname\relax\def\natexlab#1{#1}\fi

\bibitem[{{Absil} {et~al.}(2006){Absil}, {di Folco}, {M{\'e}rand}, {Augereau},
  {Coud{\'e} du Foresto}, {Aufdenberg}, {Kervella}, {Ridgway}, {Berger}, {ten
  Brummelaar}, {Sturmann}, {Sturmann}, {Turner}, \& {McAlister}}]{Absil2006}
{Absil}, O., {di Folco}, E., {M{\'e}rand}, A., {et~al.} 2006, \aap, 452, 237

\bibitem[{{Acke} {et~al.}(2005){Acke}, {van den Ancker}, \&
  {Dullemond}}]{Acke2005}
{Acke}, B., {van den Ancker}, M.~E., \& {Dullemond}, C.~P. 2005, \aap, 436, 209

\bibitem[{{Allard} {et~al.}(1997){Allard}, {Hauschildt}, {Alexander}, \&
  {Starrfield}}]{Allard1997}
{Allard}, F., {Hauschildt}, P.~H., {Alexander}, D.~R., \& {Starrfield}, S.
  1997, \araa, 35, 137

\bibitem[{{ALMA Partnership} {et~al.}(2015){ALMA Partnership}, {Fomalont},
  {Vlahakis}, {Corder}, {Remijan}, {Barkats}, {Lucas}, {Hunter}, {Brogan},
  {Asaki}, \& et~al.}]{ALMA2015}
{ALMA Partnership}, {Fomalont}, E.~B., {Vlahakis}, C., {et~al.} 2015, \apjl,
  808, L1

\bibitem[{{Arun} {et~al.}(2019){Arun}, {Mathew}, {Manoj}, {Ujjwal}, {Kartha},
  {Viswanath}, {Narang}, \& {Paul}}]{Arun2019}
{Arun}, R., {Mathew}, B., {Manoj}, P., {et~al.} 2019, \aj, 157, 159

\bibitem[{{Augereau} {et~al.}(1999){Augereau}, {Lagrange}, {Mouillet}, \&
  {M{\'e}nard}}]{Augereau1999}
{Augereau}, J.~C., {Lagrange}, A.~M., {Mouillet}, D., \& {M{\'e}nard}, F. 1999,
  \aap, 350, L51

\bibitem[{{Augereau} \& {Papaloizou}(2004)}]{Augereau2004}
{Augereau}, J.~C. \& {Papaloizou}, J.~C.~B. 2004, \aap, 414, 1153

\bibitem[{{Beckwith} \& {Sargent}(1993)}]{Beckwith1993}
{Beckwith}, S. V.~W. \& {Sargent}, A.~I. 1993, \apj, 402, 280

\bibitem[{{Benisty} {et~al.}(2017){Benisty}, {Stolker}, {Pohl}, {de Boer},
  {Lesur}, {Dominik}, {Dullemond}, {Langlois}, {Min}, {Wagner}, {Henning},
  {Juhasz}, {Pinilla}, {Facchini}, {Apai}, {van Boekel}, {Garufi}, {Ginski},
  {M{\'e}nard}, {Pinte}, {Quanz}, {Zurlo}, {Boccaletti}, {Bonnefoy}, {Beuzit},
  {Chauvin}, {Cudel}, {Desidera}, {Feldt}, {Fontanive}, {Gratton}, {Kasper},
  {Lagrange}, {LeCoroller}, {Mouillet}, {Mesa}, {Sissa}, {Vigan}, {Antichi},
  {Buey}, {Fusco}, {Gisler}, {Llored}, {Magnard}, {Moeller-Nilsson}, {Pragt},
  {Roelfsema}, {Sauvage}, \& {Wildi}}]{Benisty2017}
{Benisty}, M., {Stolker}, T., {Pohl}, A., {et~al.} 2017, \aap, 597, A42

\bibitem[{{Berger} \& {Segransan}(2007)}]{Berger2007}
{Berger}, J.~P. \& {Segransan}, D. 2007, \nar, 51, 576

\bibitem[{{Beuzit} {et~al.}(2019){Beuzit}, {Vigan}, {Mouillet}, {Dohlen},
  {Gratton}, {Boccaletti}, {Sauvage}, {Schmid}, {Langlois}, \&
  {Petit}}]{Beuzit2019}
{Beuzit}, J.~L., {Vigan}, A., {Mouillet}, D., {et~al.} 2019, arXiv e-prints
  [\eprint[arXiv]{1902.04080}]

\bibitem[{Biller {et~al.}(2015)Biller, Liu, Rice, Wahhaj, Nielsen, Hayward,
  Kuchner, Close, Chun, Ftaclas, \& Toomey}]{Biller2015}
Biller, B.~A., Liu, M.~C., Rice, K., {et~al.} 2015, Monthly Notices of the
  Royal Astronomical Society, 450, 4446

\bibitem[{{Bjorkman} \& {Wood}(2001)}]{Bjorkman2001}
{Bjorkman}, J.~E. \& {Wood}, K. 2001, \apj, 554, 615

\bibitem[{{Brittain} {et~al.}(2007){Brittain}, {Simon}, {Najita}, \&
  {Rettig}}]{Brittain2007}
{Brittain}, S.~D., {Simon}, T., {Najita}, J.~R., \& {Rettig}, T.~W. 2007, \apj,
  659, 685

\bibitem[{{Casassus} {et~al.}(2013){Casassus}, {van der Plas}, {Perez}, {Dent},
  {Fomalont}, {Hagelberg}, {Hales}, {Jord{\'a}n}, {Mawet}, {M{\'e}nard},
  {Wootten}, {Wilner}, {Hughes}, {Schreiber}, {Girard}, {Ercolano}, {Canovas},
  {Rom{\'a}n}, \& {Salinas}}]{Casassus2013}
{Casassus}, S., {van der Plas}, G.~M., {Perez}, S., {et~al.} 2013, \nat, 493,
  191

\bibitem[{{Clampin} {et~al.}(2003){Clampin}, {Krist}, {Ardila}, {Golimowski},
  {Hartig}, {Ford}, {Illingworth}, {Bartko}, {Ben{\'\i}tez}, {Blakeslee},
  {Bouwens}, {Broadhurst}, {Brown}, {Burrows}, {Cheng}, {Cross}, {Feldman},
  {Franx}, {Gronwall}, {Infante}, {Kimble}, {Lesser}, {Martel}, {Menanteau},
  {Meurer}, {Miley}, {Postman}, {Rosati}, {Sirianni}, {Sparks}, {Tran},
  {Tsvetanov}, {White}, \& {Zheng}}]{Clampin2003}
{Clampin}, M., {Krist}, J.~E., {Ardila}, D.~R., {et~al.} 2003, \aj, 126, 385

\bibitem[{{Currie} {et~al.}(2016){Currie}, {Grady}, {Cloutier}, {Konishi},
  {Stassun}, {Debes}, {van der Marel}, {Muto}, {Jayawardhana}, \&
  {Ratzka}}]{Currie2016}
{Currie}, T., {Grady}, C.~A., {Cloutier}, R., {et~al.} 2016, \apjl, 819, L26

\bibitem[{{Dent} {et~al.}(2005){Dent}, {Greaves}, \& {Coulson}}]{Dent2005}
{Dent}, W.~R.~F., {Greaves}, J.~S., \& {Coulson}, I.~M. 2005, \mnras, 359, 663

\bibitem[{{Di Folco} {et~al.}(2020){Di Folco}, {P{\'e}ricaud}, {Dutrey},
  {Augereau}, {Chapillon}, {Guilloteau}, {Pi{\'e}tu}, \&
  {Boccaletti}}]{DiFolco2020}
{Di Folco}, E., {P{\'e}ricaud}, J., {Dutrey}, A., {et~al.} 2020, \aap, 635, A94

\bibitem[{{Draine} \& {Li}(2001)}]{Draine2001}
{Draine}, B.~T. \& {Li}, A. 2001, \apj, 551, 807

\bibitem[{{Eisner} {et~al.}(2009){Eisner}, {Graham}, {Akeson}, \&
  {Najita}}]{Eisner2009}
{Eisner}, J.~A., {Graham}, J.~R., {Akeson}, R.~L., \& {Najita}, J. 2009, \apj,
  692, 309

\bibitem[{{Eisner} {et~al.}(2014){Eisner}, {Hillenbrand}, \&
  {Stone}}]{Eisner2014}
{Eisner}, J.~A., {Hillenbrand}, L.~A., \& {Stone}, J.~M. 2014, \mnras, 443,
  1916

\bibitem[{{Eisner} {et~al.}(2004){Eisner}, {Lane}, {Hillenbrand}, {Akeson}, \&
  {Sargent}}]{Eisner2004}
{Eisner}, J.~A., {Lane}, B.~F., {Hillenbrand}, L.~A., {Akeson}, R.~L., \&
  {Sargent}, A.~I. 2004, \apj, 613, 1049

\bibitem[{{Eisner} {et~al.}(2015){Eisner}, {Rieke}, {Rieke}, {Flaherty},
  {Stone}, {Arnold}, {Cortes}, {Cox}, {Hawkins}, {Cole}, {Zajac}, \&
  {Rudolph}}]{Eisner2015}
{Eisner}, J.~A., {Rieke}, G.~H., {Rieke}, M.~J., {et~al.} 2015, \mnras, 447,
  202

\bibitem[{{Ertel} {et~al.}(2014){Ertel}, {Absil}, {Defr{\`e}re}, {Le Bouquin},
  {Augereau}, {Marion}, {Blind}, {Bonsor}, {Bryden}, {Lebreton}, \&
  {Milli}}]{Ertel2014}
{Ertel}, S., {Absil}, O., {Defr{\`e}re}, D., {et~al.} 2014, \aap, 570, A128

\bibitem[{{Fairlamb} {et~al.}(2015){Fairlamb}, {Oudmaijer}, {Mendigut{\'\i}a},
  {Ilee}, \& {van den Ancker}}]{Fairlamb2015}
{Fairlamb}, J.~R., {Oudmaijer}, R.~D., {Mendigut{\'\i}a}, I., {Ilee}, J.~D., \&
  {van den Ancker}, M.~E. 2015, \mnras, 453, 976

\bibitem[{{Fisher} {et~al.}(2000){Fisher}, {Telesco}, {Pi{\~n}a}, {Knacke}, \&
  {Wyatt}}]{Fisher2000}
{Fisher}, R.~S., {Telesco}, C.~M., {Pi{\~n}a}, R.~K., {Knacke}, R.~F., \&
  {Wyatt}, M.~C. 2000, \apjl, 532, L141

\bibitem[{{Flaherty} {et~al.}(2016){Flaherty}, {Hughes}, {Andrews}, {Qi},
  {Wilner}, {Boley}, {White}, {Harney}, \& {Zachary}}]{Flaherty2016}
{Flaherty}, K.~M., {Hughes}, A.~M., {Andrews}, S.~M., {et~al.} 2016, \apj, 818,
  97

\bibitem[{{Folsom} {et~al.}(2012){Folsom}, {Bagnulo}, {Wade}, {Alecian},
  {Landstreet}, {Marsden}, \& {Waite}}]{Folsom2012}
{Folsom}, C.~P., {Bagnulo}, S., {Wade}, G.~A., {et~al.} 2012, \mnras, 422, 2072

\bibitem[{{Foreman-Mackey} {et~al.}(2013){Foreman-Mackey}, {Hogg}, {Lang}, \&
  {Goodman}}]{Foreman-Mackey2013}
{Foreman-Mackey}, D., {Hogg}, D.~W., {Lang}, D., \& {Goodman}, J. 2013, \pasp,
  125, 306

\bibitem[{{Garcia Lopez} {et~al.}(2006){Garcia Lopez}, {Natta}, {Testi}, \&
  {Habart}}]{Garcia2006}
{Garcia Lopez}, R., {Natta}, A., {Testi}, L., \& {Habart}, E. 2006, \aap, 459,
  837

\bibitem[{{Gonzalez} {et~al.}(2017){Gonzalez}, {Laibe}, \&
  {Maddison}}]{Gonzalez2017}
{Gonzalez}, J.~F., {Laibe}, G., \& {Maddison}, S.~T. 2017, \mnras, 467, 1984

\bibitem[{{Goto} {et~al.}(2006){Goto}, {Usuda}, {Dullemond}, {Henning}, {Linz},
  {Stecklum}, \& {Suto}}]{Goto2006}
{Goto}, M., {Usuda}, T., {Dullemond}, C.~P., {et~al.} 2006, \apj, 652, 758

\bibitem[{{Gravity Collaboration} {et~al.}(2017){Gravity Collaboration},
  {Abuter}, {Accardo}, {Amorim}, {Anugu}, {{\'A}vila}, {Azouaoui}, {Benisty},
  {Berger}, {Blind}, {Bonnet}, {Bourget}, {Brandner}, {Brast}, {Buron},
  {Burtscher}, {Cassaing}, {Chapron}, {Choquet}, {Cl{\'e}net}, {Collin},
  {Coud{\'e} Du Foresto}, {de Wit}, {de Zeeuw}, {Deen},
  {Delplancke-Str{\"o}bele}, {Dembet}, {Derie}, {Dexter}, {Duvert}, {Ebert},
  {Eckart}, {Eisenhauer}, {Esselborn}, {F{\'e}dou}, {Finger}, {Garcia}, {Garcia
  Dabo}, {Garcia Lopez}, {Gendron}, {Genzel}, {Gillessen}, {Gonte}, {Gordo},
  {Grould}, {Gr{\"o}zinger}, {Guieu}, {Haguenauer}, {Hans}, {Haubois}, {Haug},
  {Haussmann}, {Henning}, {Hippler}, {Horrobin}, {Huber}, {Hubert}, {Hubin},
  {Hummel}, {Jakob}, {Janssen}, {Jochum}, {Jocou}, {Kaufer}, {Kellner},
  {Kendrew}, {Kern}, {Kervella}, {Kiekebusch}, {Klein}, {Kok}, {Kolb}, {Kulas},
  {Lacour}, {Lapeyr{\`e}re}, {Lazareff}, {Le Bouquin}, {L{\`e}na}, {Lenzen},
  {L{\'e}v{\^e}que}, {Lippa}, {Magnard}, {Mehrgan}, {Mellein}, {M{\'e}rand},
  {Moreno-Ventas}, {Moulin}, {M{\"u}ller}, {M{\"u}ller}, {Neumann}, {Oberti},
  {Ott}, {Pallanca}, {Panduro}, {Pasquini}, {Paumard}, {Percheron}, {Perraut},
  {Perrin}, {Pfl{\"u}ger}, {Pfuhl}, {Phan Duc}, {Plewa}, {Popovic}, {Rabien},
  {Ram{\'\i}rez}, {Ramos}, {Rau}, {Riquelme}, {Rohloff}, {Rousset},
  {Sanchez-Bermudez}, {Scheithauer}, {Sch{\"o}ller}, {Schuhler}, {Spyromilio},
  {Straubmeier}, {Sturm}, {Suarez}, {Tristram}, {Ventura}, {Vincent},
  {Waisberg}, {Wank}, {Weber}, {Wieprecht}, {Wiest}, {Wiezorrek}, {Wittkowski},
  {Woillez}, {Wolff}, {Yazici}, {Ziegler}, \& {Zins}}]{Gravity2017}
{Gravity Collaboration}, {Abuter}, R., {Accardo}, M., {et~al.} 2017, \aap, 602,
  A94

\bibitem[{{Gravity Collaboration} {et~al.}(2020{\natexlab{a}}){Gravity
  Collaboration}, {Caratti o Garatti}, {Fedriani}, {Garcia Lopez},
  {Koutoulaki}, {Perraut}, {Linz}, {Brandner}, {Garcia}, {Klarmann}, {Henning},
  {Labadie}, {Sanchez-Bermudez}, {Lazareff}, {van Dishoeck}, {Caselli}, {de
  Zeeuw}, {Bik}, {Benisty}, {Dougados}, {Ray}, {Amorim}, {Berger},
  {Cl{\'e}net}, {Coud{\'e} Du Foresto}, {Duvert}, {Eckart}, {Eisenhauer},
  {Gao}, {Gendron}, {Genzel}, {Gillessen}, {Gordo}, {Jocou}, {Horrobin},
  {Kervella}, {Lacour}, {Le Bouquin}, {L{\'e}na}, {Grellmann}, {Ott},
  {Paumard}, {Perrin}, {Rousset}, {Scheithauer}, {Shangguan}, {Stadler},
  {Straub}, {Straubmeier}, {Sturm}, {Thi}, {Vincent}, \&
  {Widmann}}]{Gravity2020IRS2}
{Gravity Collaboration}, {Caratti o Garatti}, A., {Fedriani}, R., {et~al.}
  2020{\natexlab{a}}, \aap, 635, L12

\bibitem[{{Gravity Collaboration} {et~al.}(2020{\natexlab{b}}){Gravity
  Collaboration}, {Garcia Lopez}, {Natta}, {Caratti o Garatti}, {Ray},
  {Fedriani}, {Koutoulaki}, {Klarmann}, {Perraut}, {Sanchez-Bermudez},
  {Benisty}, {Dougados}, {Labadie}, {Brandner}, {Garcia}, {Henning}, {Caselli},
  {Duvert}, {de Zeeuw}, {Grellmann}, {Abuter}, {Amorim}, {Baub{\"o}ck},
  {Berger}, {Bonnet}, {Buron}, {Cl{\'e}net}, {Coud{\'e} Du Foresto}, {de Wit},
  {Eckart}, {Eisenhauer}, {Filho}, {Gao}, {Garcia Dabo}, {Gendron}, {Genzel},
  {Gillessen}, {Habibi}, {Haubois}, {Haussmann}, {Hippler}, {Hubert},
  {Horrobin}, {Jimenez Rosales}, {Jocou}, {Kervella}, {Kolb}, {Lacour}, {Le
  Bouquin}, {L{\'e}na}, {Ott}, {Paumard}, {Perrin}, {Pfuhl}, {Ramirez}, {Rau},
  {Rousset}, {Scheithauer}, {Shangguan}, {Stadler}, {Straub}, {Straubmeier},
  {Sturm}, {van Dishoeck}, {Vincent}, {von Fellenberg}, {Widmann}, {Wieprecht},
  {Wiest}, {Wiezorrek}, {Woillez}, {Yazici}, \& {Zins}}]{Gravity2020TWH}
{Gravity Collaboration}, {Garcia Lopez}, R., {Natta}, A., {et~al.}
  2020{\natexlab{b}}, \nat, 584, 547

\bibitem[{{Gravity Collaboration} {et~al.}(2021){Gravity Collaboration},
  {Koutoulaki}, {Garcia Lopez}, {Natta}, {Fedriani}, {Caratti O Garatti},
  {Ray}, {Coffey}, {Brandner}, {Dougados}, {Garcia}, {Klarmann}, {Labadie},
  {Perraut}, {Sanchez-Bermudez}, {Lin}, {Amorim}, {Baub{\"o}ck}, {Benisty},
  {Berger}, {Buron}, {Caselli}, {Cl{\'e}net}, {Coud{\'e} Du Foresto}, {de
  Zeeuw}, {Duvert}, {de Wit}, {Eckart}, {Eisenhauer}, {Filho}, {Gao},
  {Gendron}, {Genzel}, {Gillessen}, {Grellmann}, {Habibi}, {Haubois},
  {Haussmann}, {Henning}, {Hippler}, {Hubert}, {Horrobin}, {Jimenez Rosales},
  {Jocou}, {Kervella}, {Kolb}, {Lacour}, {Le Bouquin}, {L{\'e}na}, {Linz},
  {Ott}, {Paumard}, {Perrin}, {Pfuhl}, {Ram{\'\i}rez-Tannus}, {Rau}, {Rousset},
  {Scheithauer}, {Shangguan}, {Stadler}, {Straub}, {Straubmeier}, {Sturm}, {van
  Dishoeck}, {Vincent}, {von Fellenberg}, {Widmann}, {Wieprecht}, {Wiest},
  {Wiezorrek}, {Yazici}, \& {Zins}}]{Gravity2021-51Oph}
{Gravity Collaboration}, {Koutoulaki}, M., {Garcia Lopez}, R., {et~al.} 2021,
  \aap, 645, A50

\bibitem[{{Gravity Collaboration} {et~al.}(2019){Gravity Collaboration},
  {Perraut}, {Labadie}, {Lazareff}, {Klarmann}, {Segura-Cox}, {Benisty},
  {Bouvier}, {Brandner}, {Caratti O Garatti}, {Caselli}, {Dougados}, {Garcia},
  {Garcia-Lopez}, {Kendrew}, {Koutoulaki}, {Kervella}, {Lin}, {Pineda},
  {Sanchez-Bermudez}, {van Dishoeck}, {Abuter}, {Amorim}, {Berger}, {Bonnet},
  {Buron}, {Cantalloube}, {Cl{\'e}net}, {Coud{\'e} Du Foresto}, {Dexter}, {de
  Zeeuw}, {Duvert}, {Eckart}, {Eisenhauer}, {Eupen}, {Gao}, {Gendron},
  {Genzel}, {Gillessen}, {Gordo}, {Grellmann}, {Haubois}, {Haussmann},
  {Henning}, {Hippler}, {Horrobin}, {Hubert}, {Jocou}, {Lacour}, {Le Bouquin},
  {L{\'e}na}, {M{\'e}rand}, {Ott}, {Paumard}, {Perrin}, {Pfuhl}, {Rabien},
  {Ray}, {Rau}, {Rousset}, {Scheithauer}, {Straub}, {Straubmeier}, {Sturm},
  {Vincent}, {Waisberg}, {Wank}, {Widmann}, {Wieprecht}, {Wiest}, {Wiezorrek},
  {Woillez}, \& {Yazici}}]{Perraut2019}
{Gravity Collaboration}, {Perraut}, K., {Labadie}, L., {et~al.} 2019, \aap,
  632, A53

\bibitem[{{Gray} \& {Corbally}(1994)}]{Gray1994}
{Gray}, R.~O. \& {Corbally}, C.~J. 1994, \aj, 107, 742

\bibitem[{{Hauschildt} {et~al.}(1999){Hauschildt}, {Allard}, \&
  {Baron}}]{Hauschildt1999}
{Hauschildt}, P.~H., {Allard}, F., \& {Baron}, E. 1999, \apj, 512, 377

\bibitem[{{Heiter} {et~al.}(2002){Heiter}, {Kupka}, {van't Veer-Menneret},
  {Barban}, {Weiss}, {Goupil}, {Schmidt}, {Katz}, \& {Garrido}}]{Heiter2002}
{Heiter}, U., {Kupka}, F., {van't Veer-Menneret}, C., {et~al.} 2002, \aap, 392,
  619

\bibitem[{{Khalafinejad} {et~al.}(2016){Khalafinejad}, {Maaskant},
  {Mari{\~n}as}, \& {Tielens}}]{Khalafinejad2016}
{Khalafinejad}, S., {Maaskant}, K.~M., {Mari{\~n}as}, N., \& {Tielens},
  A.~G.~G.~M. 2016, \aap, 587, A62

\bibitem[{{Klarmann} {et~al.}(2017){Klarmann}, {Benisty}, {Min}, {Dominik},
  {Berger}, {Waters}, {Kluska}, {Lazareff}, \& {Le Bouquin}}]{Klarmann2017}
{Klarmann}, L., {Benisty}, M., {Min}, M., {et~al.} 2017, \aap, 599, A80

\bibitem[{{Kluska} {et~al.}(2018){Kluska}, {Kraus}, {Davies}, {Harries},
  {Willson}, {Monnier}, {Aarnio}, {Baron}, {Millan-Gabet}, {Ten Brummelaar},
  {Che}, {Hinkley}, {Preibisch}, {Sturmann}, {Sturmann}, \&
  {Touhami}}]{Kluska2018}
{Kluska}, J., {Kraus}, S., {Davies}, C.~L., {et~al.} 2018, \apj, 855, 44

\bibitem[{{Konishi} {et~al.}(2016){Konishi}, {Grady}, {Schneider}, {Shibai},
  {McElwain}, {Nesvold}, {Kuchner}, {Carson}, {Debes}, {Gaspar}, {Henning},
  {Hines}, {Hinz}, {Jang-Condell}, {Moro-Mart{\'\i}n}, {Perrin}, {Rodigas},
  {Serabyn}, {Silverstone}, {Stark}, {Tamura}, {Weinberger}, \&
  {Wisniewski}}]{Konishi2016}
{Konishi}, M., {Grady}, C.~A., {Schneider}, G., {et~al.} 2016, \apjl, 818, L23

\bibitem[{{K{\'o}sp{\'a}l} {et~al.}(2012){K{\'o}sp{\'a}l}, {{\'A}brah{\'a}m},
  {Acosta-Pulido}, {Dullemond}, {Henning}, {Kun}, {Leinert}, {Mo{\'o}r}, \&
  {Turner}}]{Kospal2012}
{K{\'o}sp{\'a}l}, {\'A}., {{\'A}brah{\'a}m}, P., {Acosta-Pulido}, J.~A.,
  {et~al.} 2012, \apjs, 201, 11

\bibitem[{{Kral} {et~al.}(2019){Kral}, {Marino}, {Wyatt}, {Kama}, \&
  {Matr{\`a}}}]{Kral2019}
{Kral}, Q., {Marino}, S., {Wyatt}, M.~C., {Kama}, M., \& {Matr{\`a}}, L. 2019,
  \mnras, 489, 3670

\bibitem[{{Kraus} {et~al.}(2008){Kraus}, {Hofmann}, {Benisty}, {Berger},
  {Chesneau}, {Isella}, {Malbet}, {Meilland}, {Nardetto}, {Natta}, {Preibisch},
  {Schertl}, {Smith}, {Stee}, {Tatulli}, {Testi}, \& {Weigelt}}]{Kraus2008}
{Kraus}, S., {Hofmann}, K.~H., {Benisty}, M., {et~al.} 2008, \aap, 489, 1157

\bibitem[{{Lacour} {et~al.}(2019){Lacour}, {Dembet}, {Abuter}, {F{\'e}dou},
  {Perrin}, {Choquet}, {Pfuhl}, {Eisenhauer}, {Woillez}, {Cassaing},
  {Wieprecht}, {Ott}, {Wiezorrek}, {Tristram}, {Wolff}, {Ram{\'\i}rez},
  {Haubois}, {Perraut}, {Straubmeier}, {Brand ner}, \& {Amorim}}]{Lacour2019}
{Lacour}, S., {Dembet}, R., {Abuter}, R., {et~al.} 2019, \aap, 624, A99

\bibitem[{{Lapeyrere} {et~al.}(2014){Lapeyrere}, {Kervella}, {Lacour},
  {Azouaoui}, {Garcia-Dabo}, {Perrin}, {Eisenhauer}, {Perraut}, {Straubmeier},
  {Amorim}, \& {Brandner}}]{Lapeyrere2014}
{Lapeyrere}, V., {Kervella}, P., {Lacour}, S., {et~al.} 2014, in Society of
  Photo-Optical Instrumentation Engineers (SPIE) Conference Series, Vol. 9146,
  \procspie, 91462D

\bibitem[{{Lazareff} {et~al.}(2017){Lazareff}, {Berger}, {Kluska}, {Le
  Bouquin}, {Benisty}, {Malbet}, {Koen}, {Pinte}, {Thi}, {Absil}, {Baron},
  {Delboulb{\'e}}, {Duvert}, {Isella}, {Jocou}, {Juhasz}, {Kraus}, {Lachaume},
  {M{\'e}nard}, {Millan-Gabet}, {Monnier}, {Moulin}, {Perraut}, {Rochat},
  {Soulez}, {Tallon}, {Thi{\'e}baut}, {Traub}, \& {Zins}}]{Lazareff2017}
{Lazareff}, B., {Berger}, J.~P., {Kluska}, J., {et~al.} 2017, \aap, 599, A85

\bibitem[{{Le Bouquin} {et~al.}(2009){Le Bouquin}, {Absil}, {Benisty}, {Massi},
  {M{\'e}rand}, \& {Stefl}}]{Bouquin2009}
{Le Bouquin}, J.~B., {Absil}, O., {Benisty}, M., {et~al.} 2009, \aap, 498, L41

\bibitem[{{Li} \& {Lunine}(2003)}]{Li2003}
{Li}, A. \& {Lunine}, J.~I. 2003, \apj, 594, 987

\bibitem[{{Lodato} {et~al.}(2019){Lodato}, {Dipierro}, {Ragusa}, {Long},
  {Herczeg}, {Pascucci}, {Pinilla}, {Manara}, {Tazzari}, {Liu}, {Mulders},
  {Harsono}, {Boehler}, {M{\'e}nard}, {Johnstone}, {Salyk}, {van der Plas},
  {Cabrit}, {Edwards}, {Fischer}, {Hendler}, {Nisini}, {Rigliaco}, {Avenhaus},
  {Banzatti}, \& {Gully-Santiago}}]{Lodato2019}
{Lodato}, G., {Dipierro}, G., {Ragusa}, E., {et~al.} 2019, \mnras, 486, 453

\bibitem[{{Lopez} {et~al.}(2014){Lopez}, {Lagarde}, {Jaffe}, {Petrov},
  {Sch{\"o}ller}, {Antonelli}, {Beckmann}, {Berio}, {Bettonvil}, {Glindemann},
  {Gonzalez}, {Graser}, {Hofmann}, {Millour}, {Robbe-Dubois}, {Venema}, {Wolf},
  {Henning}, {Lanz}, {Weigelt}, {Agocs}, {Bailet}, {Bresson}, {Bristow},
  {Dugu{\'e}}, {Heininger}, {Kroes}, {Laun}, {Lehmitz}, {Neumann}, {Augereau},
  {Avila}, {Behrend}, {van Belle}, {Berger}, {van Boekel}, {Bonhomme},
  {Bourget}, {Brast}, {Clausse}, {Connot}, {Conzelmann}, {Cruzal{\`e}bes},
  {Csepany}, {Danchi}, {Delbo}, {Delplancke}, {Dominik}, {van Duin}, {Elswijk},
  {Fantei}, {Finger}, {Gabasch}, {Gay}, {Girard}, {Girault}, {Gitton},
  {Glazenborg}, {Gont{\'e}}, {Guitton}, {Guniat}, {De Haan}, {Haguenauer},
  {Hanenburg}, {Hogerheijde}, {ter Horst}, {Hron}, {Hugues}, {Hummel},
  {Idserda}, {Ives}, {Jakob}, {Jasko}, {Jolley}, {Kiraly}, {K{\"o}hler},
  {Kragt}, {Kroener}, {Kuindersma}, {Labadie}, {Leinert}, {Le Poole}, {Lizon},
  {Lucuix}, {Marcotto}, {Martinache}, {Martinot-Lagarde}, {Mathar}, {Matter},
  {Mauclert}, {Mehrgan}, {Meilland}, {Meisenheimer}, {Meisner}, {Mellein},
  {Menardi}, {Menut}, {Merand}, {Morel}, {Mosoni}, {Navarro}, {Nussbaum},
  {Ottogalli}, {Palsa}, {Panduro}, {Pantin}, {Parra}, {Percheron}, {Duc},
  {Pott}, {Pozna}, {Przygodda}, {Rabbia}, {Richichi}, {Rigal}, {Roelfsema},
  {Rupprecht}, {Schertl}, {Schmidt}, {Schuhler}, {Schuil}, {Spang},
  {Stegmeier}, {Thiam}, {Tromp}, {Vakili}, {Vannier}, {Wagner}, \&
  {Woillez}}]{Lopez2014}
{Lopez}, B., {Lagarde}, S., {Jaffe}, W., {et~al.} 2014, The Messenger, 157, 5

\bibitem[{{Lord}(1992)}]{Lord1992}
{Lord}, S.~D. 1992, {A new software tool for computing Earth's atmospheric
  transmission of near- and far-infrared radiation}, NASA Technical Memorandum
  103957

\bibitem[{{Maaskant} {et~al.}(2013){Maaskant}, {Honda}, {Waters}, {Tielens},
  {Dominik}, {Min}, {Verhoeff}, {Meeus}, \& {van den Ancker}}]{Maaskant2013}
{Maaskant}, K.~M., {Honda}, M., {Waters}, L.~B.~F.~M., {et~al.} 2013, \aap,
  555, A64

\bibitem[{{Maaskant} {et~al.}(2014){Maaskant}, {Min}, {Waters}, \&
  {Tielens}}]{Maaskant2014}
{Maaskant}, K.~M., {Min}, M., {Waters}, L.~B.~F.~M., \& {Tielens}, A.~G.~G.~M.
  2014, \aap, 563, A78

\bibitem[{{Malfait} {et~al.}(1998){Malfait}, {Waelkens}, {Waters},
  {Vandenbussche}, {Huygen}, \& {de Graauw}}]{Malfait1998}
{Malfait}, K., {Waelkens}, C., {Waters}, L.~B.~F.~M., {et~al.} 1998, \aap, 332,
  L25

\bibitem[{{Marsh} {et~al.}(2002){Marsh}, {Silverstone}, {Becklin}, {Koerner},
  {Werner}, {Weinberger}, \& {Ressler}}]{Marsh2002}
{Marsh}, K.~A., {Silverstone}, M.~D., {Becklin}, E.~E., {et~al.} 2002, \apj,
  573, 425

\bibitem[{Mawet {et~al.}(2017)Mawet, Choquet, Absil, Huby, Bottom, Serabyn,
  Femenia, Lebreton, Matthews, Gonzalez, Wertz, Carlomagno, Christiaens,
  Defr{\`{e}}re, Delacroix, Forsberg, Habraken, Jolivet, Karlsson, Milli,
  Pinte, Piron, Reggiani, Surdej, \& Catalan}]{Mawet2017}
Mawet, D., Choquet, {\'{E}}., Absil, O., {et~al.} 2017, The Astronomical
  Journal, 153, 44

\bibitem[{{Meeus} {et~al.}(2012){Meeus}, {Montesinos}, {Mendigut{\'\i}a},
  {Kamp}, {Thi}, {Eiroa}, {Grady}, {Mathews}, {Sandell}, {Martin-Za{\"\i}di},
  {Brittain}, {Dent}, {Howard}, {M{\'e}nard}, {Pinte}, {Roberge},
  {Vandenbussche}, \& {Williams}}]{Meeus2012}
{Meeus}, G., {Montesinos}, B., {Mendigut{\'\i}a}, I., {et~al.} 2012, \aap, 544,
  A78

\bibitem[{{Meeus} {et~al.}(2001){Meeus}, {Waters}, {Bouwman}, {van den Ancker},
  {Waelkens}, \& {Malfait}}]{Meus2001}
{Meeus}, G., {Waters}, L.~B.~F.~M., {Bouwman}, J., {et~al.} 2001, \aap, 365,
  476

\bibitem[{{Mendigut{\'\i}a} {et~al.}(2011{\natexlab{a}}){Mendigut{\'\i}a},
  {Calvet}, {Montesinos}, {Mora}, {Muzerolle}, {Eiroa}, {Oudmaijer}, \&
  {Mer{\'\i}n}}]{Mendigutia2011}
{Mendigut{\'\i}a}, I., {Calvet}, N., {Montesinos}, B., {et~al.}
  2011{\natexlab{a}}, \aap, 535, A99

\bibitem[{{Mendigut{\'\i}a} {et~al.}(2015){Mendigut{\'\i}a}, {de Wit},
  {Oudmaijer}, {Fairlamb}, {Carciofi}, {Ilee}, \& {Vieira}}]{Mendigutia2015}
{Mendigut{\'\i}a}, I., {de Wit}, W.~J., {Oudmaijer}, R.~D., {et~al.} 2015,
  \mnras, 453, 2126

\bibitem[{{Mendigut{\'\i}a} {et~al.}(2011{\natexlab{b}}){Mendigut{\'\i}a},
  {Eiroa}, {Montesinos}, {Mora}, {Oudmaijer}, {Mer{\'\i}n}, \&
  {Meeus}}]{Mendigutia2011a}
{Mendigut{\'\i}a}, I., {Eiroa}, C., {Montesinos}, B., {et~al.}
  2011{\natexlab{b}}, \aap, 529, A34

\bibitem[{{Mendigut{\'\i}a} {et~al.}(2017){Mendigut{\'\i}a}, {Oudmaijer},
  {Mourard}, \& {Muzerolle}}]{Mendigutia2017}
{Mendigut{\'\i}a}, I., {Oudmaijer}, R.~D., {Mourard}, D., \& {Muzerolle}, J.
  2017, \mnras, 464, 1984

\bibitem[{{Menu} {et~al.}(2015){Menu}, {van Boekel}, {Henning}, {Leinert},
  {Waelkens}, \& {Waters}}]{Menu2015}
{Menu}, J., {van Boekel}, R., {Henning}, T., {et~al.} 2015, \aap, 581, A107

\bibitem[{{Mer{\'\i}n} {et~al.}(2004){Mer{\'\i}n}, {Montesinos}, {Eiroa},
  {Solano}, {Mora}, {D'Alessio}, {Calvet}, {Oudmaijer}, {de Winter}, \&
  {Davies}}]{Merin2004}
{Mer{\'\i}n}, B., {Montesinos}, B., {Eiroa}, C., {et~al.} 2004, \aap, 419, 301

\bibitem[{{Miley} {et~al.}(2018){Miley}, {Pani{\'c}}, {Wyatt}, \&
  {Kennedy}}]{Miley2018}
{Miley}, J.~M., {Pani{\'c}}, O., {Wyatt}, M., \& {Kennedy}, G.~M. 2018, \aap,
  615, L10

\bibitem[{{Millan-Gabet} {et~al.}(2001){Millan-Gabet}, {Schloerb}, \&
  {Traub}}]{MillanGabet2001}
{Millan-Gabet}, R., {Schloerb}, F.~P., \& {Traub}, W.~A. 2001, \apj, 546, 358

\bibitem[{{Min} {et~al.}(2009){Min}, {Dullemond}, {Dominik}, {de Koter}, \&
  {Hovenier}}]{Min2009}
{Min}, M., {Dullemond}, C.~P., {Dominik}, C., {de Koter}, A., \& {Hovenier},
  J.~W. 2009, \aap, 497, 155

\bibitem[{{Moerchen} {et~al.}(2010){Moerchen}, {Telesco}, \&
  {Packham}}]{Moerchen2010}
{Moerchen}, M.~M., {Telesco}, C.~M., \& {Packham}, C. 2010, \apj, 723, 1418

\bibitem[{{Monnier} {et~al.}(2005){Monnier}, {Millan-Gabet}, {Billmeier},
  {Akeson}, {Wallace}, {Berger}, {Calvet}, {D'Alessio}, {Danchi}, {Hartmann},
  {Hillenbrand}, {Kuchner}, {Rajagopal}, {Traub}, {Tuthill}, {Boden}, {Booth},
  {Colavita}, {Gathright}, {Hrynevych}, {Le Mignant}, {Ligon}, {Neyman},
  {Swain}, {Thompson}, {Vasisht}, {Wizinowich}, {Beichman}, {Beletic},
  {Creech-Eakman}, {Koresko}, {Sargent}, {Shao}, \& {van Belle}}]{Monnier2005}
{Monnier}, J.~D., {Millan-Gabet}, R., {Billmeier}, R., {et~al.} 2005, \apj,
  624, 832

\bibitem[{{Mo{\'o}r} {et~al.}(2011){Mo{\'o}r}, {{\'A}brah{\'a}m}, {Juh{\'a}sz},
  {Kiss}, {Pascucci}, {K{\'o}sp{\'a}l}, {Apai}, {Henning}, {Csengeri}, \&
  {Grady}}]{Moor2011}
{Mo{\'o}r}, A., {{\'A}brah{\'a}m}, P., {Juh{\'a}sz}, A., {et~al.} 2011, \apjl,
  740, L7

\bibitem[{{Mo{\'o}r} {et~al.}(2017){Mo{\'o}r}, {Cur{\'e}}, {K{\'o}sp{\'a}l},
  {{\'A}brah{\'a}m}, {Csengeri}, {Eiroa}, {Gunawan}, {Henning}, {Hughes},
  {Juh{\'a}sz}, {Pawellek}, \& {Wyatt}}]{Moor2017}
{Mo{\'o}r}, A., {Cur{\'e}}, M., {K{\'o}sp{\'a}l}, {\'A}., {et~al.} 2017, \apj,
  849, 123

\bibitem[{{Mo{\'o}r} {et~al.}(2015){Mo{\'o}r}, {Henning}, {Juh{\'a}sz},
  {{\'A}brah{\'a}m}, {Balog}, {K{\'o}sp{\'a}l}, {Pascucci}, {Szab{\'o}},
  {Vavrek}, {Cur{\'e}}, {Csengeri}, {Grady}, {G{\"u}sten}, \&
  {Kiss}}]{Moor2015}
{Mo{\'o}r}, A., {Henning}, T., {Juh{\'a}sz}, A., {et~al.} 2015, \apj, 814, 42

\bibitem[{{Mo{\'o}r} {et~al.}(2019){Mo{\'o}r}, {Kral}, {{\'A}brah{\'a}m},
  {K{\'o}sp{\'a}l}, {Dutrey}, {Di Folco}, {Hughes}, {Juh{\'a}sz}, {Pascucci},
  \& {Pawellek}}]{Moor2019}
{Mo{\'o}r}, A., {Kral}, Q., {{\'A}brah{\'a}m}, P., {et~al.} 2019, \apj, 884,
  108

\bibitem[{{Nilsson} {et~al.}(2010){Nilsson}, {Liseau}, {Brandeker}, {Olofsson},
  {Pilbratt}, {Risacher}, {Rodmann}, {Augereau}, {Bergman}, {Eiroa},
  {Fridlund}, {Th{\'e}bault}, \& {White}}]{Nilsson2010}
{Nilsson}, R., {Liseau}, R., {Brandeker}, A., {et~al.} 2010, \aap, 518, A40

\bibitem[{{Penprase}(1992)}]{Penprase1992}
{Penprase}, B.~E. 1992, \apjs, 83, 273

\bibitem[{{P{\'e}ricaud} {et~al.}(2017){P{\'e}ricaud}, {Di Folco}, {Dutrey},
  {Guilloteau}, \& {Pi{\'e}tu}}]{Pericaud2017}
{P{\'e}ricaud}, J., {Di Folco}, E., {Dutrey}, A., {Guilloteau}, S., \&
  {Pi{\'e}tu}, V. 2017, \aap, 600, A62

\bibitem[{{Perrot} {et~al.}(2016){Perrot}, {Boccaletti}, {Pantin}, {Augereau},
  {Lagrange}, {Galicher}, {Maire}, {Mazoyer}, {Milli}, {Rousset}, {Gratton},
  {Bonnefoy}, {Brandner}, {Buenzli}, {Langlois}, {Lannier}, {Mesa}, {Peretti},
  {Salter}, {Sissa}, {Chauvin}, {Desidera}, {Feldt}, {Vigan}, {Di Folco},
  {Dutrey}, {P{\'e}ricaud}, {Baudoz}, {Benisty}, {De Boer}, {Garufi}, {Girard},
  {Menard}, {Olofsson}, {Quanz}, {Mouillet}, {Christiaens}, {Casassus},
  {Beuzit}, {Blanchard}, {Carle}, {Fusco}, {Giro}, {Hubin}, {Maurel},
  {Moeller-Nilsson}, {Sevin}, \& {Weber}}]{Perrot2016}
{Perrot}, C., {Boccaletti}, A., {Pantin}, E., {et~al.} 2016, \aap, 590, L7

\bibitem[{{Pinilla} {et~al.}(2012){Pinilla}, {Benisty}, \&
  {Birnstiel}}]{Pinilla2012}
{Pinilla}, P., {Benisty}, M., \& {Birnstiel}, T. 2012, \aap, 545, A81

\bibitem[{{Purcell}(1976)}]{Purcell1976}
{Purcell}, E.~M. 1976, \apj, 206, 685

\bibitem[{{Reche} {et~al.}(2009){Reche}, {Beust}, \& {Augereau}}]{Reche2009}
{Reche}, R., {Beust}, H., \& {Augereau}, J.~C. 2009, \aap, 493, 661

\bibitem[{{Sandell} {et~al.}(2011){Sandell}, {Weintraub}, \&
  {Hamidouche}}]{Sandell2011}
{Sandell}, G., {Weintraub}, D.~A., \& {Hamidouche}, M. 2011, \apj, 727, 26

\bibitem[{{Seok} \& {Li}(2017)}]{Seok2017}
{Seok}, J.~Y. \& {Li}, A. 2017, \apj, 835, 291

\bibitem[{{Sloan} {et~al.}(2005){Sloan}, {Keller}, {Forrest}, {Leibensperger},
  {Sargent}, {Li}, {Najita}, {Watson}, {Brandl}, {Chen}, {Green},
  {Markwick-Kemper}, {Herter}, {D'Alessio}, {Morris}, {Barry}, {Hall}, {Myers},
  \& {Houck}}]{Sloan2005}
{Sloan}, G.~C., {Keller}, L.~D., {Forrest}, W.~J., {et~al.} 2005, \apj, 632,
  956

\bibitem[{{Sylvester} {et~al.}(2001){Sylvester}, {Dunkin}, \&
  {Barlow}}]{Sylvester2001}
{Sylvester}, R.~J., {Dunkin}, S.~K., \& {Barlow}, M.~J. 2001, \mnras, 327, 133

\bibitem[{{Sylvester} {et~al.}(1996){Sylvester}, {Skinner}, {Barlow}, \&
  {Mannings}}]{Sylvester1996}
{Sylvester}, R.~J., {Skinner}, C.~J., {Barlow}, M.~J., \& {Mannings}, V. 1996,
  \mnras, 279, 915

\bibitem[{{Taha} {et~al.}(2018){Taha}, {Labadie}, {Pantin}, {Matter},
  {Alvarez}, {Esquej}, {Grellmann}, {Rebolo}, {Telesco}, \& {Wolf}}]{Taha2018}
{Taha}, A.~S., {Labadie}, L., {Pantin}, E., {et~al.} 2018, \aap, 612, A15

\bibitem[{{Thi} {et~al.}(2014){Thi}, {Pinte}, {Pantin}, {Augereau}, {Meeus},
  {M{\'e}nard}, {Martin-Za{\"\i}di}, {Woitke}, {Riviere-Marichalar}, {Kamp},
  {Carmona}, {Sandell}, {Eiroa}, {Dent}, {Montesinos}, {Aresu}, {Meijerink},
  {Spaans}, {White}, {Ardila}, {Lebreton}, {Mendigut{\'\i}a}, \&
  {Brittain}}]{Thi2014}
{Thi}, W.~F., {Pinte}, C., {Pantin}, E., {et~al.} 2014, \aap, 561, A50

\bibitem[{{Tominaga} {et~al.}(2020){Tominaga}, {Takahashi}, \&
  {Inutsuka}}]{Tominaga2020}
{Tominaga}, R.~T., {Takahashi}, S.~Z., \& {Inutsuka}, S.-i. 2020, \apj, 900,
  182

\bibitem[{{van der Plas} {et~al.}(2015){van der Plas}, {van den Ancker},
  {Waters}, \& {Dominik}}]{Vanderplas2015}
{van der Plas}, G., {van den Ancker}, M.~E., {Waters}, L.~B.~F.~M., \&
  {Dominik}, C. 2015, \aap, 574, A75

\bibitem[{{Vioque} {et~al.}(2018){Vioque}, {Oudmaijer}, {Baines},
  {Mendigut{\'\i}a}, \& {P{\'e}rez-Mart{\'\i}nez}}]{Vioque2018}
{Vioque}, M., {Oudmaijer}, R.~D., {Baines}, D., {Mendigut{\'\i}a}, I., \&
  {P{\'e}rez-Mart{\'\i}nez}, R. 2018, \aap, 620, A128

\bibitem[{{Weigelt} {et~al.}(2011){Weigelt}, {Baur}, {Reubelt}, {Sneeuw}, \&
  {Roth}}]{Weigelt2011}
{Weigelt}, M., {Baur}, O., {Reubelt}, T., {Sneeuw}, N., \& {Roth}, M. 2011, in
  ESA Special Publication, Vol. 696, 4th International GOCE User Workshop, 36

\bibitem[{{Weinberger} {et~al.}(1999){Weinberger}, {Becklin}, {Schneider},
  {Smith}, {Lowrance}, {Silverstone}, {Zuckerman}, \&
  {Terrile}}]{Weinberger1999}
{Weinberger}, A.~J., {Becklin}, E.~E., {Schneider}, G., {et~al.} 1999, \apjl,
  525, L53

\bibitem[{White {et~al.}(2016)White, Boley, Hughes, Flaherty, Ford, Wilner,
  Corder, \& Payne}]{White2016}
White, J.~A., Boley, A.~C., Hughes, A.~M., {et~al.} 2016, The Astrophysical
  Journal, 829, 6

\bibitem[{{Woitke} {et~al.}(2016){Woitke}, {Min}, {Pinte}, {Thi}, {Kamp},
  {Rab}, {Anthonioz}, {Antonellini}, {Baldovin-Saavedra}, {Carmona}, {Dominik},
  {Dionatos}, {Greaves}, {G{\"u}del}, {Ilee}, {Liebhart}, {M{\'e}nard},
  {Rigon}, {Waters}, {Aresu}, {Meijerink}, \& {Spaans}}]{Woitke2016}
{Woitke}, P., {Min}, M., {Pinte}, C., {et~al.} 2016, \aap, 586, A103

\bibitem[{{Wyatt}(2005)}]{Wyatt2005}
{Wyatt}, M.~C. 2005, \aap, 440, 937

\bibitem[{{Wyatt}(2008)}]{Wyatt2008}
{Wyatt}, M.~C. 2008, \araa, 46, 339

\bibitem[{{Wyatt} {et~al.}(2015){Wyatt}, {Pani{\'c}}, {Kennedy}, \&
  {Matr{\`a}}}]{Wyatt2015}
{Wyatt}, M.~C., {Pani{\'c}}, O., {Kennedy}, G.~M., \& {Matr{\`a}}, L. 2015,
  \apss, 357, 103

\end{thebibliography}

\appendix

\section{GRAVITY observations}

\begin{table*}[t]
\centering
\begin{small}
\caption{\centering Observation logs of the VLTI/GRAVITY HD141569 observations.}
\begin{tabular}{ccccccccc} 
\hline
\hline
\multicolumn{1}{c}{Date} & \multicolumn{1}{c}{UT} & \multicolumn{1}{c}{Configuration} & \multicolumn{1}{c}{N} & \multicolumn{1}{c}{Calibrator (diam. [mas], SpT $^{\dagger}$)} & \multicolumn{1}{c}{Seeing\,[$^{\prime\prime}$]} & \multicolumn{1}{c}{Airmass} & \multicolumn{1}{c}{$\tau_0$\,[\rm ms]} & \multicolumn{1}{c}{Frame rate [\rm Hz]} \\
\hline
%yyyy-mm-dd & & &\\
% \hline
2019-03-18 & 08:08 & D0-G2-J3-K0 & 3 & HD\,137006 (0.291$\pm0.008$, A5\,II/III) & 0.4 & 1.1 & 14 & 909\\
2019-05-24 & 04:15 & A0-G1-J2-J3 & 1 & HD\,141977 (0.251$\pm0.006$, K3\,III)& 1.4 & 1.1 & 3 & 303\\
2019-07-12 & 00:44 & D0-G2-J3-K0 & 8 & HD\,137006 (0.291$\pm0.008$, A5\,II/III) & 0.6\,--\,1.1& 1.1 & 3 & 909  \\
\hline
\end{tabular}\label{tab:ObsLog}
\tablefoot{The date format is year-month-day. N denotes the number of 5-minute  files that have been recorded on the target. Reference $^{\dagger}$: uniform disk diameter and SpT derived from the software package SearchCal from the Jean-Marie Mariotti Center (JMMC).}
%FT mode of 2 corresponds to a tracking speed of 909\,Hz (0.00085\,s) and FT mode of 7 corresponds to a tracking speed of 303\,Hz (0.0033\,s).}
\end{small}
\end{table*}

\bigskip
\bigskip

\begin{figure*}
\centering
\includegraphics[width=\textwidth]{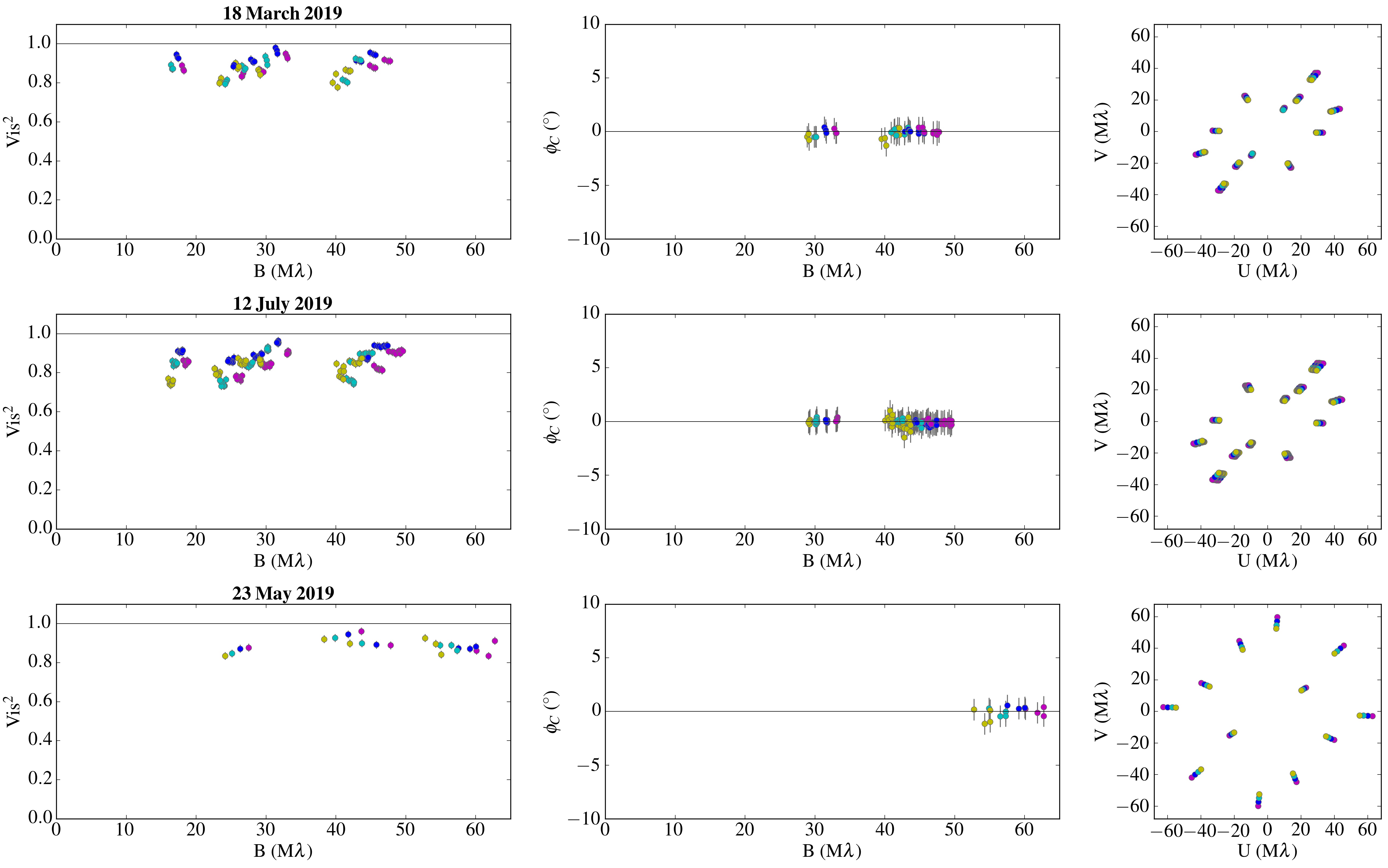}
\caption{HD\,141569 FT data, squared visibilities, closure phases, and U-V plane coverage, from the three different observation epochs. The colors refer to the different GRAVITY spectral channels.}
\label{fig:FT-Data-3epochs}
\end{figure*}

\newpage
\clearpage
\section{MCMC posterior distribution functions and visibility modeling}
\label{apx:VIS2_MCMC}

% -----
\begin{figure*}[!htbp]
\centering
\includegraphics[scale=0.13]{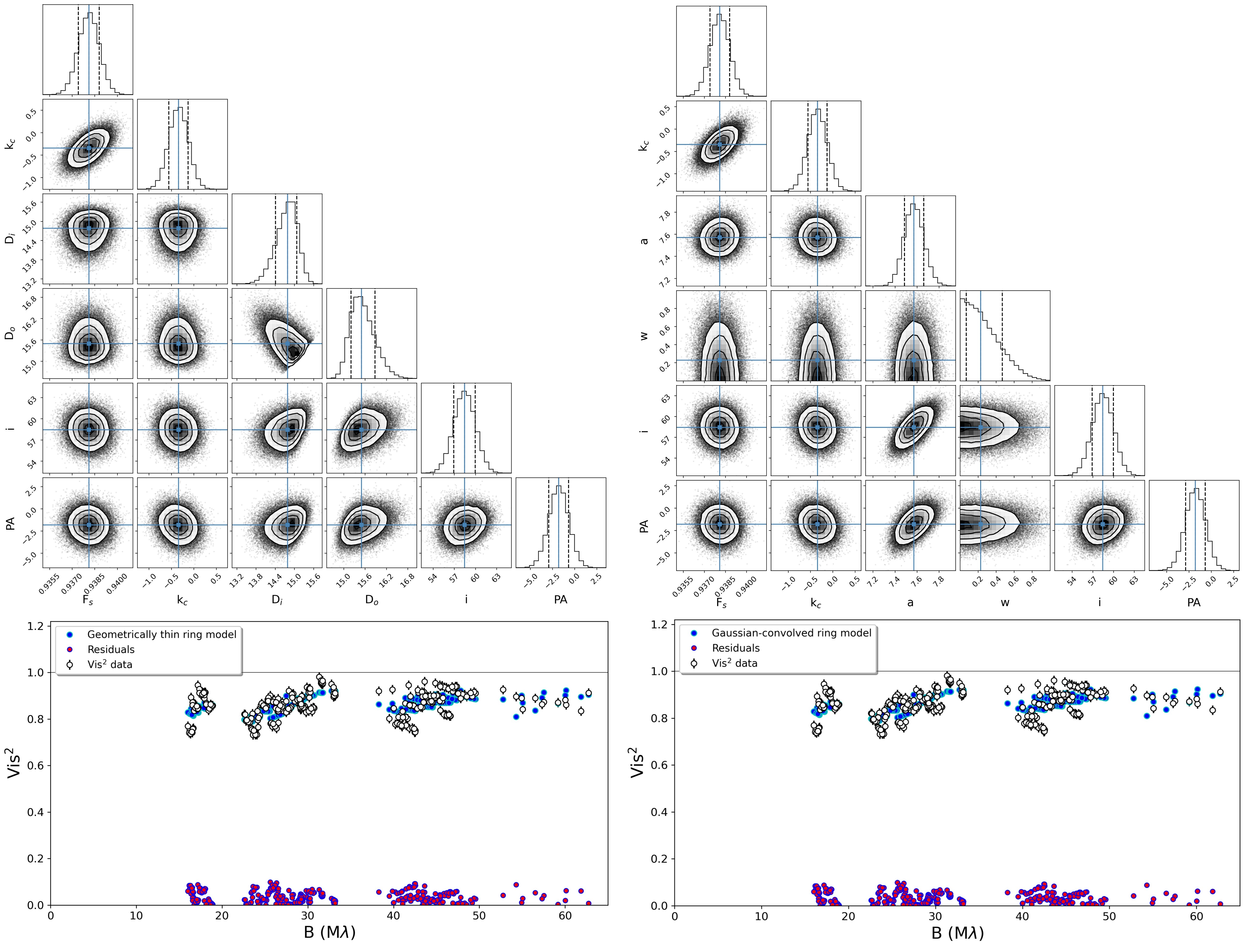}
\includegraphics[scale=0.137]{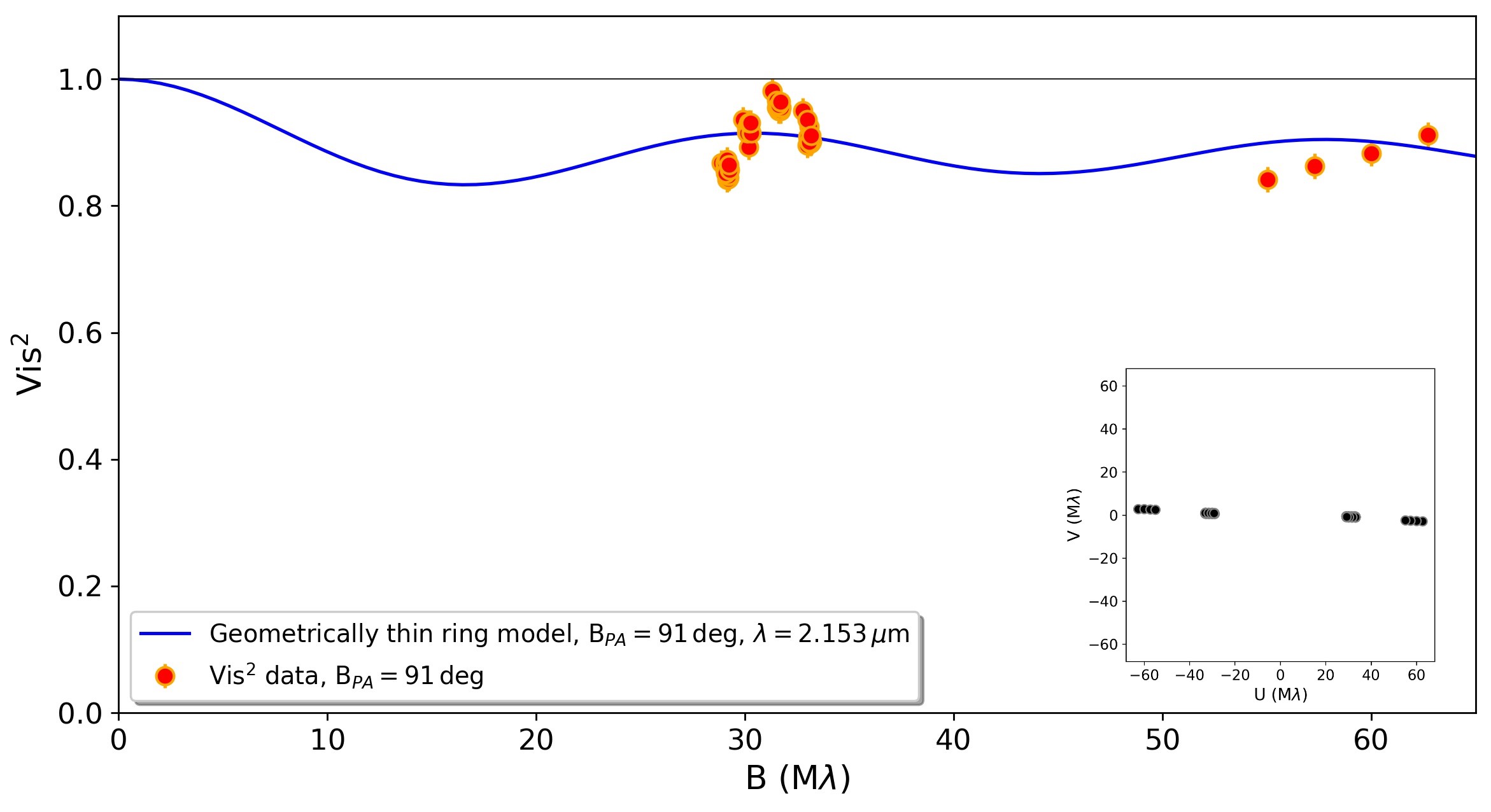} \,
\includegraphics[scale=0.137]{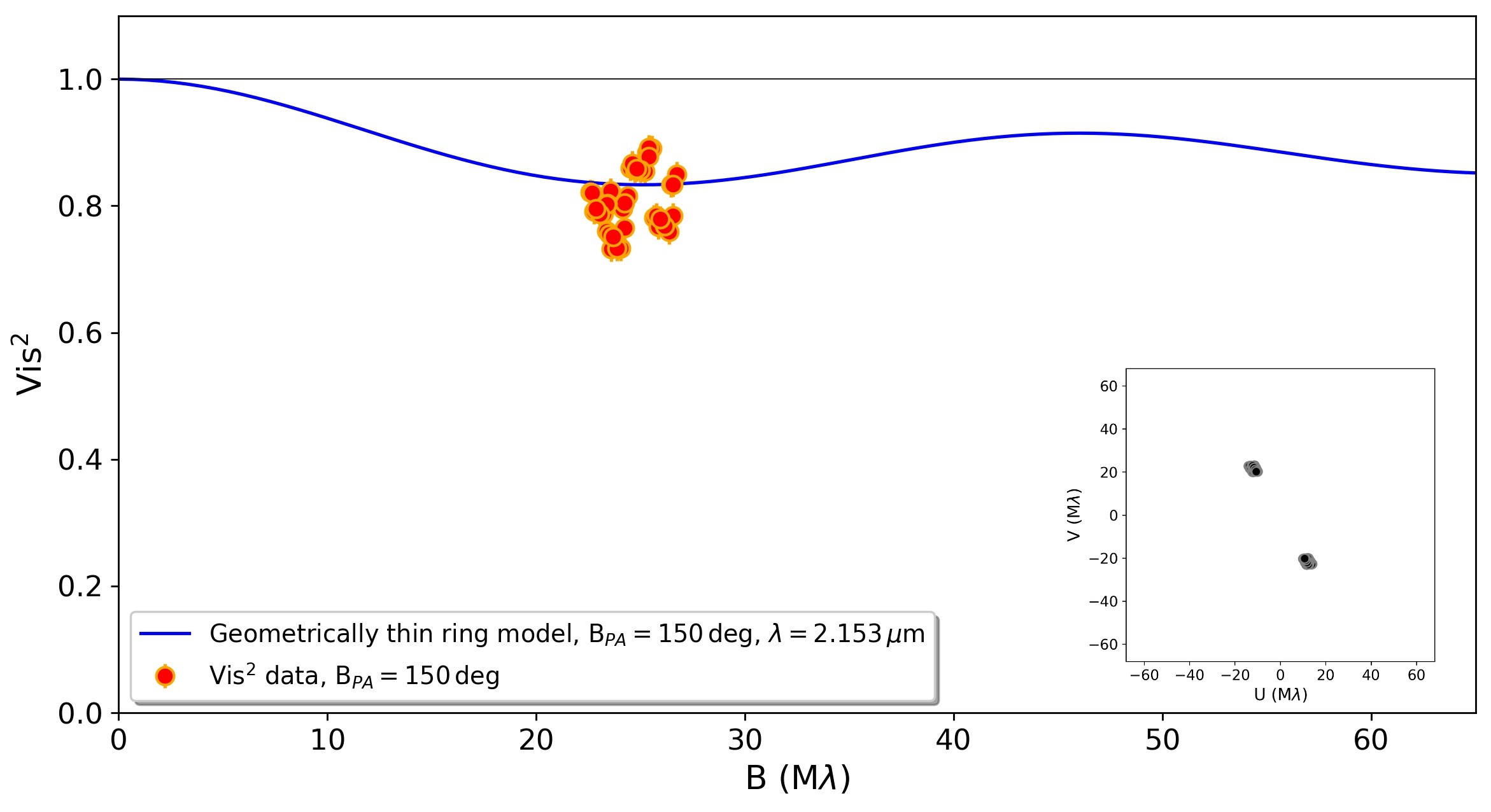}
\includegraphics[scale=0.137]{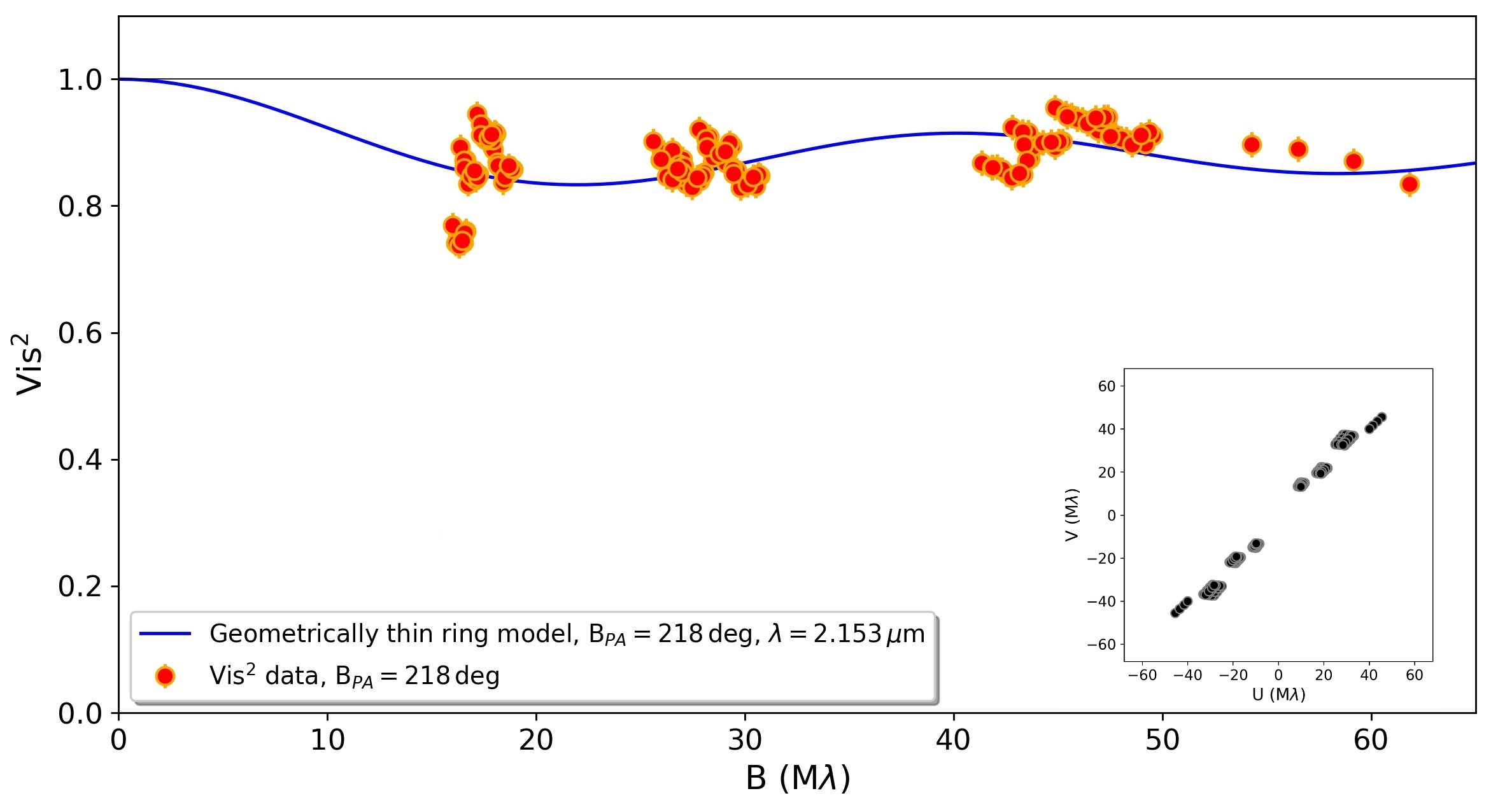}
\caption{Results of the FT data fit. Top: MCMC marginal posterior distributions of the fitted parameters. The blue lines identify the median of the distributions for the geometrically thin ring model (left) and for the Gaussian-convolved infinitesimally thin ring model (right). Center: 
Comparison between the model squared visibilities (blue dots) and the observational data (white--black dots). The red dots represent the absolute residuals.
% result of the fit to the observational data for the two models discussed and corresponding residuals. 
Bottom: Vis$^2$ best fit to the geometrically thin ring model (blue continuous line) at $\lambda$=2.153\,$\mu$m for three different position angles identified in the insets. The visibility data are shown for all spectral channels by the red circles.}
\label{fig:Continuum results}
\end{figure*}
% -----
\begin{figure*}[thpb!]
\includegraphics[width=\textwidth]{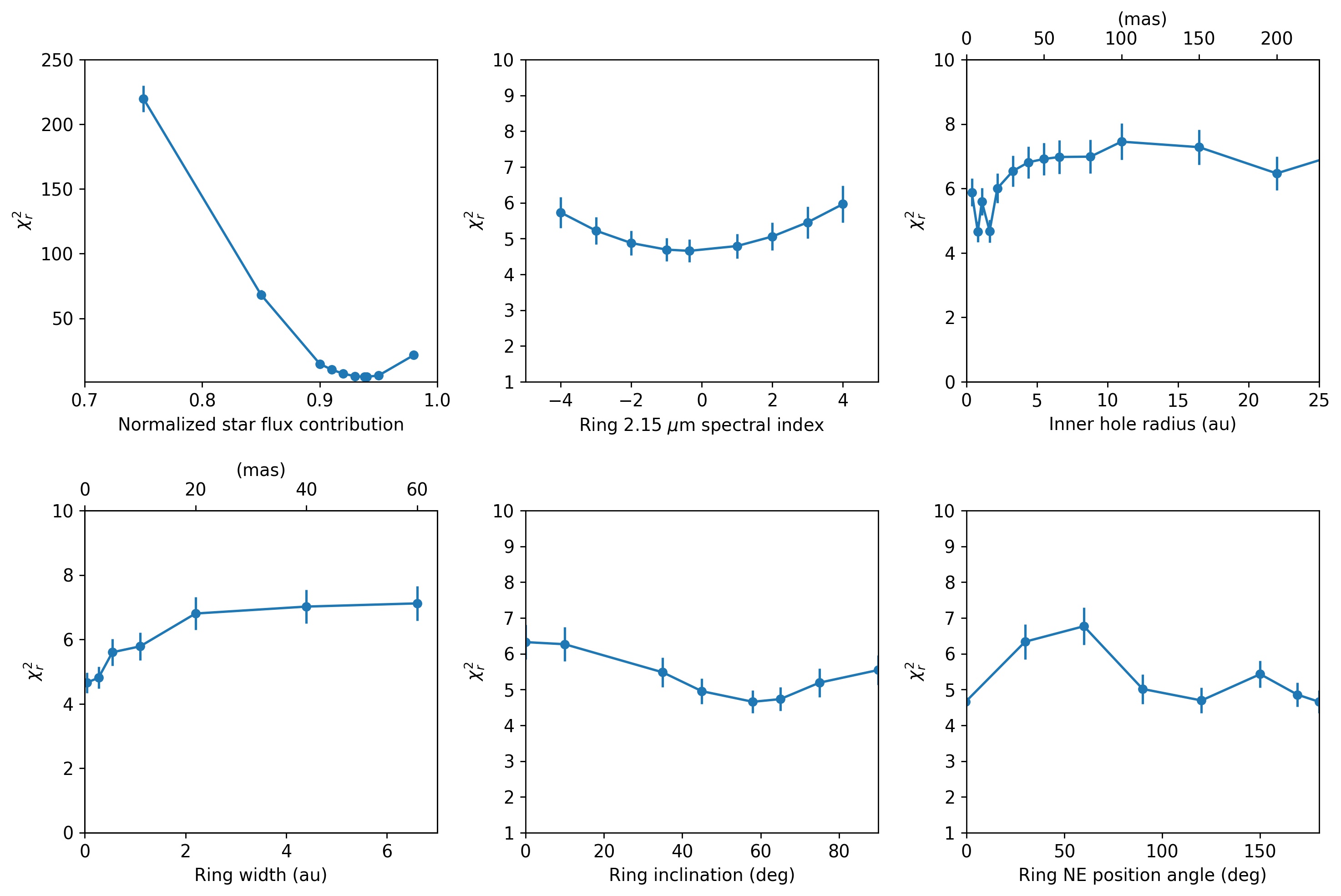}
\caption{Reduced chi-squared $\chi_r^2$ from the geometrically thin ring model squared visibility fit as a function of the different model parameters. Each point represents a squared visibility fit with the respective parameter fixed to that value while the other parameters are set free.}
\label{fig:GTRing-Chi2r_maps}
\end{figure*}
% -----

\newpage
\clearpage

\section{Spectrum wavelength calibration, telluric line correction, and photospheric absorption correction}
\label{apx:spctr-corr}

\begin{figure}
\label{fig:Transmission}
\centering
\includegraphics[width=\columnwidth]{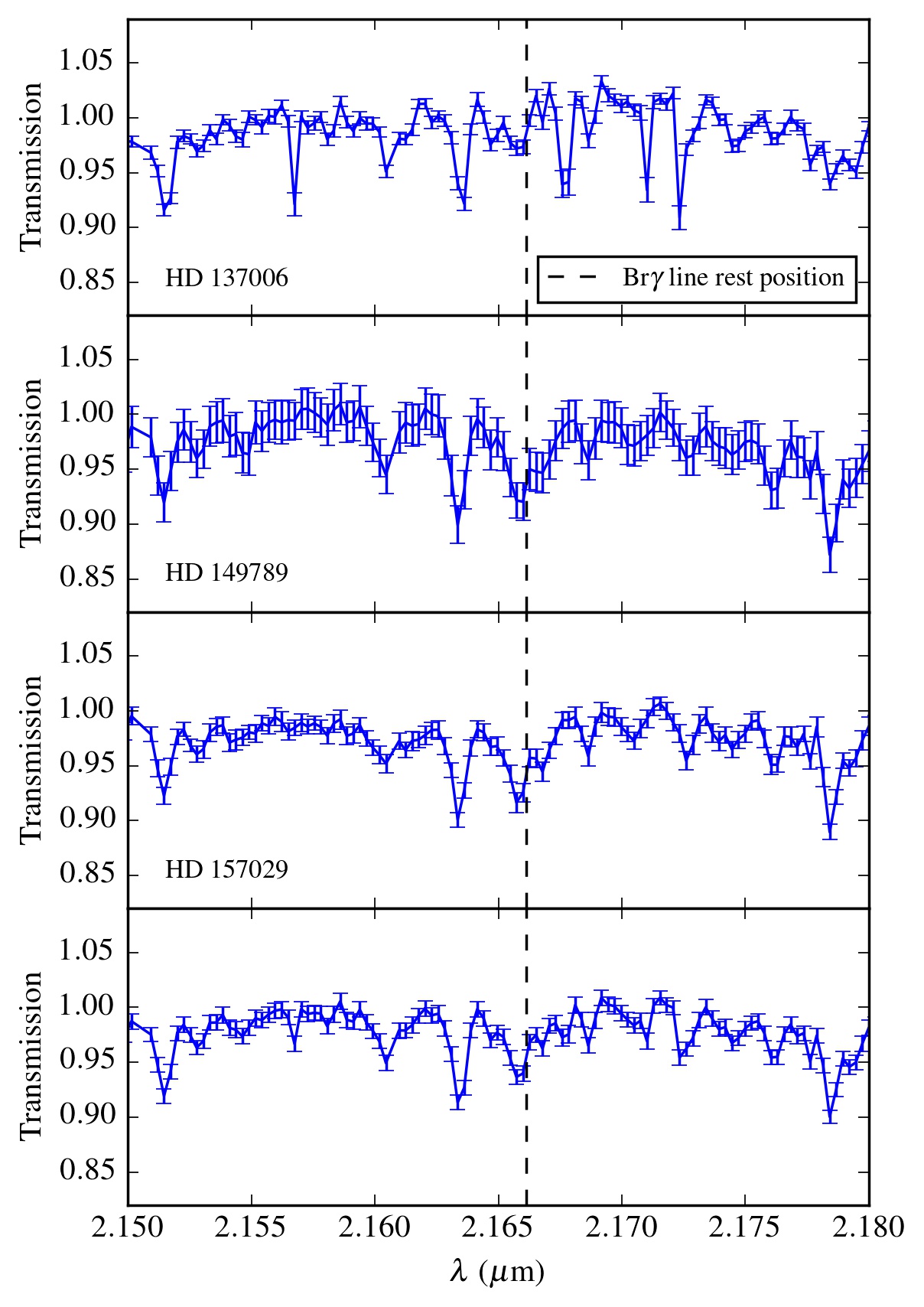}\\
\caption{Atmospheric transmission functions derived from calibrator spectra: HD\,137006 (top), HD\,149789 (center), and HD\,157029 (bottom). The last plot is the average transmission function of the  three spectra.}
\end{figure}

The 2.15-$2.18\, \mu$m part of the HD\,141569 spectrum is shown in Fig. \ref{fig:SC-Data} after continuum-normalization, wavelength calibration, telluric line correction (left panel), and photospheric absorption correction (right panel). The continuum-normalization was done in both the science object spectrum and in our inteferometric calibrator HD\,137006 spectrum ($\alpha=230.932^{\circ}$, $\delta=-01.022^{\circ}$, A5II/III SpT)  by fitting the slope of each raw spectrum and dividing the raw spectrum by the resulting fit.
The first step of the wavelength calibration was done by comparing the observed wavelength positions of the telluric lines in the HD\,141569 spectrum with respect to the positions of the telluric lines present in the IR spectrum of the atmospheric transmission above Cerro Pachon, generated using the ATRAN modeling software \citep{Lord1992} accounting for a $4.3\,$mm water vapor column and a $1.5\,$mm airmass column, available on the Gemini Observatory website. The atmospheric transmission spectrum was convolved by a Gaussian with FHWM of $6\,$\AA{} to have the same resolution as GRAVITY.
The same correction, which results in a $5\,$\AA{} blueshift, was also applied on the SC visibilities and differential phases, and on the HD\,137006 spectrum, our calibrator star. 
The calibrator was also used as a telluric spectroscopic standard. It shows a photospheric absorption feature at $2.16612\, \mu$m that was taken out before the telluric correction on the science object through a NextGen spectrum model \citep{Allard1997, Hauschildt1999}, available on the GAIA archive, with the following parameters: $7600$ K effective temperature, 6.0 surface gravity logarithm, and a $-2.0$ solar metallicity. Moreover, the spectrum shows an absorption feature at around $2.16712\, \mu$m that is not observed in the spectrum of the other calibrators observed that night and that  significantly affects the shape and the intensity of the HD\,141569 Br$\gamma$ red peak. For this reason we chose to calculate the atmospheric transmission function of the night by taking the average of three calibrator spectra, HD\,137006, HD\,149789, and HD\,157029, shown in Fig. \ref{fig:Transmission}1.
The HD\,141569 spectrum was then corrected for its radial velocity $(-6.4\,$km/s$)$ and its proper motion with respect to the local standard of rest, resulting in a correction of $-8.7\,$km/s.
Finally, the HD\,141569 spectrum was corrected for atmospheric absorption through a spectrum model from the Vienna New Model Grid of Stellar Atmospheres \citep{Heiter2002}\footnote{available on the NeMo webpage (Ch. St\"utz and E. Paunzen, \url{ http://www.univie.ac.at/nemo/})}. The model accounts for a star effective temperature of $9800$ K, a surface gravity logarithm of 4.4 \citep{Fairlamb2015},  solar metallicity, and  microturbulence of 2.0 km/s \citep{Folsom2012}. Since the star is a fast rotator having a projected linear velocity  of $v\,sin\,i = 222.0 \pm 7.0$ km/s \citep{Folsom2012}, we included the rotation broadening effect on the spectrum model using SPECTRUM \citep{Gray1994}. We decreased the intensity of the absorption model by $3\, \%$ to better fit our spectrum.

\newpage
\clearpage

\section{Pure-line visibility in presence of photospheric absorption}
\label{apx:PL-Vis}

The total visibility three-component model used to fit the SC visibilities accounts for the contributions from the star, the circumstellar dust, and the line emitting gas. The total visibility as a function of the wavelength is given by
%, and it is given by the following equation:
\begin{eqnarray}\label{eq:Apx-VL}
    V_{\rm Tot}(u,\varv,\lambda) = \nonumber \\
    \frac{\alpha(\lambda)\,F_{\rm s}(\lambda)\,V_{\rm s}(u,\varv,\lambda) + F_{\rm  c}(\lambda)\, V_{\rm  c}(u,\varv,\lambda) + F_{\rm L}(\lambda)\, V_{\rm L}(u,\varv,\lambda)}{\alpha(\lambda)\,F_{\rm s}(\lambda) + F_{\rm  c}(\lambda) + F_{\rm L}(\lambda)}.
\end{eqnarray}
The parameter $\alpha(\lambda)$ is the star continuum-normalized photospheric absorption, which implies that $\alpha(\lambda)$\,=\,1 outside the Br$\gamma$ line and $\alpha(\lambda)$\,<\,1 inside the line. This parameter is estimated from the Vienna New Model Grid of Stellar Atmospheres fitting the stellar parameters of HD\,141569, as described in Section \ref{apx:spctr-corr}. 
$F_{\rm s}(\lambda)$, $F_{\rm  c}(\lambda),$ and $F_{\rm L}(\lambda)$ are the wavelength-dependent fluxes of, respectively, the stellar continuum (i.e., outside the line and temperature-dependent), the dust ring continuum, and the 
%refers to the fluxes, V to the visibilities, the s subscript refers to the star, the r subscript to the dust ring, and the L subscript refers to the 
Br$\gamma$ line emitting gas. $F_{\rm L}(\lambda)$ varies across the line and vanishes to zero outside the line. 
$V_{\rm s}(\lambda)$, $V_{\rm  c}(\lambda),$ and $V_{\rm L}(\lambda)$ are the intrinsic visibility functions of each of the three components taken individually. 
From now on, the star is considered unresolved; therefore, $V_{\rm s}$ is equal to 1, and we drop the explicit parameter dependencies for convenience.\\
Outside the Br$\gamma$ line (i.e., in the continuum region) the total visibility is given by
\begin{equation}\label{eq:Apx-VC}
    V_{\rm Tot}^{\rm Cont} = \frac{F_{\rm s} + F_{\rm  c}\, V_{\rm  c}}{F_{\rm s} + F_{\rm  c}}
.\end{equation}
Using Eq. \ref{eq:Apx-VC} to replace $V_{\rm  c}$, we can rewrite Eq. \ref{eq:Apx-VL} as
\begin{equation}\label{eq:Apx-VL_2}
V_{\rm Tot} =  \frac{\alpha\, F_{\rm s} + V_{\rm Tot}^{\rm Cont}\, (F_{\rm s} + F_{\rm  c}) - F_{\rm s} + F_{\rm L}\, V_{\rm L}}{\alpha\, F_{\rm s} + F_{\rm  c} + F_{\rm L}}
,\end{equation}
and using the definition of the line-to-continuum flux ratio (Eq. \ref{eq:F_L/C}), Eq. \ref{eq:Apx-VL_2} becomes
\begin{equation}\label{eq:Apx-VL_3}
V_{\rm Tot} = \frac{F_{\rm s}\, (\alpha-1)}{F_{\rm Tot}}+\frac{V_{\rm Tot}^{\rm Cont}}{F_{\rm L/C}}+\frac{F_{\rm L}\, V_{\rm L}}{F_{\rm Tot}}
,\end{equation}
where $F_{\rm Tot} = \alpha\, F_{\rm s} + F_{\rm  c} + F_{\rm L}$. 
We note once again that $F_{\rm L/C}$ is the {raw} line-to-continuum ratio including the photospheric absorption. This quantity corresponds to the top left spectrum in Fig.~\ref{fig:SC-Data}. We also note that outside the line (i.e., $\alpha$=1 and $F_{\rm L}$=0), Eq.~\ref{eq:Apx-VL_3} simplifies to Eq.~\ref{eq:Apx-VC}. 
Now, making use of the parameter $\beta$ of Eq.~\ref{eq:beta}, we can write
\begin{equation}\label{eq:Apx-FL-FT}
\frac{F_{\rm L}}{F_{\rm Tot}} = 1 - \frac{\alpha + \beta}{1 + \beta} \, \frac{1}{F_{\rm L/C}}
,\end{equation}
and finally Eq. \ref{eq:Apx-VL_3} becomes
\begin{equation}\label{eq:Apx-VL_4}
V_{\rm Tot} = \frac{\alpha - 1}{ 1+ \beta}\, \frac{1}{F_{\rm L/C}} + \frac{V_{\rm Tot}^{\rm Cont}}{F_{\rm L/C}} + V_{\rm L}\, (1 - \frac{\alpha + \beta}{1 + \beta}\, \frac{1}{F_{\rm L/C}})
.\end{equation}
Solving Eq. \ref{eq:Apx-VL_4} for $V_{\rm L}$, and noting that from our data $V_{\rm Tot} = V_{\rm Tot}^{\rm Cont}$ since the SC visibilities are spectrally flat for all wavelengths and baselines, we obtain Eq. \ref{eq:Visibility-PL}:
\begin{equation}\label{eq:Apx-VL_5}
    V_{\rm L} = \frac{V_{\rm Tot}\,[(1+\beta)\,F_{\rm L/C}-1-\beta]-\alpha+1}{(1+\beta)\,F_{\rm L/C}-\alpha-\beta}.
\end{equation}
%A visualization of the function is shown in Fig. \ref{fig:PL-Vis_function} plotted as a function of $F_{\rm L/C}$ for different values of $\beta$, $\alpha$, and $V_{\rm Tot}$. There is no physical sense in considering values of $F_{\rm L/C}$ smaller than $\alpha$, since it would mean that the photospheric absorption is instead an emission.
This equation tells us that the pure-line visbility can be estimated from the total visibility in the line (which in our case is comparable to the total visibility in the continuum) if the photospheric absorption profile can be estimated, the continuum disk-to-star flux ratio $\beta$ is known and the continuum-normalized spectrum of the line is accessible. We note, in the case where $V_{\rm Tot} = V_{\rm Tot}^{\rm Cont}$, that the absence of photospheric absorption (i.e., $\alpha$=1) leads simply to $V_{\rm L}$\,=\,$V_{\rm Tot}^{\rm Cont}$=$V_{\rm Tot}$. \\
Finally, we see that the pure-line visibility $V_{\rm L}$ is lower than 1 only when $F_{\rm L/C}$ is greater than 1, which is equivalent to detecting the line above the continuum.

\newpage
\clearpage

\section{MCMax silicate model}

Here are shown three representative cases of our SED modeling through a silicate dust ring, as described in Section \ref{sec:MCMax}.
The first model accounts for particles with a lower-limit size of $1.26\,\mu$m, a dust mass of 10$^{-5}\,$M$_{\oplus}$, and $0\,$\% carbon with a resulting temperature distribution in the range  $570-1070\,$K. The second model accounts for particles with a lower-limit size of $40\,\mu$m, a mass of 1.17$\times 10^{-4}\,$M$_{\oplus}$, and $0\,$\% carbon with a resulting temperature distribution in the range  $620-840\,$K. The  last model accounts for particles with a lower-limit size of $40\,\mu$m, a mass of 3.33$\times 10^{-4}\,$M$_{\oplus}$, and $25\,$\% carbon with a resulting temperature distribution in the range  $440-770\,$K. We note that in all three cases the near-IR excess is smaller than that obtained through the GRAVITY data analysis, while the mid-IR emission exceeds the photometry data. Decreasing the mid-IR excess would at the same time decrease further the near-IR value, suggesting that a pure silicate innermost ring model is not compatible with the nature of the HD\,141569 system.

\begin{figure}[!htbp]
\centering
\includegraphics[width=\columnwidth]{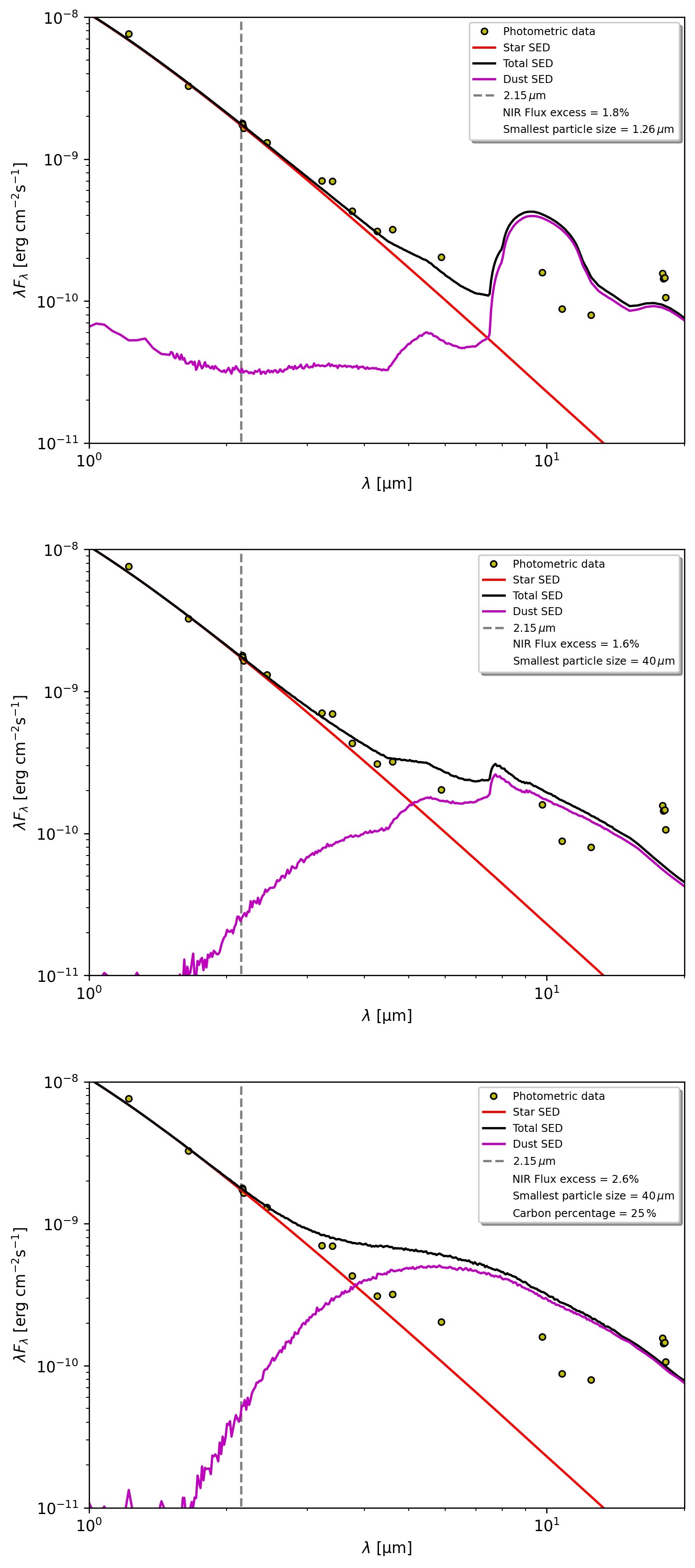}
\caption{MCMax models that account for the emission of the star and only the first innermost silicate ring located at 0.8\,au and  0.04\,au in width. The models differ for the particle size lower limit,  dust mass, and carbon percentage. The top plot model accounts for a dust mass of 10$^{-5}\,$M$_{\oplus}$ and no carbon; the center plot model accounts for a mass of 1.17$\times 10^{-4}\,$M$_{\oplus}$ and no carbon; the bottom plot model accounts for a mass of 3.33$\times 10^{-4}\,$M$_{\oplus}$ and 25\,\% carbon.}
\label{fig:MCMax-Sil-model}
\end{figure}

\newpage
\clearpage

\section{MCMax model density and temperature structure for the QHP model}

\begin{figure*}[!hb]
\centering
\includegraphics[width=\textwidth]{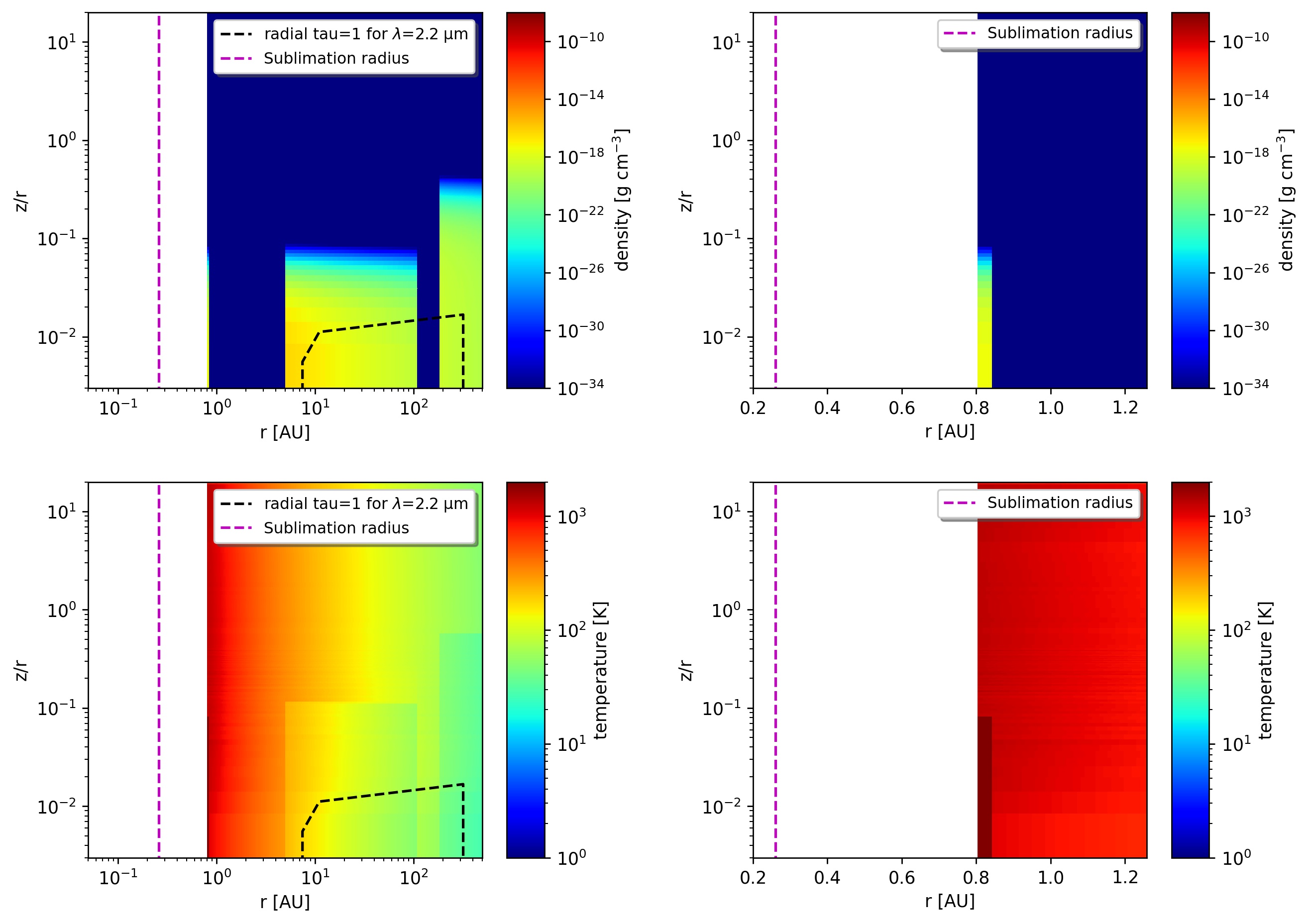}
\caption{Density and temperature structure of the QHP model described in Section \ref{sec:MCMax} and Table \ref{tab:MCMax}. The x-axis shows the radius in au, the y-axis the height divided by the radius. The colors of the top plots indicate the mass density, the colors of the bottom plot the temperature. QHPs are located at 0.8\,au in a very narrow (0.04\,au) ring. The optical depth at the observation wavelength is indicated by the radial (dotted line) $\tau = 1$ surface. The magenta dashed line represents the sublimation radius of the system. The dark red color in the temperature plot refers only to the location of the QHPs and not their temperature, since QHPs do not have an equilibrium temperature value.}
\label{fig:MCMax-model_DT}
\end{figure*}

%\begin{figure}
%\centering
%\includegraphics[width=\columnwidth]{Images/SC/Vl.jpg}\\
%\caption{Eq. \ref{eq:Apx-VL_5} plotted as a function of $F_{\rm L/C}$ for different values of $\beta$ (top), $\alpha$ (center), and $V_{\rm Tot}$ (bottom).}
%\label{fig:PL-Vis_function}
%\end{figure}

\newpage
\clearpage

\section{Photometric data}

\begin{table}[htbp!]
\centering
\footnotesize
\caption{\centering HD\,141569 photometric data.}
\begin{tabular}{lccc} 
\hline
\hline
\multicolumn{1}{l}{Band} & \multicolumn{1}{c}{$\lambda$} & \multicolumn{1}{c}{Flux} & \multicolumn{1}{c}{Beam size and reference}\\
\multicolumn{1}{l}{} & \multicolumn{1}{c}{[$\mu$m]} & \multicolumn{1}{c}{[Jy]} & \multicolumn{1}{c}{}\\
\hline

IUE       & 0.138 & 0.30               & archival data               \\
IUE       & 0.178 & 0.73               & archival data               \\
IUE       & 0.218 & 0.628              & archival data               \\
IUE       & 0.257 & 1.023              & archival data               \\
IUE       & 0.29  & 1.274              & archival data               \\
U         & 0.36  & 4.19               & \cite{Sylvester1996}        \\
B         & 0.436 & 8.37               & \cite{Sylvester1996}        \\
V         & 0.55  & 7.36               & \cite{Sylvester1996}        \\
R         & 0.708 & 5.92               & \cite{Sylvester1996}        \\
I         & 0.977 & 4.81               & \cite{Sylvester1996}        \\
J         & 1.22  & 3.1                & 2Mass                       \\
H         & 1.65  & 1.8                & 2Mass                       \\
%K         & 2.16  & 2.07$\pm 0.05$     & 2Mass                       \\
K         & 2.16  & 1.29               & 2Mass                       \\
K         & 2.16  & 1.26$\pm 0.03$     & 2Mass                       \\
K         & 2.16  & 1.25               & 2Mass                       \\
K         & 2.18  & 1.2                & 2Mass                       \\
K         & 2.19  & 1.22$\pm 0.02$     & 2Mass                       \\
K         & 2.19  & 1.23               & \cite{Penprase1992}         \\
K         & 2.24  & 1.3$\pm 0.04$      & \cite{Malfait1998}          \\
ISO       & 2.45  & 1.07               & ESA archive                 \\
ISO       & 3.23  & 0.76               & ESA archive                 \\
WISE      & 3.4   & $0.79\pm 0.025$    & 6.1" NASA archive           \\
UKIRT     & 3.76  & 0.54               & \cite{Sylvester1996}        \\
ISO       & 4.26  & 0.44               & ESA archive                 \\
WISE      & 4.6   & $0.49\pm 0.01$     & 6.4"                        \\
ISO       & 5.89  & 0.40               & ESA archive                 \\
ISO       & 6.76  & 0.43               & ESA archive                 \\
ISO       & 7.76  & 0.82               & ESA archive                 \\
ISO       & 8.70  & 0.62               & ESA archive                 \\
AKARI     & 9.0   & $0.5178\pm 0.0104$ & NASA archive                \\
ISO       & 9.77  & 0.52               & ESA archive                 \\
ISO       & 10.7  & 0.58               & ESA archive                 \\
OSCIR     & 10.8  & $0.318\pm 0.016$   & \cite{Fisher2000}           \\
Michelle  & 11.2  & $0.338\pm 0.034$   & \cite{Moerchen2010}         \\
ISO       & 11.48 & 0.635              & 14'$\times$20' ESA archive  \\
IRAS      & 12.0  & $0.55\pm 0.04$     & 1'$\times$5' NASA archive   \\
WISE      & 12.0  & $0.38\pm 0.006$    & 6.5' NASA archive           \\
MIRLIN    & 12.5  & $0.333\pm 0.022$   & \cite{Marsh2002}            \\
MIRLIN    & 17.9  & $0.936\pm 0.094$   & \cite{Marsh2002}            \\
AKARI     & 18    & $0.8655\pm 0.0168$ & NASA archive                \\
Michelle  & 18.1  & $0.883\pm 0.147$   & \cite{Moerchen2010}         \\
OSCIR     & 18.2  & $0.646\pm 0.035$   & \cite{Fisher2000}           \\
MIRLIN    & 20.8  & $1.19\pm 0.16$     & \cite{Marsh2002}            \\
WISE      & 22    & $1.44\pm 0.027$    & 12" NASA archive            \\
MIPS      & 24.0  & $1.47\pm 0.01$     & 6" Spitzer archive          \\
IRAS      & 25    & $1.87\pm 0.13$     & 1'$\times$5' NASA archive   \\
IRAS      & 60    & $5.54\pm 0.49$     & 2'$\times$5' NASA archive   \\
PACS-Spec & 63.2  & $2.98\pm 0.01$     & \cite{Thi2014}              \\
MIPS      & 70    & $4.70\pm 0.02$     & 18" Spitzer archive         \\
PACS-Spec & 72.8  & $3.91\pm 0.03$     & \cite{Thi2014}              \\
PACS-Spec & 76.4  & $3.30\pm 0.03$     & \cite{Thi2014}              \\
PACS-Spec & 90    & $2.80\pm 0.03$     & \cite{Thi2014}              \\
IRAS      & 100   & $3.48\pm 0.35$     & 4'$\times$5' NASA archive   \\
PACS-Spec & 145   & $1.1\pm 0.1$       & \cite{Thi2014}              \\
PACS-Spec & 158   & $1.18\pm 0.02$     & \cite{Thi2014}              \\
PACS-Spec & 180   & $0.83\pm 0.04$     & \cite{Thi2014}              \\
SCUBA     & 450   & $0.0649\pm 0.0133$ & \cite{Sandell2011}           \\
SCUBA     & 850   & $0.0140\pm 0.0020$ & \cite{Sandell2011}           \\
LABOCA    & 870   & $0.0126\pm 0.0046$ & \cite{Nilsson2010}          \\
MAMBO     & 1200  & $0.0047\pm 0.0005$ & \cite{Meeus2012}            \\
SCUBA     & 1350  & $0.0054\pm 0.0001$ & \cite{Sylvester2001}        \\

\hline
\end{tabular}\\
\vspace{2ex}
{\raggedright \footnotesize{ \textbf{Notes}. Data without references are  from \cite{Merin2004}.} \par}
\label{tab:K-flux}
\end{table}

\section{Inclination and position angle of outer disks from the literature}
\label{apx:inc-PA}

% ----
\begin{table}[thpb!]
\footnotesize
\centering
\caption{\centering Inclination and position angle of the HD\,141569 outer rings.}
\begin{tabular}{lccc} 
\hline
\hline
Reference & Radius & i & PA$_{NE}$ \\
\hline
 & [au] & [deg] & [deg] \\
\hline
Dust components & & & \\
\hline
\cite{Augereau1999}             & $361\pm10$    & $52.5\pm4.5$ & $-4.6\pm1.0$                 \\
\cite{Weinberger1999}   & $206$                         & $51\pm3$              & $-4.0\pm5.0$            \\
\cite{Biller2015}               & $232\pm3$     & $44.9\pm0.5$ & $-8.9\pm1.3$                 \\
                                                                        & $385\pm13$      & $47.3\pm3.3$ & $-11.3\pm6.1$  \\
\cite{Perrot2016}                       & $185\pm1$     & $56.9\pm1.0$ & $-3.9\pm0.4$            \\
                                                                        & $88\pm2$        & $57.6\pm1.3$ & $-4.0\pm2.0$           \\
                                                                        & $61\pm1$        & $56.0\pm2.2$ & $-5.5\pm1.0$           \\
                                                                        & $45\pm1$        & $57.9\pm1.3$ & $-6.3\pm1.1$           \\
\cite{Currie2016}               & $37\pm2$      & $56.0\pm4.0$ & $-1.2\pm2.4$           \\
\cite{White2016}                & $81$                  & $55$                          & $-8.8$                          \\
\cite{Mawet2017}                & $37\pm4$      & $53.0\pm6.0$ & $-11.0\pm8.0$         \\
This work                       & $\sim 1$      & $58.5\pm1.6$ & $-1.8\pm1.1$         \\
\hline
Gas components & & & \\
\hline
\cite{White2016}                        & $81-199$              & $53.4\pm1.0$ & $-3.4\pm0.6$  \\
\cite{DiFolco2020}                      & $17-277$              & $56-58$                 & $-4\pm1$              \\
                                                        &       $35-232$                 & $53\pm2$              & $0\pm2$               \\
This work                       & $0.01-0.09$       & $58.5$ & $-10 \pm 7$         \\      
\hline
\end{tabular}
\label{tab:i-PA}
\end{table}
% -----

\newpage
\clearpage

%\section{Keplerian ring model photo-center shifts and differential phases}
%
%\begin{figure*}
%\centering
%\includegraphics[width=\textwidth]{Images/SC/RP_PhShift.jpg}
%\caption{Monochromatic images (from $2.16352$ to $2.16872\,\mu$m, i.e., from $-360$ to $360\,$km/s) %of the Keplerian ring model described in Section \ref{sec:SC-Results}. Colored dashed lines refer to %the different GRAVITY baselines. Circles represent the 2-D photo-center shift for each baseline.}
%\label{fig:KM_Ph-shift}
%\end{figure*}

\end{document}